\documentstyle[epsf]{mnusa}
\input{epsf.tex}

  \def\degree{{$^\circ$}}

  \def\puncspace{\ifmmode\,\else{\ifcat.\C{\if.\C\else%
  \if,\C\else\if?\C\else\if:\C\else\if;\C\else\if-\C\else%
  \if)\C\else\if/\C\else\if]\C\else\if'\C%
  \else\space\fi\fi\fi\fi\fi\fi\fi\fi\fi\fi}%
  \else\if\empty\C\else\if\space\C\else\space\fi\fi\fi}\fi}%
  \def\SP{\let\\=\empty\futurelet\C\puncspace}

  \def\ee#1{\ifmmode {} \times 10^{#1} \else ${} \times 10^{#1}$\fi}
  \def\sub#1{\ifmmode _{#1} \else $_{#1}$\fi}
  \def\sup#1{\ifmmode ^{#1} \else $^{#1}$\fi}

  \def\about{\ifmmode \sim \else {$\sim\,$}\fi}
  \def\lta{\ifmmode {\,\mathbin{\lower 3pt\hbox   
      {$\,\rlap{\raise 5pt\hbox{$\char'074$}}\mathchar"7218\,$}}}
      \else {${\mathbin{\lower 3pt\hbox
      {$\rlap{\raise 5pt\hbox{$\char'074$}}\mathchar"7218\,$}}}
      $}\fi}
  \def\gta{\ifmmode {\mathbin{\lower 3pt\hbox   
      {$\,\rlap{\raise 5pt\hbox{$\char'076$}}\mathchar"7218\,$}}}
      \else {${\mathbin{\lower 3pt\hbox
      {$\rlap{\raise 5pt\hbox{$\char'076$}}\mathchar"7218\,$}}}
      $}\fi}

   \mathcode`*="002A   

  \def\degree{{\ifmmode ^\circ \else $^\circ$\fi}}

  \newcommand{\cir}[1]{\leavevmode\hbox{Cir~X-#1}\SP}
  
  \newcommand{\fu}[1]{\leavevmode\hbox{4U~#1}\SP}
  \newcommand{\gx}[1]{\leavevmode\hbox{GX~#1}\SP}

  \newcommand{\sco}[1]{\leavevmode\hbox{Sco~X-#1}\SP}

  \def\mdot{{\ifmmode \dot M \else {$\dot M$}\fi}}
  \def\mdote{{\ifmmode \dot M_E \else {$\dot M_E$}\fi}}
  \def\mdoti{{\ifmmode \dot M_i \else {$\dot M_i$}\fi}}
  \def\msun{{\ifmmode M_\odot \else {$M_{\odot}$}\fi}}

  \def\agt{\,\raise.3ex\hbox{$>$\kern-.75em\lower1ex\hbox{$\sim$}}\,}
  \def\alt{\,\raise.3ex\hbox{$<$\kern-.75em\lower1ex\hbox{$\sim$}}\,}

  \def\nub{\nu_{\rm b}}
  \def\nuk{\nu_{\rm K}}
  \def\nuphi{\nu_{\phi}}
  \def\nur{\nu_{\rm r}}
  \def\rp{r_{\rm p}}
  \def\Rp{r_{\rm p}}
  
  \def\ra{r_{\rm a}}

  \def\re{R_{\rm e}}
  \def\rs{R_{\rm s}}
  \def\risco{r_{\rm isco}}
  \def\rmb{r_{\rm mb}}
  
  \def\nupa{\nu_{_{\rm AP}}}
  \def\nus{\nu_{_{\rm s}}}
  
  \def\nunp{\nu_{_{\rm NP}}}

  \def\chisq{{\hat\chi}^2}
  \def\chisqdof{{\chi^2/{\rm dof}}}
  
  \def\nuap{{\nu_{_{\rm AP}}}}
  \def\nub{\nu_{_{\rm b}}}
  \def\nuhbo{{\nu_{\rm HBO}}}
  \def\nuhboi{{\nu_{_{{\rm HBO},i}}}}
  
  \def\sighboisq{{\sigma^2_{_{{\rm HBO},i}}}}

  \def\nuki{{\nu_{_{{\rm K},i}}}}
  
  \def\nunpi{{\nu_{_{{\rm NP},i}}}}
  \def\nuapi{{\nu_{_{{\rm AP},i}}}}
  \def\nur{{\nu_{\rm r}}}

  \def\nuorb{{\nu_{_{\rm orb}}}}
  \def\nuh{{\nu_{_{\rm H}}}}
  \def\nuk{{\nu_{_{\rm K}}}}
  \def\nul{{\nu_{_{\rm L}}}}
  
  \def\Dnu{\Delta\nu}

 \def\rai{{r_{{\rm a},i}}}
 \def\rpi{{r_{{\rm p},i}}}
 \def\ramax{{r_{\rm a,max}}}
 \def\ramax{{r_{\rm a,max}}}
 \def\ramin{{r_{\rm a,min}}}

\title{Eccentric orbits and QPOs in neutron star X-ray binaries}
\author[D. Markovi\'c and F.~K. Lamb]
 {D. Markovi\'c and F.~K. Lamb\thanks{Also Department of Astronomy.}\\
 Center for Theoretical Astrophysics and Department of Physics,
 University of Illinois at Urbana-Champaign,
 1110 West Green Street, \\ Urbana, IL 61801, USA;
 markovic@mail.physics.uiuc.edu, f-lamb@uiuc.edu}
\date{Submitted 2000 March 23}
\pagerange{000--000}
\pubyear{2000}

\begin{document}
\maketitle

\begin{abstract}
We investigate the suggestion that the frequencies of the
kilohertz and other QPOs in the emission from neutron stars
in low-mass X-ray binaries are generated by geodesic motion
of gas clumps around the star. First we assume, following
previous work, that the dominant frequencies produced by
such motions are the azimuthal frequency $\nuk$, the apsidal
precession frequency $\nuap$, and the first overtone of the
nodal precession frequency $\nunp$. We investigate whether
geodesics can be found for which these frequencies agree
with the observed frequencies. Correcting calculational
errors made when the geodesic precession hypothesis was
first proposed, we find that $\nuk$ and $\nuap$ disagree
qualitatively with the frequencies of the kilohertz QPOs for
infinitesimally or even moderately eccentric geodesics.
These frequencies are similar to the frequencies of the
kilohertz QPOs only for highly eccentric geodesics, with
apastron to periastron ratios $\sim$3--4; for these
geodesics, $2\nunp$ differs qualitatively from the
frequencies of the low-frequency QPOs. Next we investigate
whether these frequencies would be the dominant frequencies
produced by orbiting clumps. We find that they are not and
that the dominant frequencies are instead harmonics and
sidebands of the radial epicyclic frequency $\nur$. Finally,
we show that gas dynamical constraints restrict the radii of
orbiting clumps to be $\ll 10^{-3}$ of the stellar radius;
the fractional modulation that could be produced by clumps
is therefore $\ll 10^{-6}$. We conclude that there are
significant difficulties in interpreting the kilohertz and
low-frequency QPOs as consequences of geodesic motion of gas
clumps.
\end{abstract}

\begin{keywords}
accretion -- relativity -- stars: neutron -- X-rays: stars
\end{keywords}

\section{Introduction}
\label{intro}

Observations of low mass X-ray binaries (LMXBs) using the
{\em Rossi X-ray Timing Explorer (RXTE)\/} satellite have
led to the discovery of prominent, narrow features in power
spectra of the accretion-powered X-ray emission of these
sources, with frequencies ranging from $\sim$5~Hz to
$\sim$1300~Hz (for a review, see van der Klis 2000).
High-frequency quasi-periodic oscillations (QPOs) have been
observed with frequencies ranging from $\sim$500~Hz to
$\sim$1300~Hz. These `kilohertz QPOs' often appear as a pair
of peaks with frequencies $\nu_1$ and $\nu_2$ ($>\nu_1$)
that move together and can shift up and down in frequency by
up to a factor of two within a few hundred seconds, probably
because of changes in the accretion rate. As they shift,
$\nu_1$ and $\nu_2$ follow a narrow track in the
$\nu_1$-$\nu_2$ plane that is fixed for a given source. The
kilohertz QPOs have frequencies in the range expected for
orbital motion near neutron stars and are very likely a
strong-field general-relativistic phenomenon.

In six kilohertz QPO sources, oscillations with frequencies
of hundreds of Hertz have been observed during X-ray bursts.
These `burst oscillations' are almost certainly caused by
rotation of brighter regions of the stellar surface at the
spin frequency of the star (Strohmayer et al.\ 1996, 1998;
Bildsten 1998; Miller 1999; Strohmayer 1999). Observations
indicate that the principal burst oscillation frequency
$\nub$ is the stellar spin frequency $\nus$ or its first
overtone (see, e.g., Strohmayer \& Markwardt 1999; Miller
1999).

In four sources, burst oscillations and two simultaneous
kilohertz QPOs have both been detected with high confidence.
In 4U~1702$-$43 (Markwardt, Strohmayer \& Swank 1998),
4U~1728$-$34 (Strohmayer et al.\ 1996; M\'endez \& van der
Klis 1999), and 4U~1636$-$53 (Zhang et al.\ 1996; Wijnands
et al.\ 1997a), the difference $\Dnu \equiv \nu_2-\nu_1$
between the frequencies of the two kilohertz QPOs is equal
(to within 4--15\%) to $\nub$. In KS~1731$-$26 (Wijnands \&
van der Klis 1997), $\Dnu$ is equal (to within $\alt0.7$\%)
to one-half $\nub$. Careful analysis has shown that in
\hbox{4U~1636$-$536}, $\Dnu$ is slightly but significantly
smaller than the spin frequency inferred from its burst
oscillations (M\'endez, van der Klis \& van Paradijs 1998).

In some kilohertz QPO sources, the separation frequency
$\Dnu$ decreases systematically by 30--100~Hz, depending on
the source, as $\nu_2$ increases by a much larger amount
(\hbox{Sco~X-1}: van der Klis et al.\ 1997;
\hbox{4U~1608$-$52}: M\'endez et al.\ 1998;
\hbox{4U~1735$-$44}: Ford et al.\ 1998; \hbox{4U~1728$-$34}:
M\'endez \& van der Klis 1999; see also Psaltis et al.\
1998).

The fact that $\nub$ is close to one or two times $\Dnu$ in
four sources as well as other evidence motivated the
development of the sonic-point beat-frequency (SPBF) model
(Miller, Lamb \& Psaltis 1998; Lamb \& Miller 2000). In this
model the frequency $\nu_2$ of the upper kilohertz QPO is
close to but less than the general-relativistic orbital
frequency $\nuorb$ at the sonic radius, where the flow in
the disc changes from nearly circular to rapidly
inspiraling, whereas the frequency $\nu_1$ of the lower
kilohertz QPO is comparable to but less than the beat
frequency $\nu_{\rm B} \equiv \nuorb-\nus$ of the neutron
star's spin frequency with $\nuorb$. The transition to
hypersonic radial inflow that occurs at the sonic radius
$\rs$ in this model is a strong-field general relativistic
effect. In the SPBF model, the low-frequency QPOs seen in
the Z and atoll sources are produced by interaction of the
stellar magnetic field with the disc flow well outside the
sonic point, via the magnetospheric beat-frequency mechanism
(Alpar \& Shaham 1985; Lamb et al.\ 1985).

The SPBF model explains naturally the existence of only two
principal kilohertz QPOs in a given source, the near
commensurability of $\Dnu$ and $\nub$, and the high
frequencies, coherence, and amplitudes of these QPOs. The
SPBF model predicts that $\Dnu$ is less than but close to
$\nus$ and explains naturally the decrease in $\Dnu$ with
increasing $\nu_2$ seen in some sources (Lamb \& Miller
2000). The inward drift of the accretion flow near $\re$
causes $\nu_1$ to increase faster than $\nu_2$ as the
accretion rate increases. Depending on the source, a
differential increase of 7\% is sufficient to explain the
largest observed decrease in $\Delta\nu$.

A strong prediction of the SPBF model is that the neutron
star in 4U~1636$-$536 is not spinning at the $\sim$580~Hz
frequency of the principal burst oscillation in this source
but instead at $\sim$290~Hz. This prediction ran counter to
what was widely believed at the time (see, e.g., Stella \&
Vietri 1999), but was confirmed by detection of a
$\sim$290~Hz subharmonic with an amplitude equal to 40\% of
the amplitude of the $\sim$580~Hz oscillation (Miller 1999).
The SPBF model also predicts that the $\nu_2$-$\dot M$
relation may flatten in some sources when the region of
nearly circular flow reaches the innermost stable circular
orbit (ISCO). This signature of the ISCO may have
subsequently been observed in 4U~1820$-$30 (Zhang et al.\
1998; Kaaret et al.\ 1999). In the SPBF model, the kilohertz
QPOs are a strong-field general relativistic effect and are
therefor sensitive probes of the properties of the spacetime
near the neutron star, including whether there is an ISCO
and the gravitomagnetic torque produced by the spin of the
star. Hence, if the model is shown to be correct,
measurements of the kilohertz QPOs can be used not only to
determine the properties of neutron stars but also to
explore gravitational effects in the strong-field regime.

Following the development of the SPBF model, Stella \&
Vietri (1998; see also Morsink \& Stella 1999) proposed that
the 10--60~Hz low-frequency QPOs (LFQPOs) and bumps observed
in power spectra of the X-ray emission from the atoll
sources and, possibly, the horizontal branch oscillations
(HBOs) observed in the Z sources (which are observed
simultaneously with the kilohertz QPOs; see van der Klis
2000) might be the first or second harmonics of the nodal
precession frequency $\nunp$ of gas clumps moving on
slightly inclined, {\em infinitesimally eccentric
geodesics\/} (IEGs) with azimuthal frequencies equal to
$\nu_2$, around neutron stars with the 250--350~Hz spin
rates inferred from the burst oscillations and the SPBF
model. For such stars, nodal precession is driven by the
prograde gravitomagnetic (Lense-Thirring) torque, but is
partially offset by the retrograde torque produced by the
star's rotation-induced quadrupolar distortion.

Recent work on the global modes of viscous accretion discs
(Markovi\'c \& Lamb 1998) has demonstrated that such discs
have a dense spectrum of gravitomagnetically precessing
modes that are localised near the inner boundary of the
region of nearly circular flow. The highest-frequency of
these modes precesses with a frequency only slightly lower
than the nodal precession frequency $\nunp$ of a test
particle orbiting at the same radius. Contrary to what had
been expected since the pioneering work of Bardeen \&
Petterson (1975), the first few of these modes are very
weakly damped and may therefore be excited, producing
detectable variations in the X-ray emission (Markovi\'c \&
Lamb 1998; see also Armitage \& Natarajan 1999). However,
further work (Psaltis et al.\ 1999a; Morsink \& Stella 1999;
Kalogera \& Psaltis 1999) has shown that the frequencies and
frequency behaviors of the LFQPOs and HBOs are inconsistent
with the proposal of Stella \& Vietri (1998) that these
frequencies are the first or second harmonics of the nodal
precession frequency of the IEG with azimuthal frequencies
$\sim\nu_2$ around stars with the 250--350~Hz spin rates
inferred from the burst oscillations and the SPBF model.

\begin{table}
\begin{center}
\begin{minipage}{130mm}
\caption{Proposed QPO frequency identifications.
\label{table.models}} 
\begin{tabular}{ll}
   \hline \hline
   QPO Frequency
   &Geodesic Frequency\\
   \hline
   $\nuhbo$ or $\nul$
     & Nodal precession frequency $\nunp$ or $2\nunp$
    \\
   $\nu_1$ 
      & Apsidal precession frequency $\nuap$ 
   \\
   $\nu_2$ 
      & Azimuthal frequency $\nuk$ 
   \\
   \hline \hline
\end{tabular}
\end{minipage}
\end{center}
\end{table}

Stella \& Vietri (1999; hereafter SV) subsequently proposed
that $\nu_2$ and $\nu_1$ are the azimuthal and apsidal
precession frequencies $\nuk$ and $\nuap$ of gas clumps
orbiting the star on eccentric geodesics; $\nuap = \nuk -
\nur$ in terms of the radial epicyclic frequency $\nur$.
They again proposed that the frequencies $\nuhbo$ and $\nul$
of the low-frequency QPOs and bumps seen in power spectra of
the Z and atoll sources are the first or second harmonics of
the nodal precession frequency $\nunp$ of these same
geodesics, but argued that $\nus$ is may be $\gg300\,$Hz to
give better agreement of $\nunp$ with $\nuhbo$ and $\nul$.
The frequency identifications proposed by SV are listed in Table~\ref{table.models}. SV simply assumed that
radiating or obscuring gas clumps with the required
properties form near the star; that hydrodynamic, radiation,
and magnetic forces are negligible, so that the gas clumps
move on geodesics; that $2\nunp$, $\nuap$, and $\nuk$ are
the dominant frequencies generated by clumps in purely
geodesic motion about a neutron star; and that all the
clumps move on geodesics with the same periastron and
apastron radii $\rp$ and $\ra$. The large variations of
$\nuhbo$ or $\nul$, $\nu_1$, and $\nu_2$ that occur in the
individual sources are attributed to gas clumps being fed
onto different geodesics at different times.

SV first pointed out that the frequencies $\nuap$ and $\nuk$
of infinitesimally eccentric geodesics (IEGs) around
nonrotating stars with masses of 1.8--$2.2\,\msun$ are
comparable to the values of $\nu_1$ and $\nu_2$ observed in
\sco1, although they do not track the tightly correlated
variation of $\nu_1$ and $\nu_2$ (see their Fig.~1). Stella
\& Vietri then argued that sequences of moderately eccentric
geodesics (MEGs) would give apsidal precession and azimuthal
frequencies more consistent with the observed frequencies of
the kilohertz QPOs.

Once geodesics of finite eccentricity are allowed, the
geodesic precession hypothesis does not by itself predict a
relation between $\nuap$ and $\nuk$. The reason is that the
characteristic frequencies of such geodesics are functions
of two parameters: $\rp$ and $\ra$. The set of geodesics
that satisfy $\ra \ge \rp > \re$ generate frequencies that
cover a wide area in the two-dimensional $\nuap$-$\nuk$
plane, rather than a one-dimensional curve. In order to
obtain a $\nuap$-$\nuk$ relation that could explain the
tightly correlated variation of $\nu_1$ and $\nu_2$, some
physical constraint is needed that will convert the
two-parameter set of allowed geodesics to a one-parameter
allowed sequence of geodesics that generates a curve in the
$\nuap$-$\nuk$ plane relating $\nuap$ and $\nuk$.

SV addressed this issue by proposing that some physical
mechanism keeps $\rp$ constant in a given source while $\ra$
varies. They suggested that $\rp$ may remain constant
because it is equal to the radius $\risco$ of the innermost
stable circular orbit, to the radius $r_m$ at which the
magnetic field of the star first couples to the accreting
gas, or to the equatorial radius $\re$ of the neutron star.
However, the periastron radius of an eccentric geodesic can
be significantly smaller than $\risco$ (see Markovi\'c 2000)
and the magnetic coupling radius $r_m$ varies as the
accretion rate varies (see Lamb 1989).

SV reported that they were able to construct MEG sequences
with apsidal precession and azimuthal frequencies that agree
approximately with the kilohertz QPO frequencies observed in
\sco1, assuming that $\rp$ is constant but can have any
value greater than $\risco$. However, as we explain in
Section~\ref{CompareMEGs} and the Appendix, SV computed
$\nuk$ incorrectly. Karas (1999) noticed this error (see his
footnote 2), but thought that its effect was small because
the geodesics that SV considered have small eccentricities
$\epsilon \sim 0.1$, which corresponds to $\ra/\rp = 1.2$.
However, in order to fit the data, $\nur$ must be
$\lta0.3\,\nu_2$. A given clump therefore makes only a small
fraction of its full radial excursion during each orbit of
the star and hence---in contrast to the behavior of
particles in similar Newtonian orbits---the variation with
radius of its rate of azimuthal phase advance does not
average out. The error made in SV therefore produced a 10\%
($\approx100\,$Hz) error in $\nuk$, which is very large
compared to the uncertainties in $\nu_2$ for \sco1. When the
correct values of $\nuk$ are used, the MEG sequences cited
by SV give $\nuap$-$\nuk$ relations that disagree
qualitatively with the \sco1 frequency data (see
Section~\ref{CompareMEGs}). SV did not consider the
frequency data available on other sources.

Motivated by the proposal of SV, Karas (1999) investigated
whether geodesic sequences can be constructed which give
$\nur$-$\nuk$ relations that agree with the $\Dnu$-$\nu_2$
correlations observed in \sco1 and \hbox{4U~1608$-$52}, if
$M$ is treated as a free parameter for each source and $\rp$
and $\ra$ are freely chosen for each measurement of $\nu_1$
and $\nu_2$. Karas also investigated whether geodesic
sequences can be constructed which give $\nur$-$\nuap$
relations that agree with the $\Dnu$-$\nu_2$ correlations
observed. This assumes that the principal kilohertz QPO
frequencies are $\nuk-2\nur$ and $\nuk-\nur$; its physical
motivation is unclear. Karas considered only nonrotating
masses (he did not construct any stellar models) and allowed
highly eccentric geodesics ($\epsilon$ up to 0.4, which
corresponds to $\ra/\rp = 2.3$). Not surprisingly, given
that varying $\rp$ and $\ra$ allows coverage of a large area
of the $\nur$-$\nuk$ plane (see above), he found that fits
are possible, although he did not report the parameters or
the value of $\chisqdof$ for any individual fits. Such fits
imply that $\ra$ is always a function of $\rp$, i.e., that
specifying $\rp$ uniquely specifies a particular eccentric
geodesic. Moreover, the function $\ra=\ra(\rp)$ must be
different for every source. The physical motivation for this
is unclear. For some fits, Karas assumed that a clump only
needs to complete 92\%--97\% of the radial extent of its
geodesic, because of the clump's finite size. The physical
basis of this assumption is unclear.

Psaltis, Belloni \& van der Klis (1999b) have pointed to a
possible correspondence between features in the power
spectra of black holes and neutron stars in LMXBs,
emphasising the similar properties of the 0.1--100~Hz QPOs
observed in the atoll sources, \cir1, and some black hole
sources and the 15--60~Hz HBOs seen in the Z sources.
Somewhat more speculatively, Psaltis et al.\ suggested that
the power spectral `humps' sometimes observed at 10--60~Hz
in black hole sources may be related to the lower-frequency
members of the two kilohertz QPOs seen in the Z and atoll
sources, which have frequencies $\nu_1 \sim 300$--800~Hz.

Motivated by the results of Psaltis et al.\ (1999b), Stella,
Vietri \& Morsink (1999; hereafter SVM) compared the
frequencies of IEGs around neutron stars and black holes
with the frequencies of QPOs and other features observed in
neutron star and black hole LMXBs. Assuming that $\nuk$,
$\nuap$, and $2\nunp$ are the dominant frequencies produced
by gas clumps moving on such IEGs, that the spin rates of
the neutron stars in LMXBs range from 300 to 900~Hz, and
that the dimensionless angular momenta $j \equiv cJ/GM^2$ of
the black holes in LMXBs range from 0.1 to 0.3 (here $J$ and
$M$ are the angular momentum and mass of the black hole),
SVM showed that there are IEGs around $1.95\,\msun$ neutron
stars which give values of $\nuk$, $\nuap$, and $2\nunp$
that agree within factors $\sim$2 with the frequencies of
the kilohertz and lower-frequency QPOs observed in neutron
star LMXBs and that there are IEGs around black holes which
give values of $2\nunp$ and $\nuap$ that agree within
factors $\sim$2 with the frequencies of the 0.1--10~Hz and
2--300~Hz features observed in the power spectra of some
black hole LMXBs. However, SVM did not attempt to fit any
frequency data.

Recently, Psaltis \& Norman (2000) proposed a model of
QPOs in neutron star and black hole LMXBs based on the
assumption that there is a ``resonant ring'' in the inner
accretion disc. A ring with the properties proposed has a
dense spectrum of resonances, including resonances at the
frequencies $2\nunp$, $\nuap$, and $\nuk$ of an IEG at the
radius of the ring. Psaltis \& Norman assumed that these
three frequencies are the dominant frequencies produced by
the resonant ring and identified them with the frequencies
of the low-frequency and kilohertz QPOs in the neutron star
LMXBs.

\begin{table*}
\begin{center}
\begin{minipage}{162mm}
\caption{Types of the geodesics considered in the present
work.\label{table.geodesicTypes}}
\begin{tabular}{lllr}
   \hline \hline
   Quantity & Infinitesimally Eccentric Geodesics (IEGs)
   & Geodesics with Finite Eccentricities$^a$\\
   \hline
   Periastron radius $\rp$
      & Equal to $\ra$
      & Fixed for a given source; 
        varies from source to source
   \\
   Apastron radius $\ra$
      & Freely chosen for each frequency pair
      & Freely chosen for each frequency pair
   \\
   \hline \hline
\end{tabular}
$^a$We refer to geodesics with finite eccentricities as
moderately eccentric geodesics (MEGs) if $\ra<1.5\,\rp$
($\epsilon<0.2$) or as highly eccentric geodesics (HEGs) if
$\ra>1.5\,\rp$ ($\epsilon>0.2$).
\end{minipage}
\end{center}
\end{table*}

If it could be established that the frequencies of the most
prominent QPOs and other features observed in power spectra
of the X-ray emission from neutron star and black hole LMXBs
are simply the characteristic frequencies of test particle
geodesics around these objects, measurements of these
frequencies would provide a very simple way to explore the
effects of strong-field gravity. Moreover, as discussed by
Markovi\'c (2000), the periastron radius of an eccentric
geodesic can be smaller than the radius of the ISCO. Hence,
if the highest frequency QPOs could be shown to be produced
by clumps of gas moving on geodesics with substantial
eccentricities, it might be possible to derive even tighter
upper bounds on the radii of the neutron stars in the
kilohertz QPO sources and hence on the hardness of the
equation of state of neutron star matter than have been
derived from the models in which the frequencies of the
upper kilohertz QPOs are the orbital frequencies of gas in
nearly circular orbits (Miller et al.\ 1998; Schaab \&
Weigel 1999). On the other hand, by rejecting any causal
connection between the frequency difference $\Dnu$ and the
stellar spin frequency $\nus$, such a model would leave
unexplained the origin of the burst oscillations and the
closeness of $\Dnu$ to the burst oscillation frequency.

In the present paper we investigate further the possibility
that the frequencies of various QPOs and other features in
power spectra of the X-ray emission from neutron stars in
LMXBs are certain frequencies of special geodesics around
these stars. For the purposes of this analysis, we simply
assume that gas clumps with the required properties can be
formed and destroyed in the required way; that they are
injected onto the required geodesics; and that once
injected, the motion of the clumps is unaffected by
radiation, magnetic, or gas pressure forces, so that their
motion is purely geodesic. We then explore whether geodesic
frequencies are consistent with the observed QPO
frequencies; compute the oscillation frequencies that would
be produced by emitting, reflecting, or obscuring clumps
moving on geodesics around a neutron star; and analyze the
constraints imposed on such models by basic gas dynamics. We
consider the geodesics and frequency identifications
proposed by SV, SVM, Karas (1999), and Psaltis \& Norman
(2000), but we do not restrict ourselves to these models.
The various types of geodesics we study are listed in
Table~\ref{table.geodesicTypes}.

In Section~\ref{preliminaries} we describe how we compute
the neutron star models, spacetimes, and geodesics that we
need and the procedure we use to construct the geodesic
frequency relations that fit best the measured QPO
frequencies.

In Section~\ref{frequencyRelations} we investigate whether
certain geodesic frequencies are consistent with the
frequencies of the power spectral features observed in the
atoll and Z sources. For the purposes of this analysis, we
simply assume that clumps moving on geodesics produce X-ray
oscillations with the frequencies $\nuk$, $\nuap$, and
$2\nunp$, and then examine whether one can construct
geodesics for which these frequencies agree with the
observed frequencies. We find that the $\nuap$-$\nuk$
relations given by the best-fitting sequences of IEGs and
MEGs are qualitatively different from the observed
$\nu_1$-$\nu_2$ correlations, for any acceptable neutron
star models and spin rates. We therefore explore whether
dropping the requirement that geodesics be only moderately
eccentric would allow acceptable fits. We show that
$\nuap$-$\nuk$ relations similar to the observed kilohertz
QPO frequency correlations are possible using highly
eccentric geodesics (HEGs), but only if the geodesic
sequences include HEGs with apastron to periastron ratios
$\agt3$; most of the best-fitting frequency relations are,
however, formally unacceptable. We find that frequencies
other than $\nuap$ and $\nuk$ produce frequency relations
that disagree qualitatively with observed kilohertz QPO
frequency correlations.

Next, we examine whether sequences of HEGs can be found that
give $2\nunp$-$\nuk$ and $\nuap$-$\nuk$ relations that are
simultaneously consistent with the frequencies of the
low-frequency QPOs and the kilohertz QPOs. We find that the
best-fitting $2\nunp$-$\nuk$ and $\nuap$-$\nuk$ relations
disagree qualitatively with the observed correlations, for
realistic neutron star models and physically allowed spin
frequencies.

We also investigate whether there are sequences of geodesics
that give $2\nunp$-$\nuap$ relations consistent with the
frequencies of the low- and high-frequency spectral features
observed in \cir1. Geodesics can be found that give
relations consistent with most of the \cir1 frequency data,
but only geodesics around neutron stars with very low masses
($M \sim 1.3\,\msun$) and high spin rates (900$\,$Hz) have
frequencies that also approach the frequencies of the
highest-frequency QPOs observed in \cir1.

In Section~\ref{spectra} we compute numerically the X-ray
oscillation frequencies that would be generated by
occulting, reflecting, and radiating clumps moving on
geodesics around a neutron star by simulating the waveforms
produced by such clumps. We find that contrary to what has
been assumed in previous work, most of the power generated
by orbiting clumps appears in peaks of roughly equal total
power at the radial epicyclic frequency $\nur$, at
$\nuk+\nur$, and at $\nuk+2\nur$. There is power at $\nuk$,
but it is always $\alt25$\% of the power in each of the
three dominant peaks and there are often many other peaks as
strong or stronger than the peak at $\nuk$. For the relevant
geodesics, the power at $\nuap$ and at $\nunp$ generated by
most mechanisms is $\alt10$\% of the power in the dominant
peaks. We were unable to find any effect or geometry that
produces significant power at $2\nunp$.

In Section~\ref{dynamics} we explore the constraints on
orbiting clump models and on the amplitude of the X-ray
modulation that can be generated in such models imposed by
basic gas dynamics. We find that gas dynamical constraints
require the size of an individual clump to be a fraction
$\alt10^{-3}$ of the stellar radius and the density of the
gas in the clumps to be $\gg10^5$ times the interclump
density and $\gg10^6$ times the mean density near the star.
As a result, the maximum X-ray modulation that clumps could
produce is $\ll 10^{-6}$.

In Section~\ref{discussion} we discuss our principal results
and conclude that there are significant difficulties with
the geodesic precession and resonant ring models of the
kilohertz and other QPOs.

\section{Computation of geodesics and frequency comparison
procedure}
\label{preliminaries}

\subsection{Computation of geodesics}
\label{CompGeodesics}

We first computed numerically a large suite of neutron star
models and spacetimes for a range of neutron-star matter
equations of state (EOS), stellar spin rates, and masses. We
then integrated the equations of bound geodesic motion (see
Markovi\'c 2000) in the exterior spacetimes of these neutron
star models to determine the characteristic frequencies and
other properties of possibly relevant geodesics.

Modern numerical codes (Komatsu, Eriguchi \& Hachisu 1989a,
1989b; Cook, Shapiro \& Teukolsky 1992, 1994a, 1994b;
Stergioulas \& Friedman 1995; Nozawa et al.\ 1998) allow
accurate construction of uniformly spinning, equilibrium
stellar models using any tabulated EOS. We use the
particular variant developed by Stergioulas \& Friedman
(1995). With the exception of \cir1, the neutron star models
that give the best fits to the frequency correlations we
consider in this paper all have masses $>1.7\,\msun$ and
spin rates $<600\,$Hz, substantially lower than their
mass-shedding limits.

We considered a range of neutron star matter equations of
state. We find that stars constructed using the modern,
realistic EOS A18$+$UIX$+\delta v_b$ (Akmal, Pandharipande
\& Ravenhall 1998; Pandharipande, Akmal \& Ravenhall 1998)
give geodesics that are almost identical to those given by
the earlier realistic UU EOS of Wiringa, Fiks \& Fabrocini
(1988). In order to facilitate comparison with previous
work, we mostly report results based on the UU EOS. In order
to explore the effects of uncertainties in the EOS, we also
studied stars constructed using the early EOS of Bethe \&
Johnson (1974; EOS~C in the compilation by Arnett \& Bowers
1977), which is softer than the UU EOS at high densities. In
order to allow comparison with previous work that used this
EOS (e.g., SV), we also investigated stars based on an early
mean-field EOS constructed by Pandharipande \& Smith (1975;
EOS~L in Arnett \& Bowers 1977), even though this EOS is now
considered unrealistically hard. These three equations of
state are discussed in more detail in the companion paper
(Markovi\'c 2000).

\subsection{Construction of frequency relations}
\label{FittingSequences}

The frequencies $\nunp$, $\nuap$, and $\nuk$ of a geodesic
are functions of its periastron and apastron radii $\rp$ and
$\ra$ (we use circumferential radii everywhere), the EOS of
neutron star matter, and the mass $M$ and dimensionless
angular momentum $j \equiv cJ/GM^2$ of the neutron star. For
a given EOS and $M$, $j$ is a function of the neutron star
spin rate $\nu_s$, which may be directly observable. We
therefore parameterise the effects of stellar rotation by
$\nu_s$.

For infinitesimally eccentric geodesics, we assume that the
sequence $\{\rai\}$ of apastron radii is the same as the
sequence $\{\rpi\}$ of periastron radii and then use the
maximum likelihood method to determine the $\{\rpi\}$
sequence that gives the $\nuap$-$\nuk$ sequence that best
fits the observed $\nu_1$-$\nu_2$ sequence. We treat each
$\rpi$ as a free parameter, requiring only that it exceed
$\risco$.

For geodesics with finite eccentricities, the i{\em th\/}
members of the sequences $\{\nunpi\}$, $\{\nuapi\}$, and
$\{\nuki\}$ depend on both $\rpi$ and $\rai$. Hence the
geodesic precession hypothesis does not by itself predict a
relation between the different characteristic frequencies of
a geodesic. As explained in the Introduction, some physical
constraint is needed that will convert the two-parameter set
of geodesics specified by $\rp$ and $\ra$ to a
single-parameter sequence of geodesics that will generate a
curve in, e.g., the $\nuap$-$\nuk$ plane. We follow Stella
\& Vietri (1999) in assuming that $\rp$ is constant in a
given source but that it can be chosen freely for each
source to give the best possible fit of the $k\nunp$-$\nuk$
and $\nuap$-$\nuk$ relations to the observed frequency
correlations (we require only that $\rp$ exceed the
equatorial radius $\re$ of the star, if there is no
marginally bound geodesic, or that $\rp$ exceed $\rmb$, the
periastron radius of the marginally bound geodesic). Here
$k$ ($=1$ or 2) is the harmonic of the nodal precession
frequency that the hypothesis identifies with $\nuhbo$ or
$\nul$. For a given $\rp$, the frequencies $\nunp$, $\nuap$,
and $\nuk$ are continuous functions of $\ra$. We assume that
the $\ra$ sequence for each source can be chosen freely to
give the best possible agreement of the $k\nunp$, $\nuap$,
and $\nuk$ sequences with, e.g., the observed $\nuhbo$,
$\nu_{1}$, and $\nu_{2}$ sequences. We determine which $\ra$
sequence agrees best with the frequency data using the
maximum likelihood method. We refer to geodesics with finite
eccentricities as moderately eccentric if $\ramax/\rp < 1.5$
and as highly eccentric if $\ramax/\rp > 1.5$ (see
Table~\ref{table.geodesicTypes}).

For each EOS, we treat $M$ as a free parameter within the
mass range allowed by the EOS. The variation of $\nuap$ with
$\nuk$ is relatively insensitive to the spin rate of the
star, unless the stellar mass required is near the maximum
stable mass, in which case nonzero spin makes the fits
significantly worse. In contrast, the variation of $\nunp$
with $\nuk$ is sensitive to the spin of the star. In order
to give the geodesic precession hypothesis the best possible
chance of fitting the data, we treat $\nu_s$ as a free
parameter, allowing any value between zero and the
mass-shedding limit, even though the neutron stars in the
kilohertz QPO sources are thought to have spin frequencies
$\sim$250--350~Hz.

We assume that a necessary (although clearly not sufficient)
condition for the geodesic precession hypothesis to be
acceptable is that the best-fitting sequence of geodesics
gives frequencies consistent with the QPO frequencies
observed in all the sources to which the hypothesis is
supposed to apply. In assessing whether the frequencies
given by a particular geodesic sequence are consistent with
the observed frequencies of the relevant QPOs, we use the
frequency uncertainties reported in the literature. These
uncertainties generally include only statistical errors. We
caution that when a QPO or other feature is weak, systematic
errors introduced, for example, by uncertainties in the
continuum, are likely to be significant.

\subsection{Fits to kilohertz QPOs}
\label{KilohertzFits}

Consider first the kilohertz QPOs. Suppose there are $n$
simultaneous measurements of the two kilohertz QPO
frequencies $\nu_1$ and $\nu_2$ and let $\nu_{1,i}$ and
$\nu_{2,i}$ denote, respectively, the $i$th (simultaneous)
pair of measurements. We assume that the errors in
$\nu_{1,i}$ and $\nu_{2,i}$ are normally distributed with
standard deviations $\sigma_{1,i}$ and $\sigma_{2,i}$. Then
the likelihood of any given set of the parameters $M$,
$\nus$, $\{\rpi\}$, and $\{\rai\}$ is  \begin{eqnarray}
 \label{likelihood}
 {\cal L}(M,\nus,\{\rpi\},\{\rai\}) &\equiv&
 - \prod_{i=1}^{n}
 \frac{1}{2\pi\sigma_{1,i}\,\sigma_{2,i}}  
 \times\nonumber\\
 &&\mbox{}\hspace{-3.8 truecm}\times \exp
      \left[
-\frac{(\nu_{1,i}-\nuapi)^2}{2\sigma_{1,i}^2}
-\frac{(\nu_{2,i}-\nuki)^2}{2\sigma_{2,i}^2}
      \right]\,,\\ \nonumber 
 \end{eqnarray}
where $\{\nuapi\}$ and $\{\nuki\}$ are the apsidal
precession and azimuthal frequency sequences given by the
EOS, $M$, $\nus$, and the radius sequences $\{\rpi\}$, and
$\{\rai\}$.

We characterise the goodness of a fit by $\chisqdof \equiv
\chisq/m$, where $\chisq$ is twice the sum of the squared
terms in the exponential in equation~(\ref{likelihood}) and
$m$ is the number of degrees of freedom. For $n$ data points
($2n$ frequencies), fits of infinitesimally eccentric
geodesics have $n+2$ parameters (the $\{\rai\}$ plus $M$ and
$\nus$), or $n+1$ parameters if the spin is fixed {\em a
priori}, so $m=2n-n-2=n-2$, or $n-1$ if the spin is fixed;
fits of geodesics with finite eccentricities have $n+3$
parameters (the $\{r_{a,i}\}$ plus $\rp$, $M$, and $\nus$)
or $n+2$ if the spin is fixed, so $m=n-3$, or $n-2$ if the
spin is fixed.

We begin by picking an EOS and spin frequency $\nu_s$. The
frequencies $\nuap$ and $\nuk$ depend only on the  exterior
spacetime close to the rotation equator and are
single-valued, continuous functions of $\rp$ and $\ra$. As
shown in the companion paper (Markovi\'c 2000), the exterior
spacetime close to the rotation equator of a star with mass
$M$ and spin frequency $\nu_s$ can be described by the eight
metric functions $\rho_o(M,\nus,s)$, $\alpha_o(M,\nus,s)$,
$\beta_o(M,\nus,s)$, $\gamma_o(M,\nus,s)$,
$\bar{\rho}(M,\nus,s)$, $\bar{\alpha}(M,\nus,s)$,
$\bar{\beta}(M,\nus,s)$ and $\bar{\gamma}(M,\nus,s)$, where
$s \equiv r/(r + \re)$. We determine these metric functions
for each $\nus$ and each $M < M_{\rm max}(\nus)$ that we
consider, by interpolating in a sequence of metric functions
computed previously for this value of $\nus$ and a
sufficiently dense sequence $\{M_i\}$ of stellar masses.
Here $M_{\rm max}(\nus)$ is the maximum stable mass for
stars of spin frequency $\nus$.

Once the metric functions are known, we can compute $\nuap$
and $\nuk$ as functions of $\rp$ and $\ra$. For
infinitesimally eccentric geodesics, we determine
numerically the sequence $\{\rai\}$ of $\ra$ values that,
with $\rpi = \rai$, maximises $\cal L$. For geodesics with
finite eccentricities, we determine the sequence
$\{r_{a,i}\}$ that maximises $\cal L$ for a given $\rp$. We
then vary $\rp$ and $M$ to find the values that give the
largest maximum value of $\cal L$. As noted above, the
$\nuap$-$\nuk$ relation given by a particular sequence of
geodesics is fairly insensitive to $\nus$ and hence the
quality of the fit to the $\nu_1$-$\nu_2$ correlation
observed in a given source varies only slightly over a wide
range of spin rates. We therefore do not attempt to
determine the best-fit spin frequency accurately, but
instead quote a range of spin frequencies that have similar likelihoods.

Although we fit the $\nuap$-$\nuk$ relations allowed by a
given geodesic precession hypothesis to the $\nu_1$-$\nu_2$
data as described above, in our figures we compare the
relation between $\nu_r = \nuk-\nuap$ and $\nuk$ given by
the best-fitting sequence of geodesics with the observed
correlation between $\Delta\nu \equiv \nu_2 - \nu_1$ and
$\nu_2$, because one can assess the goodness of the fits
much more easily by examining such plots. The uncertainties
in $\Delta\nu$ shown in these plots were not used in the
fitting procedure and are shown only to give an approximate
visual impression of the goodness of fit; they were computed
by adding the uncertainties in $\nu_1$ and $\nu_2$ in
quadrature.

\subsection{Fits to kilohertz and low-frequency QPOs}
\label{LowFrequencyComparison}

Consider now the spectral features at $\nuhbo$ or $\nul$. We
wish to determine whether a sequence of geodesics can be
constructed that gives $k\nunp$-$\nuk$  and $\nuap$-$\nuk$
relations that are simultaneously consistent with, e.g., the
observed $\nuhbo$-$\nu_2$ and $\nu_1$-$\nu_2$ correlations.
For a given EOS, the parameters are the same as for the
kilohertz QPOs alone, but their likelihood is now
 \begin{eqnarray} 
 \label{likelihood.2} 
 {\cal L}(M,\nu_s;\{\rpi\},\{\rai\}) &\equiv&
  - \prod_{i=1}^{n}
 \frac{1}{
  2\pi\sigma_{1,i}\,\sigma_{2,i}\,
  \sigma_{_{{\rm HBO},i}}
         }
    \times \nonumber\\
    &&\mbox{}\hspace{-4 truecm} \times \exp 
                \left[
  -\frac{(\nu_{1,i}-\nuapi)^2}{2\sigma_{1,i}^2}
  -\frac{(\nu_{2,i}-\nuki)^2}{2\sigma_{2,i}^2}
                \right. \nonumber \\
 &&\mbox{}\hspace{-0.5 truecm} 
  -\left.
 \frac{(\nuhboi-k\,\nunpi)^2}{2\sighboisq} 
   \right] \;,\\ 
 \nonumber 
 \end{eqnarray} 
where $\nunpi \equiv \nunp(M,\nus,\rpi,\rai)$. In contrast
to $\nuap$ and $\nuk$, $\nunp$ depends sensitively on $\nus$
and hence we can determine fairly accurately the spin
frequency that gives the best fit.

We again characterise the goodness of a fit by $\chisqdof
\equiv \chisq/m$, where $\chisq$ is twice the sum of the
squared terms in the exponential in
equation~(\ref{likelihood.2}) and $m$ is the number of
degrees of freedom. For $n$ data points ($3n$ frequencies),
fits of infinitesimally eccentric geodesics have $m=2n-2$,
or $2n-1$ if the spin is fixed; fits of geodesics with
finite eccentricities have $m=2n-3$, or $2n-2$ if the spin
is fixed.

\section{Frequency relations}\label{frequencyRelations}

We now investigate whether sequences of geodesics can be
found that have characteristic frequencies which agree with
the frequencies of the kilohertz QPOs observed in the Z and
atoll sources. We first explore whether it is possible to
construct sequences of IEGs or MEGs (see
Table~\ref{table.geodesicTypes}) that give $\nuap$-$\nuk$ relations
consistent with the observed correlations of $\nu_1$ with
$\nu_2$. We find that no such sequences exist for acceptable
neutron star models and spin rates. We also show that the
behavior of other characteristic frequencies, such as $\nuk$
and $\nur+\nuk$, is inconsistent with the observed
$\nu_1$-$\nu_2$ correlations for geodesics of any
eccentricity. We therefore consider whether allowing highly
eccentric geodesics (HEGs) would make it possible to
construct geodesic sequences that give $\nuap$-$\nuk$
relations consistent with the observed $\nu_1$-$\nu_2$
correlations. We also explore whether there are sequences of
HEGs with $\nuap$-$\nuk$ and $2\nunp$-$\nuk$ relations that
are simultaneously consistent with the observed
$\nu_1$-$\nu_2$ and $\nul$-$\nuk$ or $\nuhbo$-$\nuk$
correlations. Finally, we investigate the frequency
correlations observed in \cir1.

\subsection{Frequencies of infinitesimally and moderately
eccentric geodesics} \label{CompareMEGs}

  \begin{figure} 
  \centering
  \epsfxsize=17.2truecm
  \vspace{-0.2truecm}
  \hbox{\hspace{-0.9 truecm}
  \epsfbox{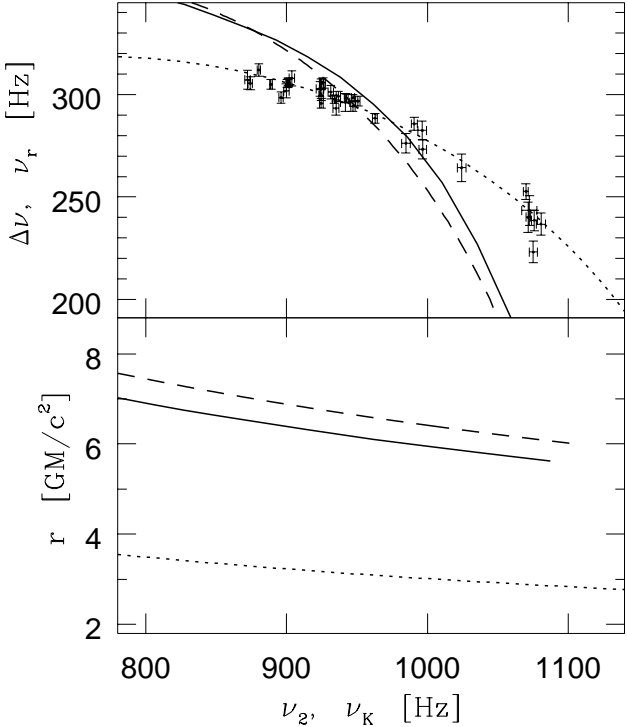}}
  \vspace{-7.2truecm}
  \caption{
{\em Upper panel\/}: $\nur$-$\nuk$ relations given by two
sequences of infinitesimally eccentric geodesics that best
fit the $\nu_1$ and $\nu_2$ data for the Z source
\hbox{Sco~X-1}. The dashed curve shows the $\nur$-$\nuk$
relation given by the best-fitting geodesic sequence around
a nonrotating star ($M = 1.99\,\msun$; $\chisqdof = 38$).
The solid curve shows the relation given by the best-fitting
geodesic sequence around a rotating star constructed using
the UU EOS($M = 2.21\,\msun$, $\nu_s = 450\,$Hz; $\chisqdof
= 35$); the relation given by the best-fitting geodesic
sequence around a rotating star constructed using EOS~L is
almost identical. The dotted curve shows the $\nur$-$\nuk$
relation given by the best-fitting sequence of geodesics
around a black hole ($M = 5.8\,\msun$, $j=0.89$; $\chisqdof
= 1.6$).
{\em Lower panel\/}: Circumferential radii of the geodesics
that give the frequencies plotted in the upper panel.
  }
  \label{ScoX1.cir.kHz} 
  \end{figure}

  \begin{figure}
  \centering
  \epsfxsize=17.2truecm
  \vspace{-0.2truecm}
  \hbox{\hspace{-0.9 truecm}\epsfbox{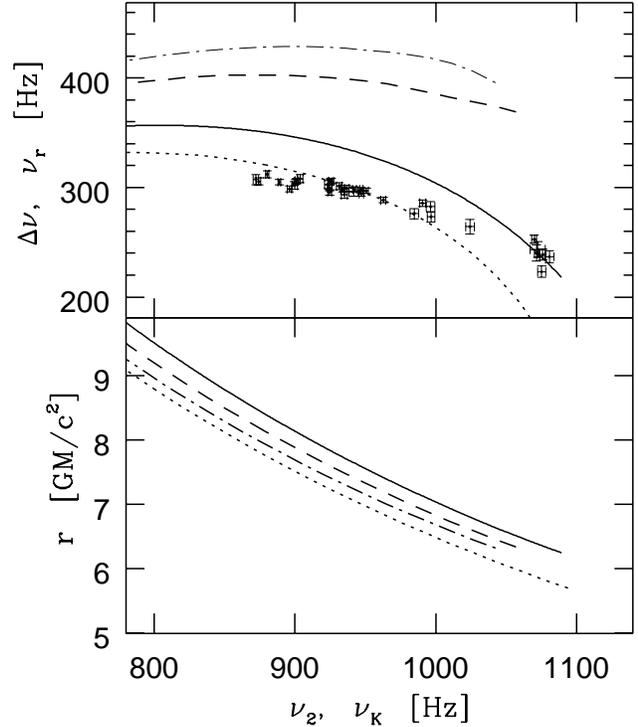}}
  \vspace{-7.2truecm}
  \caption{
{\em Upper panel\/}: Comparison of the $\nur$-$\nuk$
relations discussed in the text with measurements of $\Dnu$
and $\nu_2$ in \hbox{Sco~X-1}. The solid curve shows the
$\nur$-$\nuk$ relation given by the geodesic sequence around
a nonrotating star that SV reported as
fitting the \hbox{Sco~X-1} data; $\chisqdof$ for this
sequence is 150. The dashed and dash-dotted curves show the
$\nur$-$\nuk$ relations given by the geodesic sequences
around the stars with spin rates of $300\,$Hz and $600\,$Hz
that SV also reported as fitting the
\hbox{Sco~X-1} data. The dotted curve shows the
$\nur$-$\nuk$ relation given by the sequence of geodesics
around a rotating UU star with $\rp \geq \risco$ but any
value of $\ra$ ($\geq \rp$) that best fits the $\nu_1$ and
$\nu_2$ data for \hbox{Sco~X-1} ($M = 2.21\,\msun$, $\nus =
450\,$Hz, $\rp = \risco = 5.52$; $\chisqdof = 21$).
 {\em Lower panel\/}:
Circumferential apastron radii of the geodesics that give
the frequencies plotted in the upper panel.
  }
  \label{ScoX1.isco.kHz}
  \end{figure}

The most numerous and precise kilohertz QPO frequency
measurements are those for the Z source \sco1 (van der Klis
et al.\ 1997). We therefore consider consistency with the
frequency correlation observed in this source as a necessary
condition for any QPO model to be viable (a very similar
correlation has been observed in 4U~1608$-$52; see M\'endez
et al. 1998). The frequency relations given by geodesics
around nonrotating stars provide useful guidance because the
$\nur$-$\nuk$ relation is relatively insensitive to the
stellar spin rate and the relation for a nonrotating star
doesn't depend on the EOS of neutron star matter, provided
that it is hard enough to support the stellar mass being
considered and gives stars with $\re < \ramin$). Only
geodesics near $\risco$ are relevant, because $\nu_r$
decreases appreciably with increasing $\nuk$ only near
$\risco$.

The dashed curve in Fig.~\ref{ScoX1.cir.kHz} shows the
frequency relation constructed from IEGs around a
nonrotating star that best fits the \sco1 data; the stellar
mass derived from this fit is $1.99\,\msun$, which is a
reasonable mass for a neutron star in an LMXB. The dashed
curve is very similar to the curve shown in Fig.~1 of SV for
IEGs around a nonrotating star with a mass of $2.0\,\msun$.
As Fig.~\ref{ScoX1.cir.kHz} shows, the frequency relation
for the best-fitting sequence of IEGs around a nonrotating
star is much steeper than the observed frequency
correlation; the qualitative disagreement between the
best-fitting frequency relation and the frequency data is
indicated by the large $\chisqdof$ for this fit, which is 38.

As noted by SV, stellar rotation has little effect on the
relation between $\nuap$ and $\nuk$. This is illustrated in
Fig.~\ref{ScoX1.cir.kHz} by the similarity of the frequency
relation for the rotating star constructed using the UU EOS
that best fits the \sco1 data to the relation for the
nonrotating star that fits best. The frequency relation for
the rotating star constructed using EOS~L that best fits the
data is also very similar to the relation for the
nonrotating star that fits best. Although prograde spin does
decrease slightly the slope of the predicted $\nu_r$-$\nuk$
relation for a given mass, the effect is small. The reason
is that the dragging of inertial frames causes $\risco$ to
decrease, which significantly increases the azimuthal
frequency of geodesics near $\risco$ for a given stellar
mass. Hence the mass of the star must be increased in order
to keep the azimuthal frequencies in the observed range (all
frequencies scale approximately as $1/M$). This drives the
stellar mass close to the maximum mass allowed by the EOS.
Indeed, the mass of a rotating UU star that fits best is
$2.21\,\msun$, which is the maximum stable mass for this EOS
and the best-fitting spin rate (450~Hz). The $\chisqdof$ for
this fit is 35.5, only slightly smaller than the $\chisqdof$
for the nonrotating star that fits best. The fit is not any better for harder equations of state. Although they give larger maximum masses, $\nur$ falls more steeply because of their larger size and hence larger quadrupolar deformation (Markovi\'c 2000).

SV argued that $\nuap$ and $\nuk$ would agree with the
observed frequencies of the kilohertz QPOs if clumps form on
MEGs rather than IEGs. To support their argument, they
showed in their Fig.~2 a frequency relation supposedly for a
sequence of MEGs with $\rp = 6.25M$ around a $M =
1.9\,\msun$ nonrotating star; this relation appears to agree
fairly well with the \sco1 frequency data. However, as
explained in detail in the Appendix, the $\nur$-$\nuk$
relation plotted by SV was computed
incorrectly. The solid curve in Fig.~\ref{ScoX1.isco.kHz}
shows the actual $\nur$-$\nuk$ relation for the sequence of
MEGs specified by SV. It is similar to the relation for the
best-fitting sequence of IEGs and is much steeper than the
data ($\chisqdof = 150$).

SV also reported that sequences of MEGs with $\rp \approx
6.18M$ and $6.17M$ around rotating neutron stars with masses
of $1.94\msun$ and $1.98\msun$ constructed using EOS~L with,
respectively, $\nus=300\,$Hz and 600$\,$Hz also give
$\nur$-$\nuk$ relations similar to the $\Dnu$-$\nu_2$
correlation observed in \sco1. We have computed the
$\nur$-$\nuk$ relations for these geodesic sequences. The
results are shown as the dashed and dashed-dotted curves in
Fig.~\ref{ScoX1.isco.kHz}; these curves lie 100--200$\,$Hz
above the \sco1 frequency correlation.

The dotted curve in Fig.~\ref{ScoX1.isco.kHz} shows the
frequency relation for the best-fitting sequence of
geodesics around a rotating star constructed using
geodesics that meet the requirement $\rp\geq\risco$ proposed
by SV but are otherwise unrestricted. With this freedom, the
low-frequency geodesics become more eccentric ($\ramax/\rp
\approx 1.5$), but the frequency relation remains much
steeper than the data ($\chisqdof = 21$). For the reasons
discussed above, the stellar mass that fits best is equal to
the maximum mass for the best-fit spin rate.

The frequency relations for sequences of MEGs around stars
constructed using other equations of state are also
inconsistent with the \sco1 data. The frequency relations
given by sequences of IEGs and MEGs also disagree with the
kilohertz QPO frequencies of other sources, although
frequency measurements for sources other than \sco1 are
sparser and less precise. For example, the best-fitting
sequence of MEGs with $\rp\geq\risco$ around a UU star with
the 363~Hz spin rate inferred from the burst oscillations of
\fu{1728$-$34} (Strohmayer et al.\ 1996; M\'endez \& van der
Klis 1999) has $M = 1.92\,\msun$ and gives $\chisqdof
\approx 15$. If instead the stellar spin rate is treated as
a free parameter, the best-fitting MEG sequence with
$\rp\geq\risco$ has $\nus \approx 900\,$Hz and $M =
2.20\,\msun$ and gives $\chisqdof = 9.4$. This mass is again
very close to the maximum stable mass for this EOS.

Almost all the best-fitting neutron star masses given by MEG
sequences are very close to the maximum stable mass for the
assumed EOS, which is implausible. More seriously, the
$\nu_r$-$\nuk$ relations given by sequences of MEGs deviate
by large fractional amounts ($\Delta\nu_r/\nu_r$ as much as
30\% or $\Delta\nuk/\nuk$ as much as 15\%) from the
$\Dnu$-$\nu_2$ correlations observed. Any forces strong
enough to bring these frequency relations into agreement
with the data would invalidate the geodesic motion
hypothesis.

Fig.~\ref{ScoX1.cir.kHz} shows that one can construct a
sequence of IEGs around a $5.8\,\msun$, rapidly spinning
($j=0.89$) black hole that gives a frequency relation which
agrees fairly well with the kilohertz QPO frequencies
observed in \sco1 ($\chisqdof = 1.6$). There is compelling
evidence that \sco1 is not a black hole (see, e.g., van der
Klis 2000). The fairly good agreement of IEGs in the Kerr
spacetime with the \sco1 frequency data is a reminder that
successfully fitting QPO frequencies does not by itself
validate a model. The wealth of other information that is
available about the kilohertz QPOs and the kilohertz QPO
sources must also be taken into acount.

These results show that regardless of the neutron star model
and spin rate assumed, the MEG hypothesis gives frequency
relations that are qualitatively different from the observed
correlations between the kilohertz QPO frequencies. As
explaining this correlation was the primary motivation for
introducing the MEG hypothesis (SV), we
consider this failure serious enough to set aside the MEG
hypothesis and we therefore do not consider it further.

\subsection{Frequencies of highly eccentric geodesics}
\label{CompareHEGs}

We consider now whether allowing highly eccentric geodesics
(HEGs) would make it possible to construct sequences that
give values of $\nuap$ and $\nuk$ consistent with the values
of $\nu_1$ and $\nu_2$ observed in the kilohertz QPO
sources. As before, we assume that the periastron radius
$\rp$ has a constant value $>\re$, if there is no marginally
bound geodesic, or $>\rmb$, but we allow the apastron radius
$\ra$ to have any value $>\rp$.

\begin{figure} 
\centering
\epsfxsize=17.2truecm
\vspace{-0.2truecm}
\hbox{\hspace{-0.9 truecm}\epsfbox{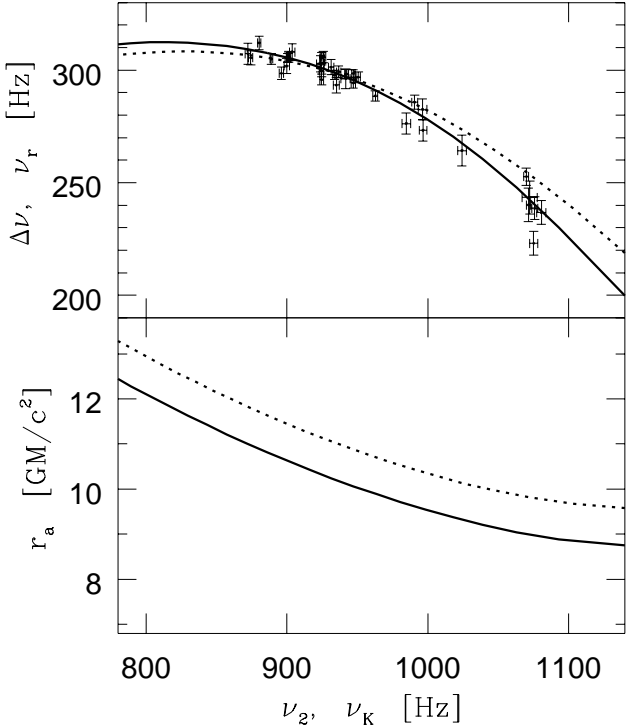}}
\vspace{-7.2 truecm}
\caption{
{\em Upper panel\/}: $\nur$-$\nuk$ relations given by two
geodesic sequences that best fit the $\nu_1$ and $\nu_2$
data for \hbox{Sco~X-1}. The solid curve shows the
$\nur$-$\nuk$ relation given by the geodesic sequence that
fits best overall ($M = 1.89\,\msun$, $\nus=0$, $\rp =
5.3M$; $\chisqdof = 1.4$). This is the best-fitting sequence
for any EOS, such as UU, that can support a nonrotating star
with $M > 1.89\,\msun$ and an equatorial radius $\re <
5.3M$. The dotted curve shows the $\nur$-$\nuk$ relation
given by the best-fitting sequence for EOS~C ($M =
1.86\,\msun$, $\nus=0$, $\rp = 5.1M$; $\chisqdof = 2.2$);
the fit would be worse for a rotating star (see text). To
achieve fits this good, both sequences must include highly
eccentric geodesics. 
 {\em Lower panel\/}: Circumferential apastron radii of the
highly eccentric geodesics that give the frequencies
displayed in the upper panel.} \label{ScoX1.kHz}
\end{figure}

\subsubsection{Sco~X-1 kilohertz QPO frequencies}

Fig.~\ref{ScoX1.kHz} shows the $\nuap$-$\nuk$ relations
given by the sequences of geodesics around stars constructed
using EOS~UU and EOS~C that fit best the $\nu_1$ and $\nu_2$
data for \sco1.

The best-fitting sequence for stars constructed using the UU
EOS is for a nonrotating star with $M=1.89\,\msun$. This fit
gives $\chisqdof = 1.4$, which is unacceptable at the 94\%
confidence level. To achieve fits this good, the sequence
must include HEGs with $\ra/\rp = 2.1$. Such highly
eccentric geodesics are required in order to generate
$\nu_r$-$\nuk$ relations that are sufficiently flat.
Physically, clumps must be formed or injected on a special
sequence of highly eccentric geodesics with a common
periastron radius that is 40\% larger than the stellar
radius and 15\% smaller than $\risco$ (as shown in
Fig.~\ref{ScoX1.isco.kHz}, if $\rp > \risco$  is required,
as suggested by SV, the fits would be much poorer). The
best-fitting mass is about $0.3\,\msun$ less than the
maximum stable mass and hence the fit is relatively
insensitive to the stellar spin rate ($\chisqdof$ would be
about the same for any spin frequency $\alt600\,$Hz). This
geodesic sequence is the best-fitting sequence for any EOS
that can support a nonrotating star with $M > 1.89\,\msun$
and an equatorial radius $\re < 5.3M$.

The best-fitting geodesic sequence for stars constructed
using the softer EOS~C is for a nonrotating star with
$M=1.86\,\msun$, the maximum stable mass for this EOS. The
smaller maximum mass of C stars requires geodesics that are
more eccentric than for UU stars, in order to keep $\nu_r$
within the observed range of $\Delta\nu$. This causes the
slope of the best-fitting $\nur$-$\nuk$ relation to be
flatter than for UU stars and flatter than the observed
correlation ($\chisqdof=2.2$). Requiring the star to have a
spin rate $\agt300\,$Hz, like that expected for the neutron
stars in the kilohertz QPO sources, would increase all
orbital frequencies, making the fit significantly worse (the
maximum mass is not increased significantly for the relevant
spin rates). We conclude that the geodesic precession
hypothesis is inconsistent with the kilohertz QPO
frequencies observed in \sco1 if neutron star matter is as
soft as EOS C.

\begin{figure} 
\centering
\epsfxsize=17.2truecm
\vspace{-0.2truecm}
\hbox{\hspace{-0.9 truecm}\epsfbox{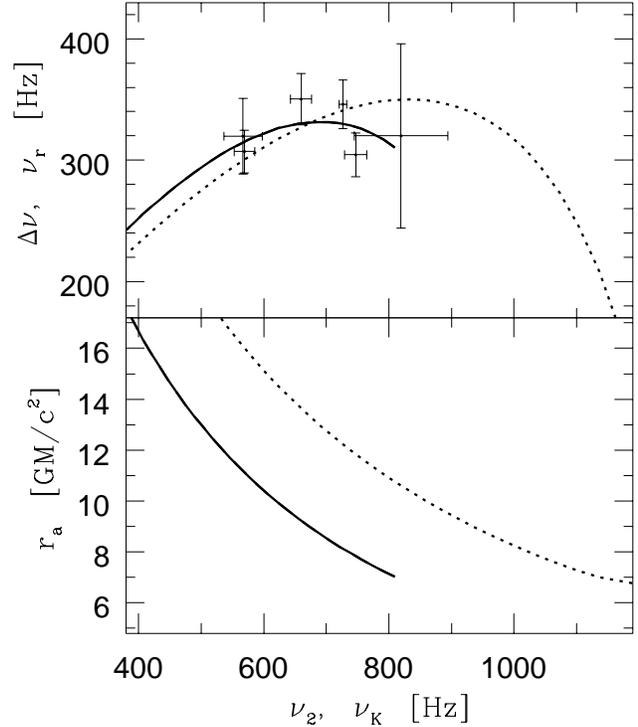}}
\vspace{-7.2 truecm}
\caption{
{\em Upper panel\/}: $\nur$-$\nuk$ relations given by two
geodesic sequences that best fit the $\nu_1$ and $\nu_2$
data for the Z source \hbox{GX~340$+$0}. The solid and
dotted curves show respectively the $\nur$-$\nuk$ relations
given by the best-fitting sequences for neutron stars
constructed using EOS~UU ($M = 2.12\,\msun$, $\nus=0$, $\rp
= 7.0M$; $\chisqdof = 0.84$) and EOS~C ($M = 1.86\,\msun$,
$\nus=0$, $\rp = 5.8M$; $\chisqdof = 2.3$). Both sequences
require highly eccentric geodesics ($\ramax/\rp = 1.6$--2.8).
{\em Lower panel\/}: Circumferential apastron radii of the
highly eccentric geodesics that give the frequencies
displayed in the upper panel.} 
\label{GX340.kHz}
\end{figure}

\subsubsection{\gx{340$+$0} kilohertz QPO frequencies}

Fig.~\ref{GX340.kHz} shows the $\nur$-$\nuk$ relations
given by the sequences of geodesics for neutron stars
constructed using EOS~C and EOS~UU that best fit the $\nu_1$
and $\nu_2$ data for the Z source \gx{340$+$0} \cite{GX340}.
The fit is again better for higher mass stars, because the
observed frequencies of the upper kilohertz QPO in
\gx{340$+$0} are relatively low when interpreted as
azimuthal frequencies near the ISCO. Hence stars constructed
using EOS~C fit significantly worse than stars constructed
using EOS~UU, because the former have smaller maximum stable
masses.

The best-fitting geodesic sequence for stars constructed
using EOS C is for a nonrotating star with a mass equal to
the maximum stable mass for this EOS, requires a very high
eccentricity ($\ramax/\rp=2.8$), and does not fit the data
adequately ($\chisqdof=2.3$), even though the uncertainties
in the measured frequency differences $\Delta\nu$ are rather
large. The disagreement would be worse for a rotating star.
Hence we conclude that the geodesic precession hypothesis is
inconsistent with the kilohertz QPO frequency correlation
observed in \gx{340$+$0} if neutron star matter is as soft
as EOS C. We caution, however, that a better understanding
of the uncertainties in measurements of $\nu_2$ above
$700\,$Hz caused by the higher noise power density at these
frequencies~\cite{GX340} could affect this conclusion.

The best-fitting geodesic sequence for the UU EOS again
requires highly eccentric ($\ramax/\rp=1.6$) geodesics with
a fixed periastron radius more than twice as large as the
star's equatorial radius. The sequence of geodesics around a
nonrotating star that fits best has a mass only about
$0.1\,\msun$ less than the maximum stable mass for this EOS;
$\chisqdof$ for the best-fitting sequence is 0.84. The
best-fitting geodesic sequence and the $\chisqdof$ would be
approximately the same for any spin rate $\alt200\,$Hz.

\subsubsection{\fu{1728$-$34} kilohertz QPO frequencies}

\begin{figure} 
\centering
\epsfxsize=17.2truecm
\vspace{-0.2truecm}
\hbox{\hspace{-0.9 truecm}
\epsfbox{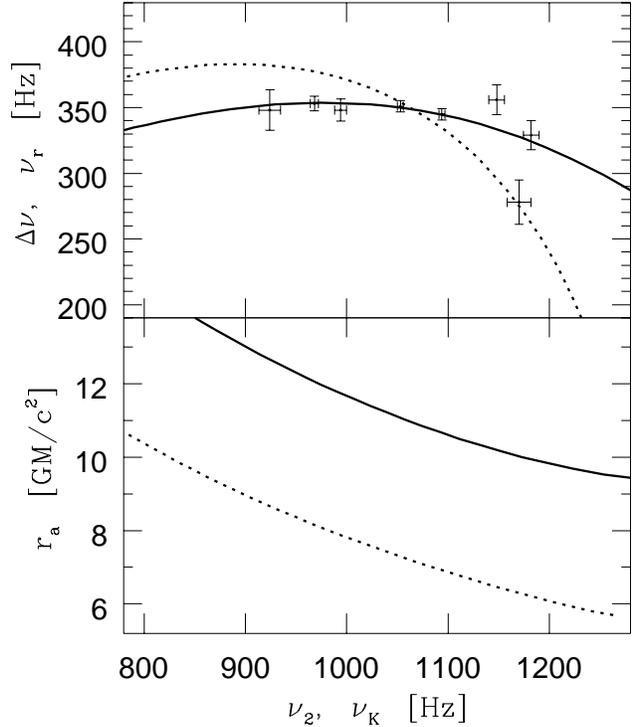}}
\vspace{-7.2 truecm}
\caption{
{\em Upper panel\/}: $\nur$-$\nuk$ relations given by two
geodesic sequences that best fit the $\nu_1$-$\nu_2$
data for the atoll source \hbox{4U~1728$-$34}. The solid and
dotted curves show respectively the $\nur$-$\nuk$ relations
given by the best-fitting sequences for neutron stars
constructed using EOS UU ($M = 1.76\,\msun$, $\Rp = 4.7M =
1.09\re = 12.4\,$km; $\chisqdof = 1.7$) and EOS~L ($M =
1.97\,\msun$, $\Rp = \re = 5.3M = 15.4\,$km; $\chisqdof =
8.6$), assuming $\nus = 363\,$Hz, the burst oscillation
frequency. Both sequences require highly eccentric geodesics
($\ramax/\rp = 2.7$ for the best-fitting UU sequence).
{\em Lower panel\/}: Circumferential apastron radii of the
highly eccentric geodesics that give the frequencies
displayed in the upper panel.
    }
\label{4U1728.kHz}
\end{figure}

The observed frequencies of the kilohertz QPOs in the atoll
source \fu{1728$-$34} narrow further the range of equations
of state consistent with the geodesic precession hypothesis.
Fig.~\ref{4U1728.kHz} shows the $\nur$-$\nuk$ relation given
by the sequence of geodesics for stars constructed using the
UU EOS that best fit the $\nu_1$ and $\nu_2$ data for
\fu{1728$-$34}. This fit assumes that the stellar spin
frequency is the 363~Hz frequency of the burst oscillations
in this source. The fit would be very similar for any spin
frequency $\lta500\,$Hz.

Again, a $\nur$-$\nuk$ relation qualitatively similar to the
observed $\nu_1$-$\nu_2$ correlation can be achieved only by
using HEGs. The $\nur$-$\nuk$ relation given by the
best-fitting geodesic sequence for stars constructed using
the UU EOS has $\ramax/\rp = 2.7$ and gives $\chisqdof=1.7$.
The large scatter in the observed $\Delta\nu$-$\nu_2$
relation for upper kilohertz QPO frequencies in the range
$1150 \alt \nu_2 \alt 1200\,$Hz (see M\'endez \& van der
Klis 1999) contributes significantly to the $\chisqdof$.
This scatter appears inconsistent with the simple, smooth
frequency relations characteristic of geodesic precession
models. The best-fitting geodesic sequence for stars
constructed using the modern, realistic EOS
\hbox{A18$+$UIX$+$$\delta v_b$} gives the same model
parameters, except that the best-fitting value of $\rp$ is
$1.05\re$ instead of $1.09\re$ (see
Table~\ref{table.geodesics} below).

Stars constructed using harder equations of state, such as
the early EOS L, which are now thought to be unrealistically
hard, have larger maximum masses but also larger equatorial
radii. This eliminates some of the highly eccentric
geodesics that are available if the EOS is softer, making
the geodesic precession hypothesis inconsistent with the
kilohertz QPO data for such hard equations of state. The
\fu{1728$-$34} frequency data illustrate this. The flatness
of the $\Delta\nu$-$\nuk$ correlation for $\nu_2 \alt
1100\,$Hz (see Fig.~\ref{4U1728.kHz}) constrains the mass of
the neutron star to be $\alt 2\msun$. The large equatorial
radii of stars constructed using hard equations of state
leave room only for less highly eccentric geodesics, which
produce $\nu_r$-$\nuk$ relations that fall steeply at high
$\nuk$ and are therefore inconsistent with the observed
$\Dnu$-$\nu_2$ correlation.

Our findings up to this point show that $\nuap$-$\nuk$
relations similar to the $\nu_1$-$\nu_2$ correlations
observed in the kilohertz QPO sources require highly
eccentric geodesics and are not possible for geodesics
around neutron stars constructed with equations of state
that are either significantly softer or significantly harder
than the UU EOS, i.e., the geodesic precession hypothesis is
inconsistent with softer or harder equations of state.
Hence, from now on we consider only EOS~UU. The realistic
modern EOS A18$+$UIX$+$$\delta v_b$ gives results that
differ only very slightly from those obtained with the UU
EOS.

\begin{figure} 
\centering
\epsfxsize=17.2truecm
\vspace{-0.2truecm}
\hbox{\hspace{-0.9 truecm}\epsfbox{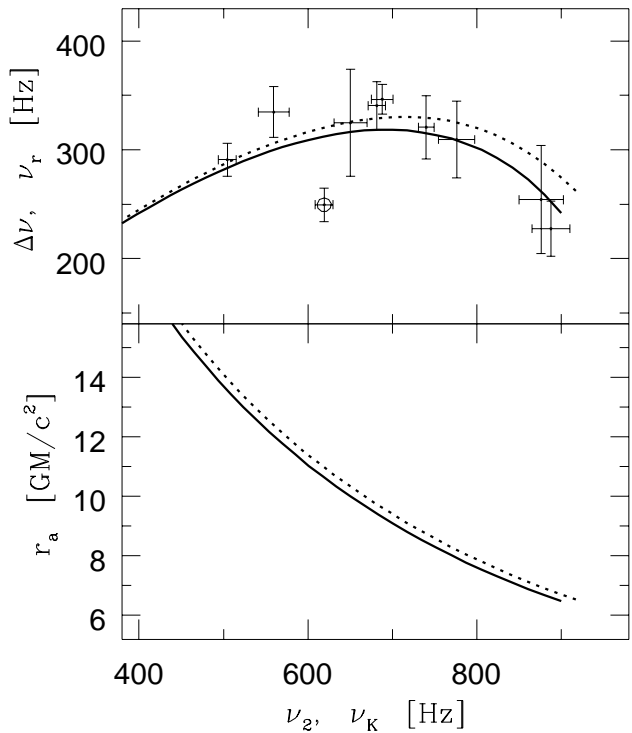}}
\vspace{-7.2 truecm}
\caption{
{\em Upper panel\/}: $\nur$-$\nuk$ relations given by two
geodesic sequences that best fit the $\nu_1$-$\nu_2$
data for the Z source \hbox{GX~5$-$1}. The solid and dotted
curves show respectively the $\nur$-$\nuk$ relations given
by the best-fitting sequences for neutron stars constructed 
using the UU EOS if all the frequency measurements are
included ($M = 2.17\,\msun$, $\rp = 6.5M$; $\chisqdof =
3.6$) or, alternatively, if all but the circled measurement
are included ($M = 2.09\,\msun$, $\rp = 6.5M$; $\chisqdof =
1.2$). Fits this good require highly eccentric geodesics
with $\ra/\rp \approx 2$.
{\em Lower panel\/}: Circumferential apastron radii of the
highly eccentric geodesics that give the frequencies
displayed in the upper panel.
	}
\label{GX5-1.kHz}
\end{figure}

\subsubsection{\gx{5$-$1} and \gx{17$+$2} kilohertz QPO
frequencies}

Fig.~\ref{GX5-1.kHz} shows the ten pairs of kilohertz QPO
frequencies measured so far in the Z source
\gx{5$-$1}\cite{GX5}. Most have relatively large
uncertainties and one (the circled point in the figure)
appears discordant. If all ten frequency pairs are included,
the $\nur$-$\nuk$ relation given by the best-fitting
sequence of geodesics for stars constructed using EOS UU has
a rather large $\chisqdof$; discarding the discordant
frequency pair gives a smaller $\chisqdof$ of 1.2. However,
values of $\chisqdof$ this small are possible only if the
star has a mass $M\approx2.1$--$2.2\,\msun$ near the maximum
stable mass and highly eccentric geodesics are included
($\ramax/\rp \approx 2.1$).

Only eight pairs of kilohertz QPO frequencies have been
measured so far in the Z source \gx{17$+$2} (Wijnands et al.
1997; see Fig.~\ref{GX17+2.comb.NP} below). The
$\nur$-$\nuk$ relation given by the best-fitting sequence of
geodesics for stars constructed using EOS UU is acceptable
($\chisqdof \approx 0.4$), but only if all the geodesics are
highly eccentric ($\ramax/\rp \approx 3$--4). However, the
uncertainties in these frequency measurements are large.

\subsection{Other geodesic frequencies}
\label{otherFreqs}

We have also explored whether the frequencies of the two
kilohertz QPOs are consistent with geodesic frequencies
other than $\nuap$ and $\nuk$. As discussed in
Section~\ref{spectra}, the most prominent frequencies
generated by clumps moving on geodesics around neutron stars
are $\nur$, $2\nur$, $\nuk+\nur$, and $\nuk+2\nur$. We find
that for realistic neutron star models, these frequencies
are not in the same range as the frequencies of the
kilohertz QPOs. The two frequencies $\nuk$ and $\nuk+\nur$
are in the same range as the frequencies of the kilohertz
QPOs only for unrealistically hard neutron-star equations of
state, such as EOS~L; however, even for such hard equations
of state, the best-fitting geodesic sequences give frequency
relations that have qualitatively different shapes from
those observed and give $\chisqdof>10$ for \gx{340$+$0} and
\gx{5$-$1}. We therefore conclude that geodesic frequencies
other than $\nuap$ and $\nuk$ are not viable explanations of
the frequencies of the kilohertz QPOs.

\subsection{LFQPO and HBO frequencies}
\label{CompareNP}

In the previous two sections we investigated whether
sequences of geodesics can be found that give $\nuap$-$\nuk$
relations consistent with the kilohertz QPO frequencies
observed in the Z and atoll sources. There we showed that
the best-fitting sequences of infinitesimally or moderately
eccentric geodesics give $\nuap$-$\nuk$ relations
qualitatively different from the kilohertz QPO frequency
correlations in all the sources considered. We showed
further that $\nuap$-$\nuk$ relations similar to the
kilohertz QPO frequencies observed in the sources considered
are possible only if highly eccentric geodesics ($\ra/\rp
\sim 2$--4) are included in all the sequences. HEG sequences
can be found that give $\nuap$-$\nuk$ relations roughly
consistent with the $\nu_1$-$\nu_2$ correlations observed in
\sco1 and \fu{1728$-$34}, although the values of $\chisqdof$
are not formally acceptable. HEG sequences can also be
found that formally fit the $\nu_1$-$\nu_2$ correlations
observed in \gx{17$+$2} and \gx{340$+$0}, but there are only
a few frequency measurements for each these sources and the
uncertainties are large. However, the geodesic sequence that
best fits the \gx{5$-$1} frequency data appears to be
inconsistent with these data.

SV and SVM proposed that gas orbits the neutron stars in the
Z and atoll sources in the form of clumps moving on slightly
inclined geodesics and that the low-frequency peaks
observed in power spectra of the Z and atoll sources are
caused by nodal precession of these geodesics. As explained
in Section~\ref{intro}, $\nunp$ is too low to explain the
frequencies of these low-frequency features. SV and SVM
therefore suggested that clump nodal precession somehow
produces an X-ray oscillation with frequency $2\nunp$.
(However, as explained in Section~\ref{spectra}, we were
unable to find a clump mechanism that generates a
significant peak at $2\nunp$.)

Here we follow up the proposal that the low-frequency
peaks are generated by nodal precession by investigating
whether sequences of geodesics can be found that give
$\nuap$-$\nuk$ and $2\nunp$-$\nuk$ relations that are both
{\em simultaneously \/} consistent with the $\nul$-$\nu_2$
and $\nu_1$-$\nu_2$ correlations observed in the Z and atoll
sources. In contrast to the $\nuap$-$\nuk$ relations studied
in the previous section, the $2\nunp$-$\nuk$ relations
studied here are sensitive to the stellar spin rate, because
of the gravitomagnetic effect.

Before proceeding, we caution that measurements of the
centroid frequencies of the bumps observed in atoll-source
power spectra can be affected significantly by systematic
errors that arise from uncertainties in modeling the noise
continuum in the vicinity of these features (M.\ M\'endez
and M. van der Klis 1999, private communication). The
systematic errors are often difficult to estimate and
typically are not reported in the literature. Instead, only
statistical errors are reported. We use the reported errors
to make a preliminary assessment of whether the fits below
are acceptable, but one should bear in mind that the total
errors may be somewhat larger.

\begin{figure} 
\centering
\epsfxsize=17.2truecm
\vspace{-0.2truecm}
\hbox{\hspace{-0.9 truecm}
\epsfbox{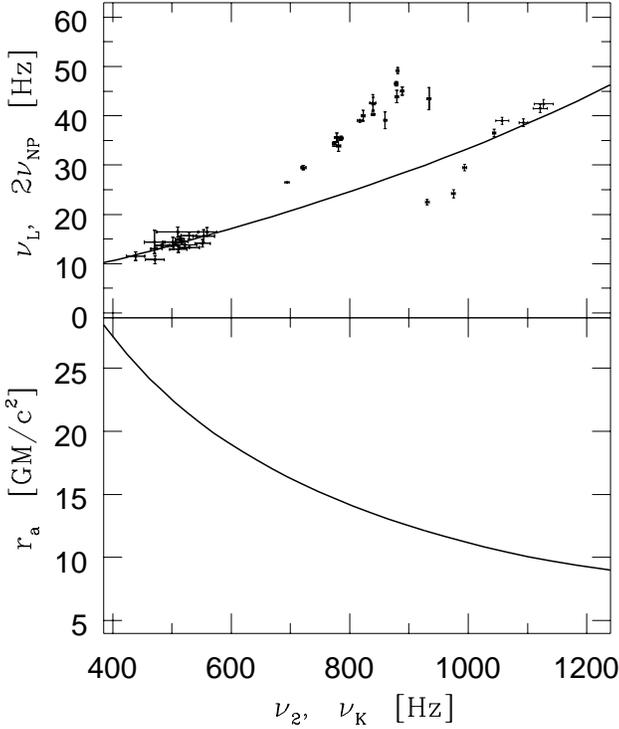}}
\vspace{-7.2 truecm}
\caption{
{\em Upper panel\/}: Comparison of the $\nu_{\rm
L}$-$\nu_{\rm K}$ behaviour seen in \hbox{4U~1728$-$34} with
the $2\nu_{\rm NP}$-$\nu_{\rm K}$ relation predicted by the
geodesic sequence that gives the best fit to the
$\nu_1$-$\nu_2$ data for this source (see Fig.~5).
{\em Lower panel\/}: Circumferential apastron radii of the
highly eccentric geodesics that give the frequencies
displayed in the upper panel.
    }
\label{4U1728.NP}
\end{figure}

\subsubsection{\fu{1728$-$34} LFQPO and kilohertz QPO
frequencies}

The $2\nunp$-$\nuk$ relation given by the geodesic sequence
which gives the $\nuap$-$\nuk$ relation that best fits the
$\nu_1$-$\nu_2$ data in \fu{1728$-$34} (see
Section~\ref{CompareHEGs} and Fig.~\ref{4U1728.kHz}) has a
shape that is qualitatively different from the observed
$\nul$-$\nuk$ correlation in this source (see
Fig.~\ref{4U1728.NP}). The predicted relation shown is for a
stellar model constructed using the UU EOS and a spin rate
equal the 363~Hz burst oscillation frequency. The fit cannot
be improved significantly by increasing or decreasing the
spin rate.

We also considered stellar models constructed using the
modern realistic EOS A18$+$UIX$+\delta v_b$ (see
Section~\ref{CompGeodesics}). The best-fitting stellar mass
is 0.1\% larger and the dimensionless angular momentum is
3\% larger ($j=0.156$ vs.\ $j=0.149$) for this EOS, because
it is slightly harder. This increases the nodal precession
frequency for geodesics with a given $\nuk$, but only by
$\alt4$\%.

The predicted $2\nunp$-$\nuk$ relation shown in
Fig.~\ref{4U1728.NP} is clearly inconsistent with the
observed $\nul$-$\nu_2$ correlation: the predicted relation
does not reproduce the abrupt drop in $\nul$ by a factor
$>2$ that occurs at $\nu_2\approx 900\,$Hz in the observed
correlation, nor is it consistent with the observed
$\nul$-$\nu_2$ correlation either before or after the abrupt
drop. The predicted $\nunp$-$\nuk$ relation misses many of
the observed values of $\nul$ by a factor $\sim$2. This
qualitative disagreement between the frequency behavior
predicted by the nodal precession hypothesis and the
frequency behaviour observed in \fu{1728$-$34} occurs for
any realistic neutron star model and spin rate.

Very recent analyses of \fu{1728$-$34} power spectra (Di
Salvo et al.\ 2000, private communication) indicate that the
break frequency of the low-frequency broad-band noise
component and the frequency of the LFQPO are related: the
frequencies of the break and the LFQPO both increase as the
frequency of the upper kilohertz QPO increases from 400 to
900 Hz, but at this point the existing LFQPO disappears and
the noise feature peaks up into a new LFQPO. The frequency
of the old and new LFQPOs are not harmonically related. This
behaviour appears inconsistent with any simple
interpretation in terms of nodal precession of HEGs.

\subsubsection{\gx{5$-$1} HBO and kilohertz QPO frequencies}

\begin{figure} 
\centering
\epsfxsize=17.2truecm
\vspace{-0.3truecm}
\hbox{\hspace{-0.9 truecm}\epsfbox{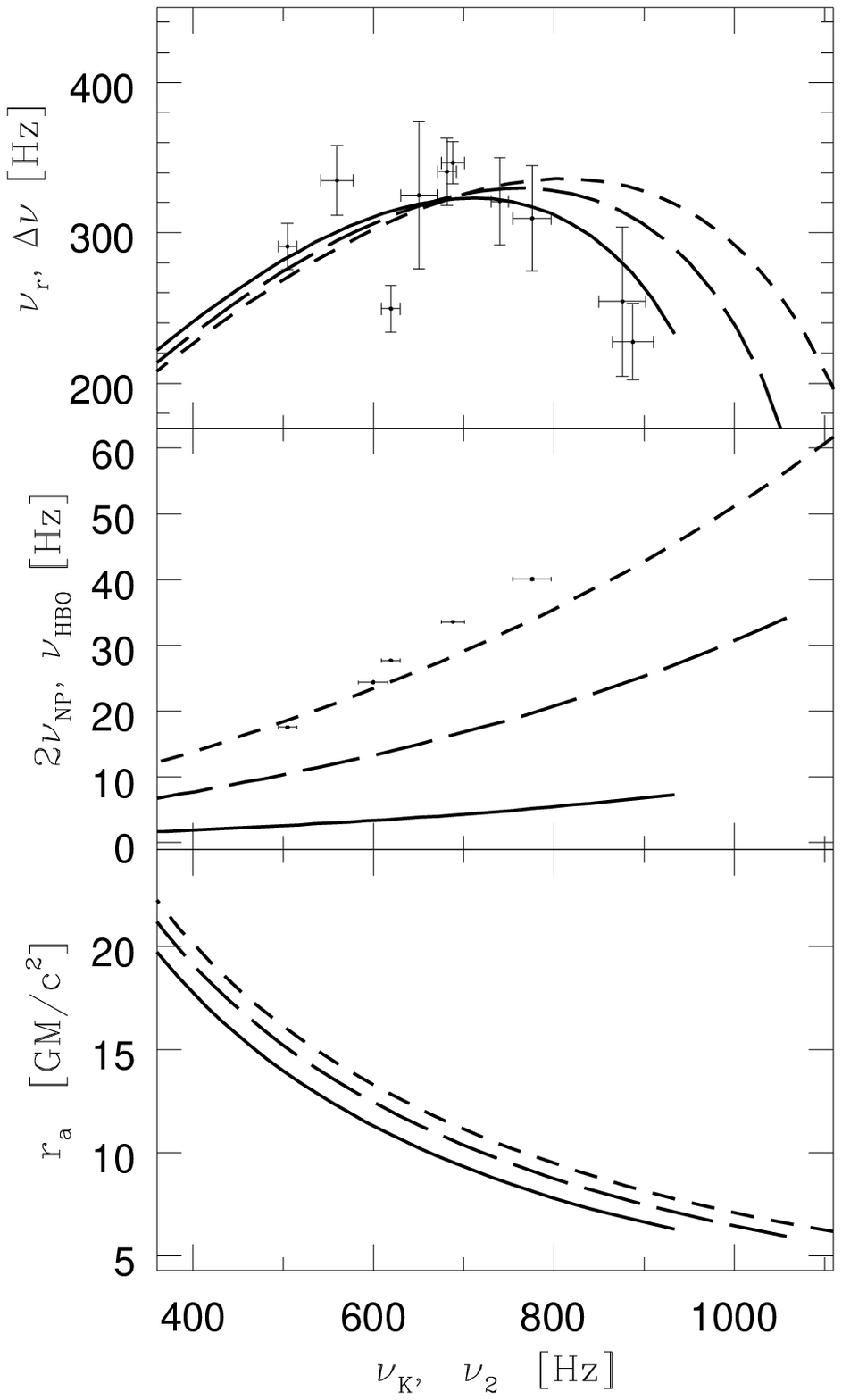}}
\vspace{-3.0truecm}
\caption{Geodesic sequences around UU stars that best fit
the $\nu_1$-$\nu_2$ data for \hbox{GX~5$-$1}, for different
stellar spin rates. The solid, long-dashed, and short-dashed
curves show the relations given by the geodesic sequences
that fit best for $\nus = 100\,$Hz ($M = 2.20\,\msun$, $\rp
= 6.3M$; $\chisqdof = 3.6$), $\nus = 363\,$Hz ($M =
2.21\,\msun$, $\rp = 5.8M$; $\chisqdof = 4.5$) and $\nus =
600\,$Hz ($M = 2.22\,\msun$, $\rp = 5.3M$; $\chisqdof =
5.6$). $\re = 3.02M$ for all three stellar models.
{\em Top panel\/}: Comparison of the best-fitting
$\nuap$-$\nuk$ relations with the observed $\nu_1$-$\nu_2$
correlation. 
{\em Middle panel\/}: Comparison of the predicted
$\nunp$-$\nuk$ relations with the observed $\nuhbo$-$\nu_2$
correlation.
{\em Bottom panel\/}: Circumferential apastron radii of the
highly eccentric geodesics that give the frequencies plotted
in the upper and middle panels.
        }
\label{GX5-1.NP}
\end{figure}

The behaviour of $2\nunp$ is also qualitatively inconsistent
with the behaviour of the HBO frequency in \gx{5$-$1}. This
is illustrated in Fig.~\ref{GX5-1.NP}, which compares the
$\nuhbo$-$\nu_2$ correlation observed in \gx{5$-$1} with the
$2\nunp$-$\nuk$ relations predicted by several geodesic
sequences constructed to give $\nuap$-$\nuk$ relations that
fit as well as possible the $\nu_1$-$\nu_2$ data.

\begin{figure} 
\centering
\epsfxsize=17.2truecm
\vspace{-0.3truecm}
\hbox{\hspace{-0.9 truecm}\epsfbox{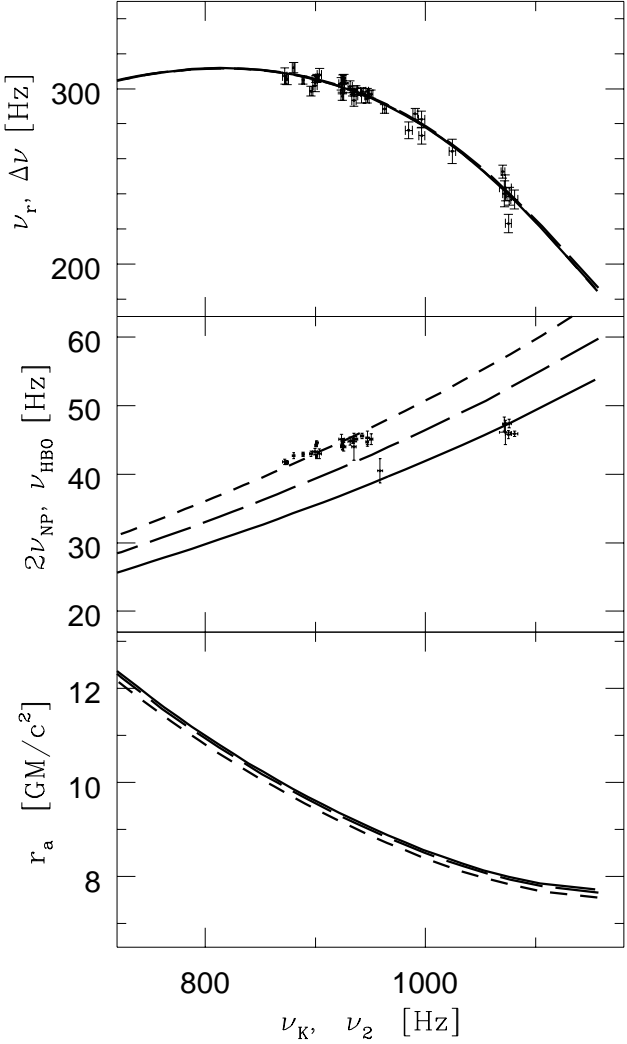}}
\vspace{-3.0truecm}
\caption{Geodesic sequences around UU stars that best fit
the $\nu_1$-$\nu_2$ data for \hbox{Sco~X-1}, for different
stellar spin rates. The solid, long-dashed, and short-dashed
curves show the relations given by the geodesic sequences
that fit best for $\nus = 450\,$Hz ($M = 2.13\,\msun$, $\re
= 3.32M$, $\rp = 4.9M$), $\nus = 500\,$Hz ($M =
2.16\,\msun$, $\re = 3.27M$, $\rp = 4.8M$) and $\nus =
550\,$Hz ($M = 2.18\,\msun$, $\re = 3.19M$, $\rp = 4.8M$).
$\chisqdof = 1.4$ for all three sequences.
    {\em Top panel\/}: Comparison of the best-fitting
$\nuap$-$\nuk$ relations with the observed $\nu_1$-$\nu_2$
correlation.
{\em Middle panel\/}: Comparison of the predicted
$2\nunp$-$\nuk$ relations with the observed $\nuhbo$-$\nu_2$
correlation.
{\em Bottom panel\/}: Circumferential apastron radii of the
highly eccentric geodesics that give the frequencies plotted
in the upper and middle panels.
     }
\label{ScoX1.NP}
\end{figure}

\begin{figure} 
\centering
\epsfxsize=17.2truecm
\vspace{-0.2truecm}
\hbox{\hspace{-0.9 truecm}
\epsfbox{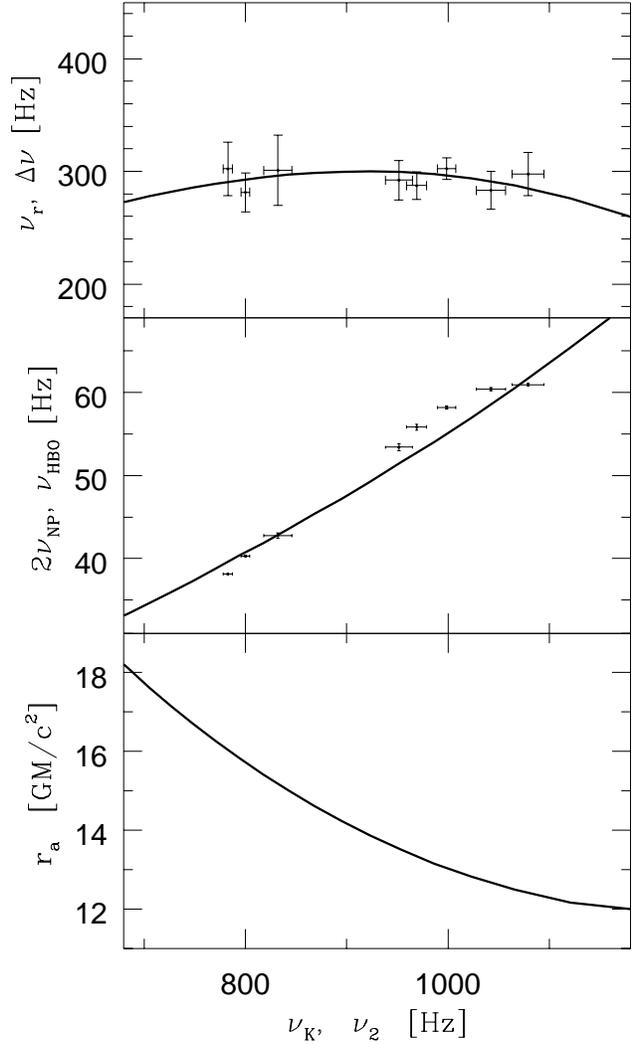}}
\vspace{-3.0truecm}
\caption{The geodesic sequence around UU stars that best
fits the $\nu_1$-$\nu_2$ data for \hbox{GX~17+2}. The
best-fit parameters are $M = 1.82\,\msun$ and $\rp = 4.2\,M
\approx 1.01\,\re$; $\chisqdof=0.39$.
{\em Top panel\/}: Comparison of the best-fitting
$\nur$-$\nuk$ relation with the observed $\Delta\nu$-$\nu_2$
correlation.
{\em Middle panel\/}: Comparison of the predicted
$2\nunp$-$\nuk$ relation with the observed $\nuhbo$-$\nu_2$
correlation, for $\nus=550\,$Hz.
{\em Bottom panel\/}: Circumferential apastron radii of the
highly eccentric geodesics that give the frequencies plotted
in the upper and middle panels.
     }
  \label{GX17+2.NP}
  \end{figure}

The basic difficulty is that in order to construct a
sequence of HEGs that gives a $\nuap$-$\nuk$ relation
similar to the observed $\nu_1$-$\nu_2$ correlation, the
stellar spin rate must be $\alt400\,$Hz, but for spin rates
this low, $2\nunp$ is much smaller than the observed HBO
frequencies. If the spin rate is increased to the values
$\agt600\,$Hz required to lift $2\nunp$ into the observed
frequency range of the HBO, $\nuk$ is shifted upward so much
that the $\nuap$-$\nuk$ relation given by the best-fitting
geodesic sequence is much flatter than the observed
$\nuhbo$-$\nuk$ correlation. Furthermore, attempting to fit
this correlation drives the stellar mass to the maximum
stable mass for the spin rate considered, which is
implausible.

\subsubsection{\sco1 HBO and kilohertz QPO frequencies}

The behaviour of $2\nunp$ is qualitatively inconsistent with
the observed behaviour of the HBO frequency in \sco1. This
is demonstrated in Fig.~\ref{ScoX1.NP}, which compares the
$\nuhbo$-$\nu_2$ correlation observed in \sco1 with the
$2\nunp$-$\nuk$ relations predicted by three sequences of
HEGs constructed to give the $\nuap$-$\nuk$ relation that
best fits the $\nu_1$-$\nu_2$ data for \sco1, for various
assumed spin rates. For the reason discussed previously, the
predicted $\nuap$-$\nuk$ relations are insensitive to the
star's spin frequency for the frequencies considered and
they therefore lie almost on top of one another.

Here the basic difficulty is that the $2\nunp$-$\nuk$
relations predicted by the geodesic sequences that give
acceptable fits to the $\nu_1$-$\nu_2$ data rise much more
steeply than the observed $\nuhbo$-$\nu_2$ correlation,
regardless of the spin rate of the star.

\begin{figure} 
\centering
\epsfxsize=17.2truecm
\vspace{-0.2truecm}
\hbox{\hspace{-0.9 truecm}
\epsfbox{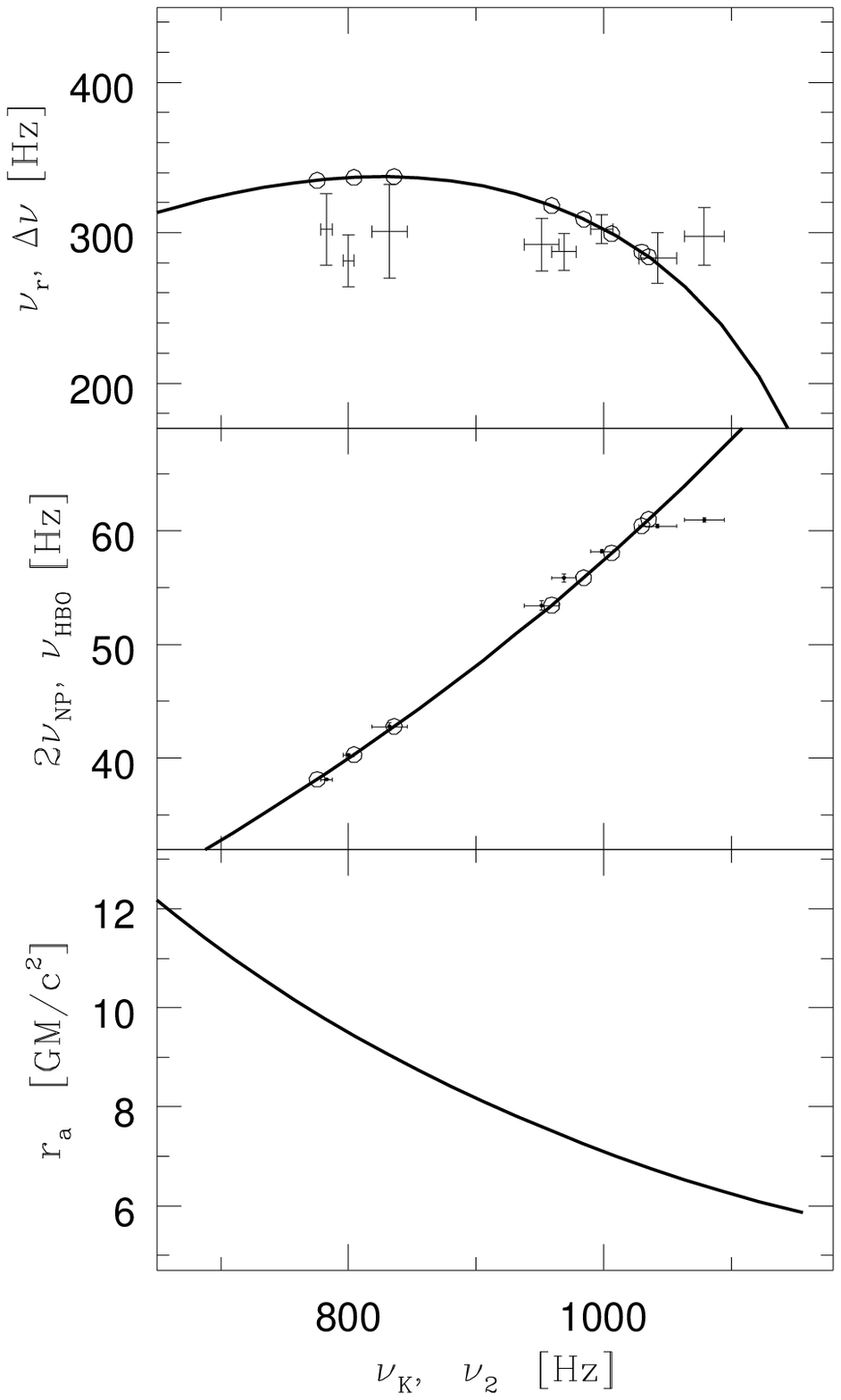}}
\vspace{-3.0truecm}
\caption{The geodesic sequence around UU stars that provides
the best {\em simultaneous\/} fit to the $\nu_1$-$\nu_2$ and
$\nuhbo$-$\nuk$ data for \hbox{GX~17+2}. The best-fit
parameters are $M = 2.21\,\msun$, $\rp = 5.0\,M \approx
1.7\,\re$, and $\nus = 660\,$Hz; $\chisqdof=4.3$. The open
circles show the frequency triplets $\{\nuapi, \nuki,
2\nunpi\}$ that fit best.
{\em Top panel\/}: Comparison of the resulting $\nur$-$\nuk$
relation with the observed $\Delta\nu$-$\nu_2$ correlation.
{\em Middle panel\/}: Comparison of the resulting
$2\nunp$-$\nuk$ relation with the observed $\nuhbo$-$\nu_2$
correlation.
{\em Bottom panel\/}: Circumferential apastron radii of the
highly eccentric geodesics that give the frequencies plotted
in the upper and middle panels.
     }
  \label{GX17+2.comb.NP}
  \end{figure}

\subsubsection{\gx{17$+$2} HBO and kilohertz QPO
frequencies}

The behaviour of $2\nunp$ predicted by the sequence of HEGs
that best fits the $\nu_1$-$\nu_2$ data for \gx{17$+$2} is
qualitatively similar to the $\nuhbo$-$\nu_2$ correlation
observed in this source, but is inconsistent in detail. This
is shown in Fig.~\ref{GX17+2.NP}. The top panel compares the
$\nur$-$\nuk$ relation given by the best-fitting sequence of
HEGs with the observed $\Delta\nu$-$\nu_2$ correlation. Very
highly eccentric geodesics ($\ra/\rp \approx 4$) are
required in order to obtain an acceptable fit. The
$\chisqdof$ is substantially less than 1 because of the
large uncertainties reported for the $\nu_1$ and $\nu_2$
measurements. The fit is relatively insensitive to the spin
frequency, for low to moderate spin frequencies.

The middle panel of Fig.~\ref{GX17+2.NP} compares the
$2\nunp$-$\nuk$ relation given by the sequence of HEGs used
in the top panel that is most similar to the observed
$\nuhbo$-$\nu_2$ correlation. The relation shown is not a
joint fit to the HBO and kilohertz QPO frequencies, but was
instead obtained by adjusting the spin frequency to give the
$2\nunp$-$\nuk$ relation that is closest to the observed
$\nuhbo$-$\nu_2$ correlation. The spin frequency that gives
the best fit is 550~Hz. Although the $2\nunp$-$\nuk$
relation passes near the $\nuhbo$-$\nu_2$ data, it misses
many of the data points by much more than their
uncertainties. The best-fitting $\nuap$-$\nu_2$ relation
clearly does not correctly predict the $\nuhbo$-$\nu_2$
correlation observed.

We then explored whether an acceptable {\em joint\/} fit to
the $\nu_1$-$\nu_2$ and $\nuhbo$-$\nu_2$ correlations is
possible. Fig.~\ref{GX17+2.comb.NP} shows the
$2\nunp$-$\nuk$ and $\nuap$-$\nuk$ relations given by the
geodesic sequence that best fits both the $\nuhbo$-$\nu_2$
and the $\nu_1$-$\nu_2$ data. The high precision of the HBO
frequency measurements forces the fit to make the $2\nunp$
values agree closely with them. However, this is possible
only if the eccentricity of the geodesics is reduced, but
this makes the $\nuap$-$\nuk$ relation steeper than the
observed $\nu_1$-$\nu_2$ correlation. The best-fit mass is
driven to the maximum stable mass in an effort to
compensate, but the compensation is insufficient to give an
acceptable fit. Also, the periastron radius is driven far
from the stellar surface ($\rp=1.7\,\re$). Although the
geodesics favoured by the attempt to fit jointly the
$\nuhbo$-$\nu_2$ and $\nu_1$-$\nu_2$ data are less eccentric
than those required by the $\nu_1$-$\nu_2$ data, they are
still highly eccentric ($\ra/\rp \approx 2$).

\begin{figure} 
\centering
\epsfxsize=17.2truecm
\vspace{-0.3 truecm}
\hbox{\hspace{-0.9 truecm}\epsfbox{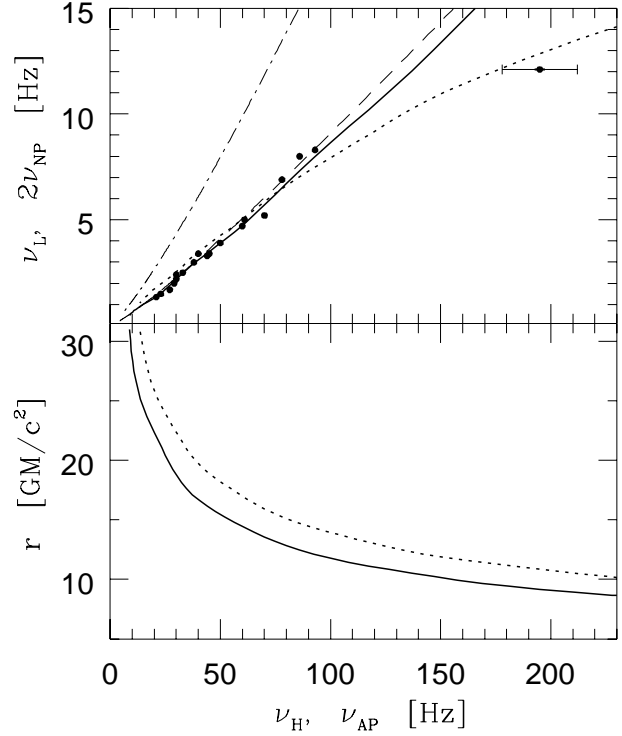}}
\vspace{-7.3 truecm}
\caption{
{\em Upper panel\/}: Comparison of the $2\nunp$-$\nuap$
relations given by sequences of infinitesimally eccentric
geodesics with the correlation between the frequencies
$\nul$ and $\nuh$ observed in \hbox{Cir~X-1}. The solid and
dotted curves show respectively the frequency relations
given by sequences of MEGs around UU stars that give
relations similar to the $\nul$-$\nuh$ correlation observed
at $\nuh < 100\,$Hz, for $\nu_{\rm s}=700\,$Hz ($M =
2.1\,\msun$, $j=0.26$) and $\nu_{\rm s}=900\,$Hz ($M =
1.3\,\msun$, $j=0.50$). The dashed and dash-dotted curves
show respectively the $2\nunp$-$\nuap$ relations given by
moderately eccentric geodesics around black holes with the
same values of $M$ and $j$ as for the high- and low-mass neutron stars.
{\em Lower panel\/}: Circumferential radii of the neutron
star geodesics that give the frequencies plotted in the
upper panel.
        }
\label{CirX1.ncg} 
\end{figure}

\subsection{\cir1 QPO frequencies}
\label{CompareCir1}

Although the peculiar accreting X-ray star \cir1 is
generally thought to be a neutron star (see Tennant, Fabian
\& Shafer 1986), Psaltis et al.\ (1999b) have suggested that
its QPOs may be related to some of the QPOs observed in
black-hole candidates, because power spectra of its X-ray
emission are similar in some ways to the power spectra of
black holes. SVM have proposed that the frequencies $\nuh$
and $\nul$ of the higher- and lower-frequency features
observed in \cir1 are the apsidal precession frequency
$\nuap$ and twice the nodal precession frequency $\nunp$ of
a sequence of slightly inclined IEGs around a neutron star,
although they did not attempt a fit of the resulting
frequency relations with the frequencies of the QPOs
observed in \cir1.

\begin{figure} 
\centering
\epsfxsize=20truecm
\vspace{-6 truecm}
\hbox{\hspace{-2.3truecm}\epsfbox{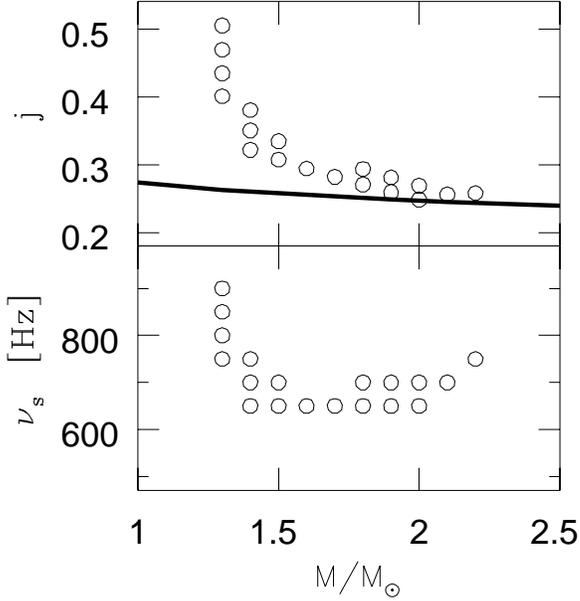}}
\vspace{-6truecm}
\caption{EOS UU-based neutron star models (open circles)
that allow $2\nunp$-$\nuap$ relations similar to the
$\nul$-$\nuh$ correlation observed in \hbox{Cir~X-1} for
$\nuh < 100\,$Hz. The solid line in the upper panel shows
for comparison the $j$-$M$ track along which sequences of
geodesics around a spinning black hole can be constructed
that give $2\nunp$-$\nuap$ relations similar to the observed
$\nul$-$\nuh$ correlation.}
\label{CirX1.survey}
\end{figure}

\begin{figure}
  \centering
  \epsfxsize=17.2truecm
  \vspace{-0.3 truecm}
  \hbox{\hspace{-0.9 truecm}\epsfbox{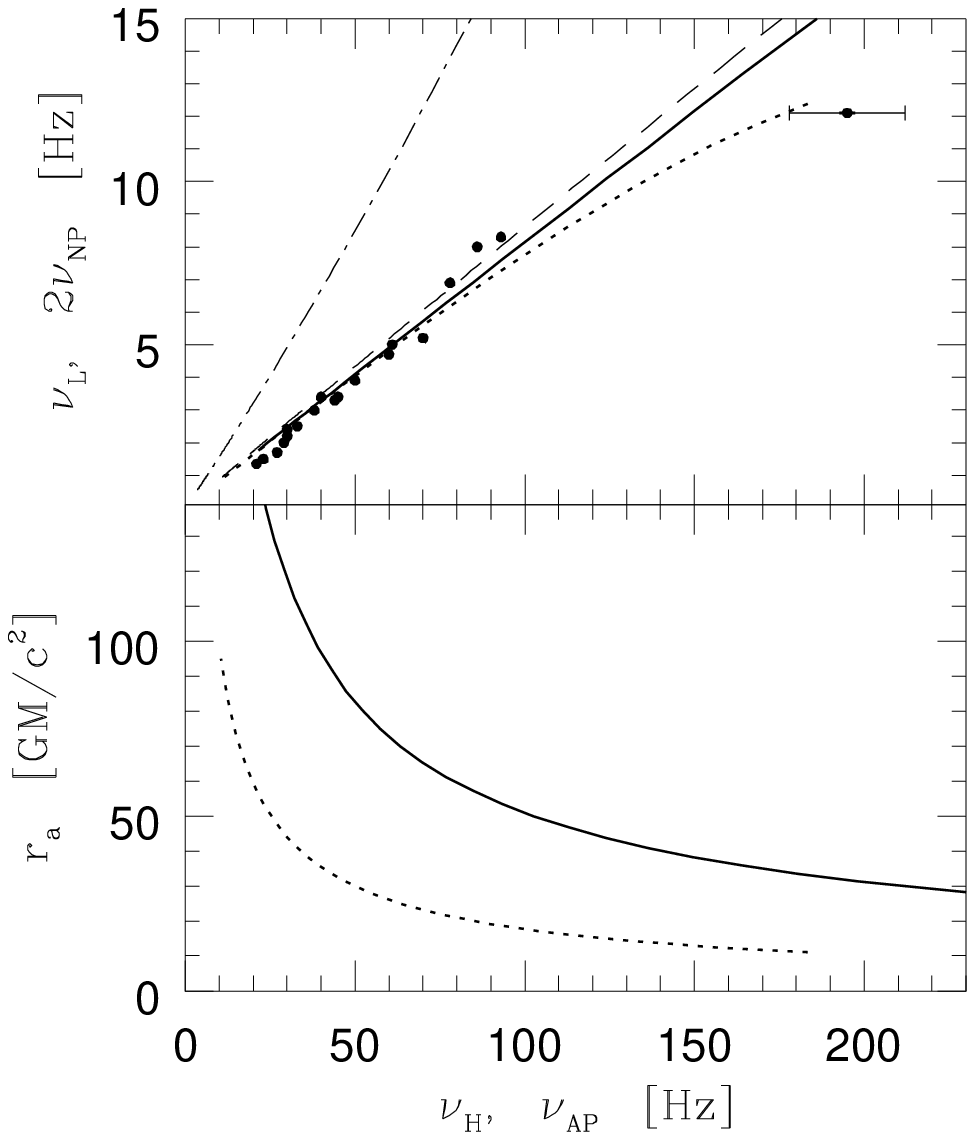}}
  \vspace{-7.3 truecm}
  \caption{
{\em Upper panel\/}: Comparison of the $2\nunp$-$\nuap$
relations given by sequences of highly eccentric geodesics
with the correlation between the frequencies $\nul$ and
$\nuh$ observed in \hbox{Cir~X-1}. The solid and dotted
curves show respectively the frequency relations given by
sequences of HEGs around UU stars that give relations
similar to the $\nul$-$\nuh$ correlation observed at $\nuh <
100\,$Hz, for $\nu_{\rm s}=550\,$Hz ($M = 2.2\,\msun$,
$j=0.18$, $\rp = 4.1M$) and $\nu_{\rm s}=900\,$Hz ($M =
1.3\,\msun$, $j=0.50$, $\rp=11.0M$).  The dashed and
dash-dotted curves show, for comparison, the
$2\nunp$-$\nuap$ relations given by geodesics around black
holes with the same values of $M$, $j$, and $\rp$ as for the
high- and low-mass neutron stars.
{\em Lower panel\/}: Circumferential radii of the neutron
star geodesics that give the frequencies plotted in the
upper panel.
  }
  \label{CirX1.heg}
  \end{figure}

Here we first investigate whether a sequence of IEGs can be
found that gives a $2\nunp$-$\nuap$ relation that is
consistent both with the QPO frequencies reported by Shirey
et al.\ (1996, 1998) and Shirey (1998), and with the
centroid frequencies of the peaked noise and QPO reported by
Tennant (1987). No error bars were reported by Shirey et
al., so our conclusions are necessarily tentative. The solid
and dashed lines in Fig.~\ref{CirX1.ncg} show the
$2\nunp$-$\nuap$ relations given by the two sequences of
IEGs that roughly bracket the range of masses and spin
frequencies that agree qualitatively with the observed
$\nul$-$\nuh$ correlation for $\nuh<100\,$Hz (see
Fig.~\ref{CirX1.survey}). In order to achieve qualitative
agreement with the observed frequency correlation, the
neutron star must be rapidly spinning ($\nus \gta 650\,$Hz)
and the radii of the geodesics in the sequence must vary
smoothly by a factor of 2.5, from $\sim25\,M$ to
$\sim10\,M$, as the frequencies increase.

Inclusion of the higher-frequency ($\nul \approx 12\,$Hz,
$\nuh \approx 200\,$Hz) measurement reported by Tennant
(1987) makes it much more difficult to find a sequence of
geodesics that gives a frequency relation similar to the
observed correlation. Only IEGs around stars with very low
masses and very high spin rates give frequency relations
that pass close to the lower-frequency data of Shirey et
al.\ (1996, 1998) and then bend over sufficiently to pass
near the higher-frequency measurement. Sufficient flattening
of the $2\nunp$-$\nuap$ relation is achieved only for stars
that are highly deformed by their rapid spins and therefore
have large mass quadrupole moments whose effect largely
cancels (at small radii) the prograde precession produced by
frame-dragging. The effect of a large mass quadrupole may be
seen clearly in Fig.~\ref{CirX1.ncg} by comparing the
$2\nunp$-$\nuap$ relations for neutron stars with the
relations for black holes with the same $M$ and $j$; notice
the particularly large difference for $M = 1.3\,\msun$ and
$j=0.50$; the neutron star model is still fairly far from
the mass-shedding limit, which is $0.93\,\msun$.

 \begin{table*}
 \begin{center}
 \begin{minipage}{190mm}
\caption{Fits of geodesic sequences to the measured
frequencies of the kilohertz and lower-frequency QPOs.
\label{table.geodesics}}
      \begin{tabular}{||cc|ccccccccccccc||}
      \hline \hline
      Source  & Model & EOS & $\nus$[Hz] & $M[\msun]$ & $j$ &  $\re$[km]
   &  $\re$  & $\risco$
       & $\rp$ & $r_{\rm a, min}$  & $r_{\rm a, max}$ & $\chisqdof$ & 
$n-p$$^a$ &
  Fig. \\
      \hline
      Sco X-1  & IEG & ---$^b$ & 0 & 1.99 & 0 & 10.7 & 3.64 & 6.0
                      &$\ra$ & 6.2 & 7.0 & 38 & $72-37$ & 
\ref{ScoX1.cir.kHz}\\
               & IEG & UU & 450 & 2.21 & 0.145 & 9.8 & 3.02 & 5.52
                      &$\ra$ & 5.7 & 6.5 & 35.5 & $72-37$ & 
\ref{ScoX1.cir.kHz}\\
               & MEG$^c$ & ---$^b$ & 0 & 1.90 & 0 & 10.8 & 3.86 & 6.00
                      & 6.25 & 6.4 & 8.4 & ---$^c$ & ---$^c$ & 
\ref{ScoX1.isco.kHz}\\
               & MEG & UU & 450 & 2.21 & 0.145 & 9.8 & 3.02 & 5.52
                      & 5.52 & 5.9 & 7.8 & 21 & $72-38$ & 
\ref{ScoX1.isco.kHz}\\
               & MEG$^c$ & L & 300 & 1.94 & 0.197 & 15.3 & 5.35 & 5.51
                      & 6.18 & 6.3 & 8.0 & ---$^c$ & ---$^c$ &
\ref{ScoX1.isco.kHz}\\
               & MEG$^c$ & L & 600 & 1.98 & 0.423 & 16.2 & 5.57 & ---$^d$
                      & 6.17 & 6.3 & 7.8 & ---$^c$ & ---$^c$ &
\ref{ScoX1.isco.kHz}\\
               & HEG & ---$^b$ & 0 & 1.89 & 0 & 10.8 & 3.88 & 6.00
                      & 5.3 & 8.9 & 11.0 & 1.4 & $72-38$ & 
\ref{ScoX1.kHz}\\
               & HEG & ---$^b$ & 0 & 1.86 & 0 & 10.0 & 3.65 & 6.00
                      & 5.1 & 9.8 & 11.8 & 2.2 & $72-38$ & 
\ref{ScoX1.kHz}\\
               & HEG & UU & 450 & 2.13 & 0.158 & 10.5 & 3.32 & 5.49
                      & 4.9 & 8.0 & 10.0 & 1.4 & $72-38$ & 
\ref{ScoX1.NP}\\
               & HEG & UU & 500 & 2.16 & 0.173 & 10.4 & 3.27 & 5.44
                      & 4.8 & 8.0 & 10.0 & 1.4 & $72-38$ & 
\ref{ScoX1.NP}\\
               & HEG & UU & 550 & 2.18 & 0.187 & 10.2 & 3.19 & 5.40
                      & 4.8 & 7.8 & 9.8  & 1.4 & $72-38$ & 
\ref{ScoX1.NP}\\
      \hline
      GX~340+0 & HEG & ---$^b$ & 0 & 1.86 & 0 & 10.0 & 3.66 & 6.00
                      & 5.8 & 10.6 & 16.2 & 2.3 & $12-8$ & 
\ref{GX340.kHz}\\
               & HEG & ---$^b$ & 0 & 2.12 & 0 & 10.3 & 3.30 & 6.00
                      & 7.0 & 7.0 & 11.1 & 0.8 & $12-8$ & 
\ref{GX340.kHz}\\
      \hline
      4U 1728$-$34
               & IEG & UU & 363 & 1.92 & 0.140 & 10.9 & 3.85 & 5.57
                      & $\ra$ & 5.9 & 6.8 & 14.7 & $16-9$ & ---\\
               & IEG & UU & 900 & 2.20 & 0.314 & 10.6 & 3.24 & 5.01
                      & $\ra$ & 5.3 & 6.1 & 9.4 & $16-9$ & ---\\
               & HEG & UU & 363 & 1.76 & 0.149 & 11.1 & 4.27 & 5.55
                      & 4.7 & 9.9 & 12.7 & 1.7 & $16-10$ & 
\ref{4U1728.kHz},\ref{4U1728.NP}\\
               & HEG & AU$\delta$ & 363 & 1.76 & 0.156 & 11.6 & 4.46 & 
5.54
                    & 4.7 & 9.9 & 12.7 & 1.7 & $16-10$ & 
\ref{4U1728.kHz}\\
               & HEG & L & 363 & 1.97 & 0.239 & 15.4 & 5.30 & 5.44
                      & 5.3 & 6.2 & 8.7 & 8.6 & $16-10$ & 
\ref{4U1728.kHz}\\
      \hline
      GX~5$-$1  & HEG & ---$^b$ & 0 & 2.17 & 0 & 10.1 & 3.16 & 6.00
                      & 6.5 & 6.5 & 13.5 & 3.6 & $20-12$ & 
\ref{GX5-1.kHz}\\
                & HEG & ---$^b$ & 0 & 2.09 & 0 & 10.4 & 3.38 & 6.00
                      & 6.5 & 6.5 & 13.9 & 1.2 & $18-11$ & 
\ref{GX5-1.kHz}\\
                & HEG & UU & 100 & 2.20 & 0.032 & 9.8 & 3.02 & 5.89
                      & 6.3 & 6.7 & 13.5 & 3.6 & $20-12$ & 
\ref{GX5-1.NP}\\
                & HEG & UU & 363 & 2.21 & 0.117 & 9.8 & 3.02 & 5.62
                      & 5.8 & 7.0 & 14.9 & 4.5 & $20-12$ & 
\ref{GX5-1.NP}\\
                & HEG & UU & 600 & 2.22 & 0.194 & 9.9 & 3.02 & 5.37
                      & 5.3 & 7.6 & 16.0 & 5.6 & $20-12$ & 
\ref{GX5-1.NP}\\
      \hline
      GX~17$+$2  & HEG & UU & 550 & 1.81 & 0.224 & 11.1 & 4.16 & 5.35
                       & 4.2 & 12.4 & 16.0 & 0.39 & $16-10$ & 
\ref{GX17+2.NP}\\
                 & HEG+NP & UU & 660 & 2.21 & 0.220 & 10.1 & 3.10 & 5.29
                       &  5.0 & 6.9 & 10.0 & 4.3 & $24-11$ & 
\ref{GX17+2.comb.NP}\\
       \hline
       Cir~X-1   & NCG & UU & 700 & 2.10 & 0.256 & 10.8 & 3.48 & 5.21 
                       & $\ra$ & 9 & 22  & ---$^e$ & ---$^e$ & 
\ref{CirX1.ncg}\\ 
                 & NCG & UU & 900 & 1.30 & 0.509 & 12.6 & 6.56 & ---$^d$
                       & $\ra$ & 11 & 26 & ---$^e$ & ---$^e$ & 
\ref{CirX1.ncg}\\
                 & HEG & UU & 550 & 2.20 & 0.184 & 10.2 & 3.13 & 5.40
                       & 4.1  & 34 & 150 & ---$^e$ & ---$^e$ & 
\ref{CirX1.heg}\\
                 & HEG & UU & 900 & 1.30 & 0.509 & 12.6 & 6.56 & ---$^d$
                       & 11   & 11 & 63  & ---$^e$ & ---$^e$ & 
\ref{CirX1.heg}\\
      \hline \hline
      \end{tabular}
$^a$Number of frequency measurements ($n$) minus the number
of parameters in the model ($p$). $^b$Fits for nonrotating
stars do not depend on the EOS, provided that it is hard
enought to support the best-fit stellar mass and that $\re <
r_{\rm p, min}$. The values of $\re$ listed for all
nonrotating stars are for models constructed using the UU
EOS, except for the $1.86\,\msun$ stars, which were
constructed using EOS~C. $^c$These models are not fits and
would give very large values of $\chisqdof$. They are the
geodesic sequences reported by SV as adequately fitting the
\sco1 kilohertz QPO frequency data (see text). $^d$No
innermost stable circular orbit exists, because the radius
of the star is too large. $^e$No formal fits were possible,
because no errors have been reported for most of the
frequency measurements used; the fit was therefore done by
eye, giving each point the same weight.
  \end{minipage}
  \end{center}
  \end{table*}

Given the difficulty of approximating the $\nul$-$\nuh$
correlation observed in \cir1 using IEG sequences, we
explored whether it would be easier to reproduce the
correlation using a sequence of HEGs. Our results are shown
in Fig.~\ref{CirX1.heg}. The lower-frequency portion of
the $\nul$-$\nuh$ correlation drives inclusion of very
highly eccentric ($\ra/\rp > 6$) geodesics that extend
far from the neutron star ($\ra/\re > 10$) in the
sequence of HEGs. Inclusion of the higher-frequency
measurement of Tenant (1987) again makes it much more
difficult to find a sequence of geodesics that gives a
frequency relation similar to the observed correlation. Only
sequences of HEGs around stars with very low masses and very
high spin rates give frequency relations that pass near all
the measurements. The geodesic sequences that work best have
infinitesimally or moderately eccentric geodesics as their
most compact members, but even these geodesics have
periastron radii much larger than the large equatorial radii
of these stellar models ($\rp/\re=1.7$).

\section{Power density spectra}
\label{spectra}

SV, SVM, and Karas (1999) simply assumed that the most
prominent frequencies in the X-ray waveforms produced by
clumps moving on geodesics around a neutron star are
$2\nunp$, $\nuap$, and $\nuk$. No attempt was made to
compute the frequencies that would be generated or the
relative amplitudes of the oscillations at these
frequencies. Here we report calculations of the frequencies
and amplitudes that would be produced.

In order to determine which are the dominant frequencies
that would be generated and the relative amplitudes of the
oscillations at these frequencies, we carried out an
extensive suite of time-dependent, numerical simulations of
the geodesic motion of clumps around spinning neutron stars.
For the purposes of this investigation we simply assumed
that all clumps are formed and move on the same geodesic,
that they are held together by some force, and that they
move forever on this same geodesic. We modeled the clumps as
spheres of constant radius $d$. In order to facilitate the
computations, we assumed $d=0.1\re$, which is about $100$
times larger than allowed by the dynamical constraints (see
Section~\ref{dynamics}). We treated the motion of the clumps
in full general relativity and took into account special
relativistic Doppler and other effects on the radiation
absorbed and emitted by the clumps, but we did not perform
ray tracing. Results using full ray tracing would be
qualitatively similar to the results reported here. We
surveyed a broad range of geodesic parameters, including
$\rp$, $\ra$, and the tilt of the geodesic, and a wide range
of viewing angles.

\begin{figure*}
  \centering
  \epsfxsize=17truecm
  \vspace{-0.2truecm}
  \hbox{\hspace{-0.2 truecm}\epsfbox{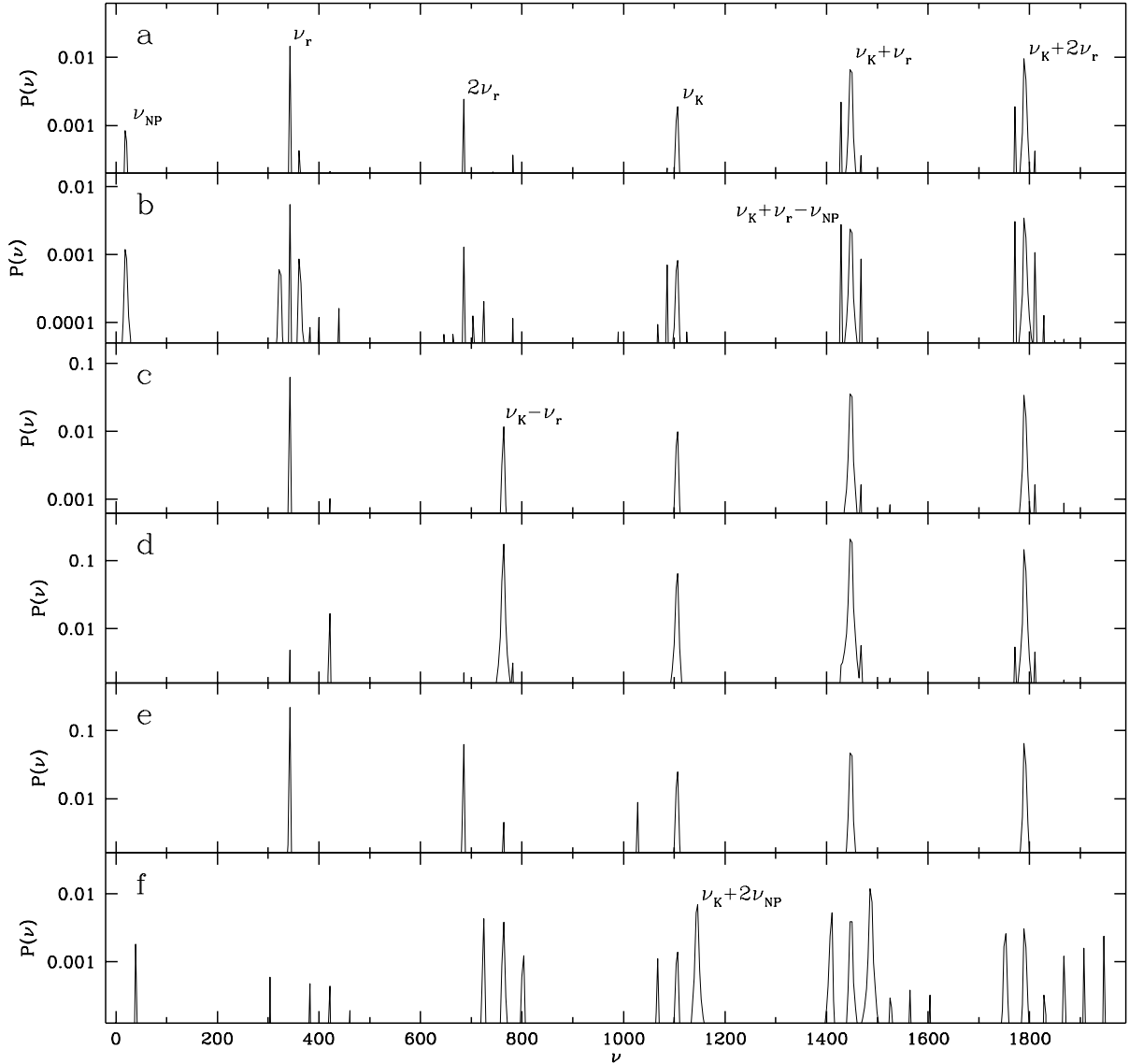}}
  \vspace{-1.6 truecm}
  \caption{
Typical power spectra produced by six different effects, for
roughly spherical clumps on highly eccentric, tilted
geodesics around a spinning neutron star. The heights of the
peaks are not proportional to their total power, because
their widths are different (total powers are listed in
Table~\ref{table.frequencies}. From top to bottom, the
effects considered are (see text for details):
    (a)~occultation of a uniformly bright neutron star
surface;
    (b)~occultation of a belt with a Gaussian brightness
distribution of semi-width $0.1\re$ around the stellar equator;
    (c)~reflection of radiation from a uniformly bright
neutron star surface;
    (d)~radiation by a clump with a radius-independent
proper luminosity;
    (e)~radiation by a clump with a radius-dependent proper
luminosity;
    (f)~radiation by a clump in a tilted orbit that emits
only while it is passing through a rarified disc of height
$h(r) = 0.05\,r$ in the equatorial plane of the system.
    In this example, $M = 1.76\,\msun$, $\nus = 363\,$Hz
$\rp = 4.7M$, and $\ra = 10.0M$, yielding $\nuk = 1107\,$Hz,
$\nur = 343\,$Hz, and $\nunp = 19\,$Hz. The geodesic was
assumed to be inclined $10^{\circ}$ relative to the
equatorial plane. The line of sight was assumed to be
inclined $45^{\circ}$.
    As explainlained in the text, the actual power spectrum
produced by a particular clump model is a linear combination
of two or more of the spectra shown.
  }
  \label{HEG.PDS}
  \end{figure*}

  \begin{figure*}
  \centering
  \epsfxsize=17truecm
  \vspace{-0.2truecm}
  \hbox{\hspace{-0.2 truecm}\epsfbox{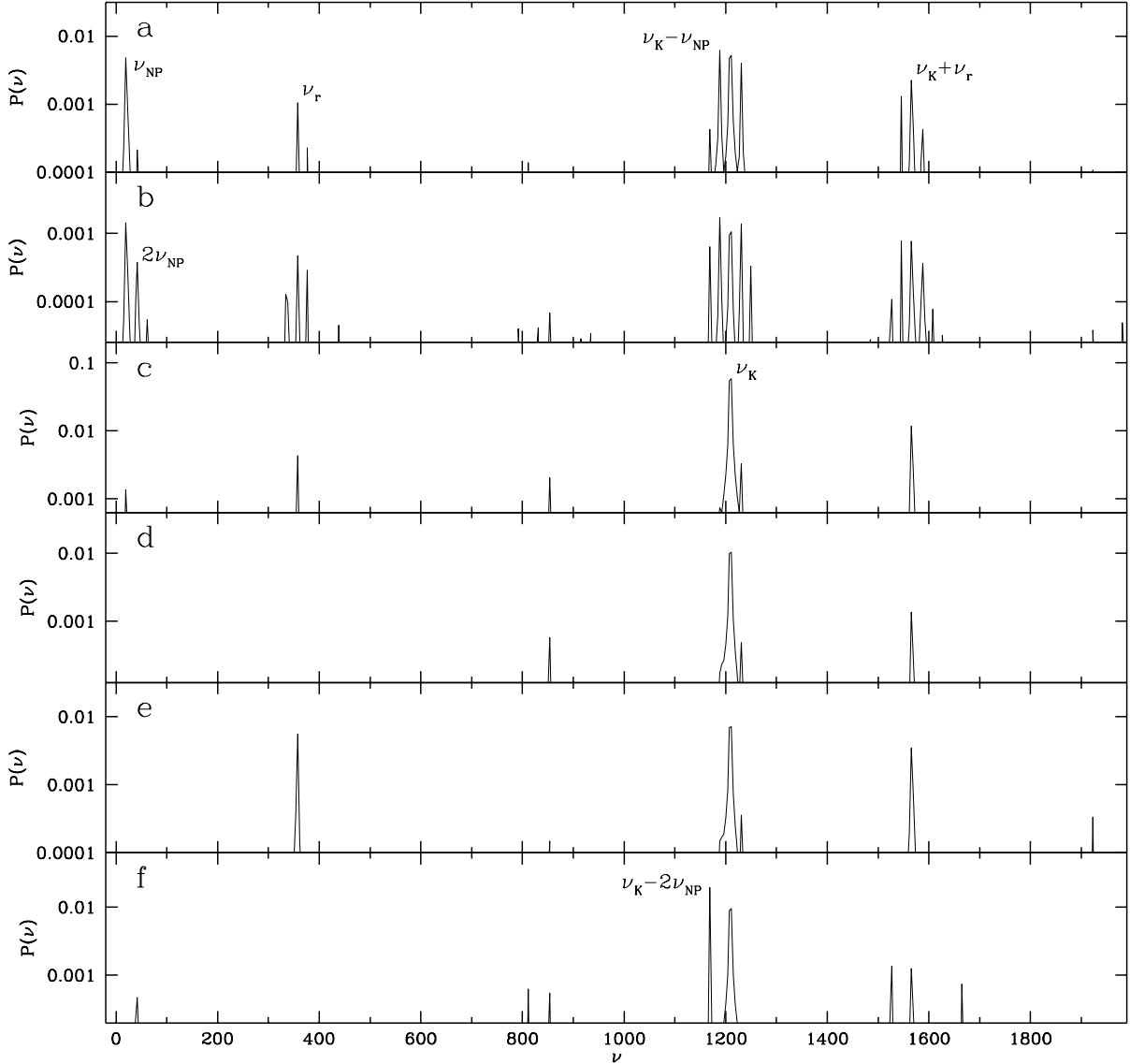}}
  \vspace{-1.6 truecm}
  \caption{
Typical power spectra produced by the same six effects
considered in Fig.~15, but for moderately eccentric
geodesics. As in Fig.~15, the heights of the peaks are not
proportional to their total power, because their widths are
different (total powers are listed below, in
Table~\ref{table.frequencies}). The stellar mass and spin
rate and the inclinations of the geodesic and line of sight
are the same as in Fig.~15, but $\rp = 5.6M$ and $\ra =
6.7M$, corresponding to an eccentricity $\varepsilon = 0.09$
and yielding $\nuk = 1211\,$Hz, $\nur = 358\,$Hz, and $\nunp
= 19\,$Hz.
    As explainlained in the text, the actual power spectrum
produced by a particular clump model is a linear combination
of two or more of the spectra shown.
  }
  \label{NCG.PDS}
  \end{figure*}

\subsection{Highly eccentric geodesics}

We simulated six different effects that could modulate the
X-ray flux seen by a distant observer. The six panels of
Fig.~\ref{HEG.PDS} show power spectra that are typical of
the spectra produced by each of these effects, for clumps
moving on geodesics with the high eccentricities required to
produce kilohertz QPO frequencies qualitatively similar to
those observed in the atoll and Z sources. The heights of
the peaks are not proportional to their total power, because
the motions involved are not periodic, which causes the
widths of the peaks to differ. The frequencies and relative
total powers of the most prominent peaks are listed below in
Table~\ref{table.frequencies}. As discussed further below,
the actual power spectrum produced by a particular clump
model is a linear combination of two or more of the spectra
shown in Fig.~\ref{HEG.PDS}. We now discuss briefly each of
six effects (the labels a--f used here match the labels of
the panels in Fig.~\ref{HEG.PDS}).

(a)~{\em Occultation of the stellar surface}. An optically
thick clump moving around a neutron star on a geodesic may
occult a portion of the stellar surface, thereby modulating
the X-ray flux from the surface seen by a distant observer.
Occultation occurs only for geodesics that satisfy $\rp <
(\re + d)/\cos i$, i.e., that closely approach the stellar
surface. Some of the best-fit geodesics listed in
Table~\ref{table.geodesics} do not satisfy this constraint
unless the observer's sightline is close to the plane of the
disc.

In the absence of apsidal and nodal precession, the star
would be occulted periodically, with frequency $\nuk =\nur$.
However, general relativistic effects near the neutron star
make $\nuk > \nu_r$. For the geodesics considered here, a
clump typically orbits the star three or four times each
time it moves radially inward and outward at the radial
epicyclic frequency $\nur$. The rate of azimuthal phase
advance varies significantly with the radial position of a
clump and hence the occultation caused by the azimuthal
motion is not periodic. The only strictly periodic motion is
the radial epicyclic motion, which causes the observed flux
to vary with frequency $\nur$ as the extent of the
occultation increases and decreases with the motion of the
clump toward and away from the star (depending on the value
of $\ra$ and the inclination of the geodesic and the line of
sight, a clump that is near $\ra$ may not occult the star at
all).

The strongest peaks in the power density spectrum of the
X-ray waveform are at the radial epicyclic frequency $\nur$
and the combination frequencies $\nuk +\nur$ and
$\nuk+2\nur$ (see Fig.~\ref{HEG.PDS}a). The next strongest
peak is at $2\nur$. Weaker nodal precession sidebands appear
at $\nuk+\nur-\nunp$ and $\nuk+2\nur-\nunp$ in this power
spectrum and the others discussed below only if the geodesic
is significantly tilted. {\em There is much less power at
the azimuthal frequency $\nuk$ and almost none at the
apsidal precession frequency $\nuap =\nuk-\nu_r$}. These are
the frequencies that SV, SVM, and Karas (1999) assumed would
be the dominant frequencies in the power spectrum. The power
in the peak at $\nuk$ is typically $\sim$4--5 times less
than the power in the peak at $\nur$; the power at $\nuap$
is too low for the peak to be visible in
Fig.~\ref{HEG.PDS}a. Nodal precession modulates the angle
between the orbital plane and the observer's line of sight,
producing a very weak peak at $\nunp$; there is no
significant power at $2\nunp$.

(b)~{\em Occultation of a bright equatorial band}. If
radiation comes predominantly from a narrow belt around the
star's rotation equator, nodal precession has a stronger
effect on the waveform. Fig.~\ref{HEG.PDS}b shows a typical
power spectrum of the waveform for this situation. In this
example, the surface brightness of the star was assumed to
vary as $\exp(-z^2/\sigma^2)$, where $z$ is the distance of
the point on the stellar surface from the equatorial plane
and $\sigma=0.1\re$.

The strongest peaks are again at $\nur$, $\nuk+\nur$, and
$\nuk+2\nur$. The next strongest peaks, which are about half
as strong, are at $\nuk+\nur-\nunp$ and $\nuk+2\nur-\nunp$.
There is a peak at $\nunp$ that is slightly weaker than
these nodal-precession sidebands. {\em Again, there is much
less power at $\nuk$ and almost none at $\nuap$}. The peak
at $\nuk$ is $\sim$3--4 times weaker than the three
strongest peaks in the power spectrum and is significantly
weaker even than the peak at $\nunp$. The next strongest
peaks, which have almost as much power as the peak at
$\nuk$, are at the nodal sideband frequencies $\nur-\nunp$,
$\nur+\nunp$, and $\nuk+2\nur+\nunp$. The peak at $\nuap$ is
too weak to be visible and there is no significant power at
$2\nunp$.

(c)~{\em Reflection of stellar radiation by clumps}.
Fig.~\ref{HEG.PDS}c shows the power spectrum for a typical
waveform generated by clumps that reflect radiation coming
from the stellar surface. In this example, we assumed that
the surface brightness of a clump varies with its distance
$r$ from the center of the star as $1/r^2$ and that the flux
is proportional to its surface brightness and the projected
area of the illuminated part of the surface seen by the
distant observer. The Doppler shifts caused by the velocity
of the clump relative to the stellar surface and to the
observer were included.

The strongest peak produced by reflection is at $\nuk+\nur$.
There are also strong peaks at $\nur$ and $\nuk+2\nur$.
There are peaks at $\nuk$ and $\nuap$, but the power in
these peaks is $\sim$4--5 times less than in the peak at
$\nuk+\nur$ and $\sim$3--4 times less than in the peaks at
$\nur$ and $\nuk+2\nur$. There is no significant power at
either $\nunp$ or $2\nunp$.

(d)~{\em Radiation by constant temperature clumps}.
Significant radiation by clumps with the small size and high
density required by gas dynamical constraints (see
Section~\ref{dynamics}) is ruled out by their very small
areas compared to that of the accreting star and by the
upper bound on their temperature imposed by their origin in
a cool disc flow and the requirement that they be in
pressure equilibrium with their environment (otherwise they
will expand and dissipate on a timescale much shorter than
the lower bound on the durations of their wavetrains imposed
by the observed coherence of the kilohertz QPOs).
Nevertheless, for completeness we explored the waveform that
would be produced by such radiation.

Even if the luminosity of a clump is constant in time in its
rest frame, the general relativistic redshift and other
effects will cause the luminosity seen by a distant observer
to increase and decrease with the clump's distance $r$ from
the center of the star, which varies at the radial epicyclic
frequency. For the illustrative calculations reported here,
we took into account only the Doppler shift caused by the
clump's motion relative to the distant observer.

A typical power spectrum is shown in Fig.~\ref{HEG.PDS}d.
The strongest peak is at $\nuk+\nur$. There is also a peak
at $\nuk+2\nur$ which is about half as strong. There is a
peak at $\nuap$, which is also about half as strong as the
peak at $\nuk+\nur$, as well as a peak at $\nuk$, which is
$\sim$4 times weaker than the peak at $\nuk+\nur$ and only
about half as strong as the peak at $\nuk+2\nur$. There are
no other significant peaks. Hence, {\em even in this case
the peaks at $\nuk+\nur$ and $\nuk+2\nur$ are as strong or
stronger than the peaks at $\nuk$ and $\nuap$}.

(e)~{\em Radiation by clumps heated by radiation from or
interaction with the star}. Again, significant radiation by
clumps is ruled out by their very small areas and by the
upper bound on their temperature imposed by the requirement
that they be in pressure equilibrium. Furthermore, any
interaction of the clumps with the stellar surface or with a
stellar magnetic field that is strong enough to release an
appreciable fraction of their kinetic energy would disrupt
them and cause their orbits to decay. Nevertheless, for
completeness we also explored the waveform that would be
produced by such radiation.

The rest-frame luminosity of a clump heated by radiation
from the star or by interaction with a stellar atmosphere or
magnetic field necessarily varies strongly with $r$. The
luminosity seen by a distant observer is also affected by
the variation of the Doppler shift with $r$. For the
illustrative calculations reported here, we assumed that
heating causes the brightness of a clump to vary with radius
as $1/[(r-\re)^2 + 0.01\re^2]$.

\begin{table*}
    \begin{center}
    \begin{minipage}{156 mm}
\caption{Total power in the prominent peaks in the twelve
typical power spectra shown in Figs.~15 and~16.$^1$\label{table.frequencies}}
      \begin{tabular}{||cc|ccccccccccccccc||}
      \hline \hline
                    \vspace{0.15cm}
                 &  &  \multicolumn{5}{c}{HEG~($\rp = 4.7$,  $\ra =
10.0$)$^3$} & &
                    &  \multicolumn{5}{c}{MEG~($\rp = 5.6$,  $\ra =
6.7$)$^3$} & \\
     Frequency $^2$  &  & a & b & c & d & e & f &  &
         & a & b & c & d & e & f & \\
     \hline
      $\nunp$  &  & 0.10 & 0.40 & --- & --- & --- & --- &  &
                  & 0.54 & 0.82 & --- & --- & --- & --- &  \\
     $2\nunp$  &  & --- & --- & ---  & --- & --- & --- &  &
                  & --- & 0.24 & --- & --- & --- & --- &  \\
                \multicolumn{17}{c}{\dotfill} \\
     $\nu_r$         &  & 0.90 & 0.90 & 0.79 & --- & 1 & --- &  &
                  & 0.10 & 0.24 & --- & --- & 0.37 & --- &  \\
      $\nu_r -\nunp$   &  & --- & 0.21 & --- & --- & --- & --- &  &
                  & --- & 0.12 & --- & --- & --- & --- & \\
     $\nu_r +\nunp$   &  & --- & 0.23 & --- & --- & --- & --- &  &
                  & --- & 0.14 & --- & --- & --- & --- &  \\
     $2\nu_r$         &  & 0.16 & 0.22 & --- & --- & 0.30  & --- &  &
                  & --- & --- & --- & --- & --- & --- &  \\
                 \multicolumn{17}{c}{\dotfill} \\
  $\nuk -\nu_r$  &  & --- & --- & 0.21 & 0.54 & --- & 0.23 &  &
                  & --- & --- & --- & --- & --- & --- &  \\
  $\nuk -\nu_r -2\nunp$  &  & --- & --- & --- & --- & --- & 0.21 &  &
                  & --- & --- & --- & --- & --- & --- &  \\
                 \multicolumn{17}{c}{\dotfill} \\
  $\nuk$         &  & 0.22 & 0.27 & 0.24 & 0.26 & 0.21 & 0.12 &  &
                  & 1 & 1 & 1 & 1 & 1 & 1 &  \\
 $\nuk -\nunp$  &  & --- & 0.12 & --- & --- & --- & --- &  &
                  & 0.63 & 0.85 & --- & --- & --- & --- &  \\
 $\nuk +\nunp$  &  & --- & --- & --- & --- & --- & --- &  &
                 & 0.40 & 0.68 & --- & --- & --- & --- &  \\
 $\nuk -2\nunp$  &  & --- & --- & --- & --- & --- & --- &  &
                  & --- & 0.30 & --- & --- & --- & 0.88 &  \\
  $\nuk +2\nunp$  &  & --- & --- & --- & --- & --- & 0.63 &  &
                  & --- & 0.17 & --- & --- & --- &  \\
           \multicolumn{17}{c}{\dotfill} \\
 $\nuk +\nu_r$  &  & 0.89 & 0.86 & 1 & 1 & 0.47 & 0.40 &  &
                  & 0.26 & 0.45 & 0.12 & --- & 0.27 & --- &  \\
 $\nuk +\nu_r -\nunp$  &  & 0.14 & 0.46 & --- & --- & --- & --- &  &
                  & 0.12 & 0.36 & --- & --- & --- & --- &  \\
 $\nuk +\nu_r +\nunp$  &  & --- & 0.15 & --- & --- & --- & --- &  &
                  & --- & 0.25 & --- & --- & --- & --- &  \\
 $\nuk +\nu_r -2\nunp$  &  & --- & --- & --- & --- & --- & 0.39 &  &
                  & --- & --- & --- & --- & --- & --- &  \\
 $\nuk +\nu_r +2\nunp$  &  & --- & --- & --- & --- & --- & 1 &  &
                  & --- & --- & --- & --- & --- & --- &  \\
              \multicolumn{17}{c}{\dotfill} \\
  $\nuk +2\nu_r$  &  & 1 & 1 & 0.72 & 0.53 & 0.50 & 0.24  & &
                  & --- & --- & --- & --- & --- & --- &  \\
 $\nuk +2\nu_r -\nunp$  &  & 0.12 & 0.51 & --- & --- & --- & --- &  &
                  & --- & --- & --- & --- & --- & --- &  \\
  $\nuk +2\nu_r +\nunp$  &  & --- & 0.19 & --- & --- & --- & --- &  &
                  & --- & --- & --- & --- & --- & --- &  \\
 $\nuk +2\nu_r -2\nunp$  &  & --- & --- & --- & --- & --- & 0.22 &  &
                  & --- & --- & --- & --- & --- & --- &  \\
      \hline \hline
      \end{tabular}
    $^1$The total power in each peak is normalised to the total power in the strongest peak produced by each effect.
    $^2$Only peaks with total power $\ge10$\% of the
strongest peak at frequencies $\le2,000\,$Hz are listed.
    $^3$The six columns for the HEG and MEG models list the
power ratios for the prominent frequencies produced by the
six different effects discussed in the text. These are
    (a) occultation of a uniformly bright neutron star surface;
    (b) occultation of a belt with a Gaussian brightness distribution of
semi-width $0.1 \re$ around the stellar equator;
    (c) reflection of radiation from a uniformly bright neutron star
surface;
    (d) radiation by a clump with a radius-independent proper luminosity;
    (e) radiation by a clump with a radius-dependent proper luminosity;
    (f) radiation by a clump in a tilted orbit that emits
only while it is passing through a rarified disc of height
$h(r) = 0.05\,r$ in the equatorial plane of the system.
    \par\parindent10pt  The actual power spectrum produced
by a particular clump model is a linear combination of two
or more of these spectra. For example, a clump that reflects
radiation from the stellar surface [effect~(e)] or produces
a flare of emission as it passes through the orbital plane
[effect~(f)] will generally also occult the star
[effects~(a) and~(b)].
  \end{minipage}
  \end{center}
  \end{table*}

A typical power spectrum is shown in Fig.~\ref{HEG.PDS}e.
The strongest peak is at $\nur$, with peaks at $\nuk+\nur$
and $\nuk+2\nur$ that are about half as strong and a peak at
$2\nur$ that is about a third as strong. {\em There is a
peak at $\nuk$, but it is $\sim$5 times weaker than the main
peak and less than half as strong as the peaks at
$\nuk+\nur$ and $\nuk+2\nur$; there is no significant power
at $\nuap$, $\nunp$, or $2\nunp$}.

(f)~{\em Interaction of clumps with a rarefied gaseous
disc}. Finally, we consider the possibility that the
luminosity increases when clumps on tilted geodesics pass
through the equatorial plane. Such flaring could be caused,
for example, by interaction of the clumps with a rarified,
remnant gaseous disc in the plane near the star. The
additional luminosity produced by enhanced dissipation would
have to be very small in order to avoid causing the clumps'
orbital motion to decay more rapidly than is allowed by the
observed coherence of the kilohertz QPOs. This scenario was
constructed specifically to produce a peak as strong as
possible at $2\nunp$. The results presented here assume
the clump brightness varies as $\exp(-z^2/h^2)$, where 
$h(r) = 0.05\,r$ is the height of the disc, but they are
insensitive to the precise value of $h$.

Even in this scenario, the two strongest peaks are at
$\nuk+\nur+2\nunp$ and $\nuk+2\nunp$. The next strongest
peaks are at $\nuk+\nur$ and $\nuk+\nur-2\nunp$ and have
$\sim$2.5 times less power than the peak at
$\nuk+\nur+2\nunp$. There are two peaks at $\nuap$ and
$\nuap-2\nunp$, but they have $\sim$4--5 times less power
than the peak at $\nuk+\nur+2\nunp$ and are no stronger than
the peaks at $\nuk+\nur$ and $\nuk+\nur-2\nunp$. There is
also a peak at $\nuk$, but it has ten times less power than
the peak at $\nuk+\nur+2\nunp$. The peak at $2\nunp$ has 12
times less power than the peak at $\nuk+\nur+2\nunp$ and 6
times less power than the peak at $\nuk+2\nunp$ (it is too
weak to be listed in Table~\ref{table.frequencies}). There
is no significant power at $\nunp$.

We emphasise that {\em the actual power spectrum produced by
a particular clump model is a linear combination of two or
more of the spectra shown in Fig.~\ref{HEG.PDS}}. For
example, a clump that reflects radiation from the stellar
surface [effect~(c)] or produces a flare of emission as it
passes through the orbital plane [effect~(f)] will generally
also occult the star [effects~(a) and~(b)]; indeed,
occultation is typically the dominant effect.

\subsection{Nearly circular geodesics}

For completeness, we also computed the waveforms that would
be produced by each of the six effects just discussed if the
clumps were moving on nearly circular geodesics, even though
such geodesics are excluded by the observed QPO frequency
relations. Fig.~\ref{NCG.PDS} shows power spectra of the
typical waveforms produced by such geodesics. The
frequencies and relative total powers of the most prominent
peaks are again listed in Table~\ref{table.frequencies}.

For orbits with eccentricities $\varepsilon \equiv (\ra
-\rp)/(\ra + \rp) \alt 0.1$, there is (almost) always a
strong peak at $\nuk$. Usually, there is also a significant
peak at $\nuk+\nur$. Occultation can produce relatively
strong peaks at $\nunp$ and $2\nunp$ and significant peaks
at the nodal sideband frequencies of $\nuk$ and $\nuk+\nur$,
but most effects do not produce peaks at these frequencies.

We emphasise that for infinitesimally or only moderately eccentric orbits, $r$ must
be $<(\re + d)/\cos i$ for the star to be even
partially occulted. Hence, in order to produce any
occultation, such geodesics must have
azimuthal frequencies
 \begin{equation}
 \nuk > 1550\,{\rm Hz}\,
 (\sqrt{2}\cos i)^{3/2}
 \left(\frac{10\,{\rm km}}{\re + d}\right)^{3/2}
 \left(\frac{M}{2\,\msun}\right)^{1/2} \;.
 \end{equation}
This requirement is inconsistent with the interpretation of
the upper kilohertz QPO frequency as an azimuthal frequency,
unless all the kilohertz QPO systems are highly inclined,
which is unlikely and contradicts other evidence concerning
their inclinations.

\section{Gas dynamical constraints}
\label{dynamics}

The geodesic precession model assumes that at a particular
radius the accretion flow onto the neutron star changes from
a disc flow to gas clumps moving on geodesics with the same
or very similar apastron and periastron radii $\ra$ and
periastron $\rp$. It assumes further that the geodesics that
are populated by clumps at any given time have the
azimuthal, apsidal precession, and nodal precession
frequencies required to explain the frequencies of the QPOs
observed in the atoll and Z sources. Finally, the hypothesis
assumes that as the accretion rate varies, the orbital
parameters of the geodesics that are populated change in the
way needed to produce the observed correlated variation of
$\nul$ or $\nuhbo$, $\nu_1$, and $\nu_2$.

As we now discuss, the behaviour of the accreting gas
required by the geodesic precession hypothesis is very
difficult to achieve in any physically consistent gas
dynamical picture. For the purposes of the present section
we set aside the results of the previous section, which
showed that the frequencies generated by orbiting clumps are
not the frequencies assumed by previous workers, and focus
on the gas dynamical requirements of the model.

\subsection{Circularity of the flow in the disc}

We begin by showing that the stream lines in the accretion
disc are necessarily nearly circular, in contrast to the
highly eccentric geodesics required to produce orbital
frequencies qualitatively similar to the observed
frequencies of the kilohertz QPOs. The reason is that the
apsidal precession rate varies rapidly with the radius of
the geodesic. Hence any eccentric, adjacent streamlines
would intersect in less than an orbital period, creating
strong shear stresses, pressure gradients, and possibly
shock waves that will quickly circularise the motion.

To see this, consider gas streaming along two nearby,
eccentric geodesics with periastron radii $r_{\rm p1}$ and
$r_{\rm p2} = r_{\rm p1} + \delta\rp$. For the present
purpose we assume the geodesics are ellipses in Euclidean
space and describe each of them using the familiar
radius-angle relation $r = \rp (1 + \varepsilon)/(1 +
\varepsilon\cos\phi)$, where the angle $\phi$ is measured
from the geodesic's periastron. Suppose the geodesics are
initially concentric and aligned and have the same, finite
eccentricity $\varepsilon$. In the absence of shear or
pressure stresses, the different precession rates of the two
geodesics will bring them into contact at the `osculation'
point after a time $\delta t$. The angle
between the axes of the two ellipses at this time can be
obtained from the osculation point conditions $r_1 = r_2$
and $[dr/d\phi]_1 = [dr/d\phi]_2$, where the subscripts
indicate the radius-angle relations for the two ellipses. 

Solving for $\delta\phi$ gives, to the lowest order in
$\delta\rp/\rp$, $\sin\delta\phi = (\delta\rp/\rp) (1 -
\varepsilon^2)^{1/2}/\varepsilon$. The differential rate of
periastron advance is $\delta\nuap \approx
(\partial\nuap/\partial\rp) \delta\rp$. Using the
lowest-order post-Newtonian expression for $\nupa$ (see
Markovic 2000) we obtain
 \begin{equation}
 \label{deltat}
 {\delta t} =
 \frac{(\delta\phi/2\pi)}
      {(\partial\nupa/\partial\rp)\delta\rp}
 \simeq \frac{1}{15\pi}
 \frac{(1 +\varepsilon)^{3/2}(1 - \varepsilon)^{1/2}}
      {\varepsilon}
 \frac{1}{\nu_{\rm K,N}},
 \end{equation}
where $\nu_{\rm K,N} \equiv (1 - \varepsilon)^{3/2}
(GM/\rp^3)^{1/2}/2\pi$ is the Newtonian Kepler frequency.
Equation~(\ref{deltat}) shows that the two streamlines will
intersect in less than one orbital period even if they are
only very slightly eccentric ($\varepsilon \agt 0.02$).
Moderately or highly eccentric streamlines wil intersect in
much less than one orbital period.

In reality, the two streamlines will not cross. Instead,
shear stresses and gas pressure forces will grow to oppose
their differential apsidal precession, forcing the two
streamlines to deviate from purely geodesic motion. The
result will be to circularise both streamlines within a time
$\sim {\delta t}$, i.e., in less than one orbital period.

\subsection{Effects of tidal forces}

The high coherence ($Q \sim 100$) of the kilohertz QPOs
requires that the postulated gas clumps survive for at least
$Q$ orbital periods. However, the strong gravitational tidal
force near the star will tidally disrupt clumps unless they
are either confined by external forces or are very small.

The rapid damping of noncircular flow in the inner disc
makes persistence of the highly eccentric geodesic motion
required by the geodesic precession model impossible unless
the clumps are isolated from the general gas flow, i.e.,
unless there is an extensive region around the star where
the density of the interclump gas is negligible. The clumps
therefore cannot be confined by the surrounding gas. Nor can
they be confined by the stellar magnetic field, because the
torque produced by a field strong enough to confine the gas
in the clumps would be strong enough to cause them to
deviate from purely geodesic motion. Therefore, the clumps
can persist for $Q$ orbital periods only if they are very
small.

To estimate how long a clump of radius $d$ can persist,
consider two elements of the clump moving on concentric
orbits with the same eccentricity and orientation but
slightly different periastra $\rp$ and $\rp + 2d$.
Neglecting apsidal precession and other relativistic
effects, the difference between the azimuthal frequencies of
the two elements is $\delta\nu_{\rm K,N} \approx
(2d/\rp)(3/2)\nu_{\rm K,N}$. Hence, after a time $t$ the
clump will be sheared out a distance $\Delta d \approx \rp
2\pi\delta\nu_{\rm K,N} t$. The X-ray modulation produced by
such a clump will certainly be markedly changed by the time
the clump has been spread a quarter of the way around the
star (i.e., over a phase interval of $\pi/2$). This will
happen in less than $Q$ orbital periods unless
 \begin{equation}
 \label{tidal}
 \frac{d}{\rp} \ll \frac{1}{12 Q}
 \approx 8\ee{-4}\left(\frac{100}{Q}\right) \;.
 \end{equation}
In the last expression on the right we have scaled $Q$ by
100, the minimum value required for consistency with the
observed coherence of the kilohertz QPOs.
Inequality~(\ref{tidal}) is a strong upper bound on the size
of a roughly spherical clump that would survive tidal
disruption for $Q$ orbital periods. As we show below, this
bound is a very significant constraint on precessing clump
models.

\begin{figure} 
\centering
\epsfxsize=8.4truecm
\vspace{+0.1truecm}
\hbox{\hspace{-0.0 truecm}\epsfbox{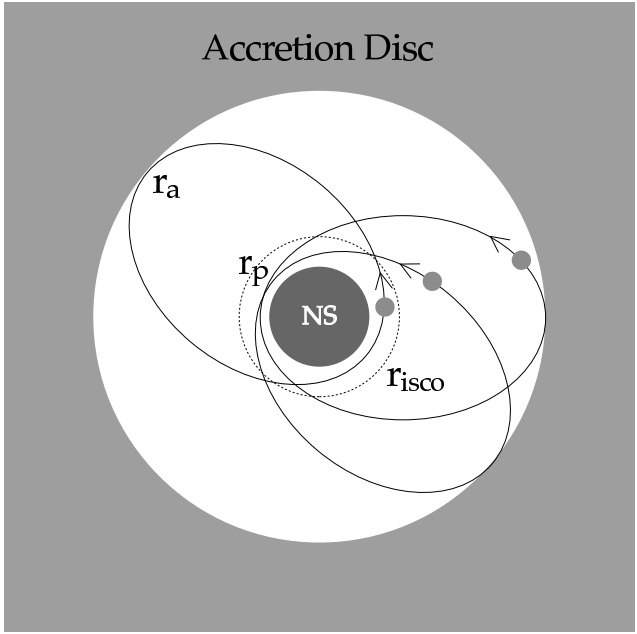}}
\vspace{-0.0 truecm}
\caption{ Schematic depiction of the breakup of the
accretion disc into dense clumps on nearly identical highly
eccentric geodesics as required by the geodesic precession
model. Here $\ra$ and $\rp$ are the apastron and periastron
radii of the geodesic, $\risco$ is the radius of the
innermost stable circular orbit, and the dark disc shows the
size of a typical neutron star. In reality, geodesics near
the star are very open rosettes rather than ellipses: a
clump on these geodesics would go around the star 3--4 times
as it moves from $\ra$ to $\rp$. The figure shows that the
clumps have relative velocities comparable to their
azimuthal velocities, even if they are on geodesics with
similar values of $\ra$ and $\rp$.}
\label{fig.geodesics} 
\end{figure}

\subsection{Constraints on the gas density in the clumps}

The requirement that the interaction of clumps with the
interclump gas not affect significantly their orbits for $Q$
orbital periods leads to a lower bound on the density
$\rho_{\rm c}$ of the gas in the clumps relative
to the density $\rho_{\rm i}$ of the interclump medium. As
shown above, the interclump gas must follow nearly circular
orbits; hence any clumps on highly eccentric geodesics must
move at supersonic speeds through the interclump gas.
Assuming that it can be held together by some means, a gas
clump of size $d$ moving with supersonic velocity ${\bf
v}_{\rm c} - {\bf v}_{\rm g}$ through interclump gas of
density $\rho_{\rm i}$ will experience a drag force
 \begin{equation}
 \label{dragf}
 F_{\rm D} =
 C_{\rm D} d^2 \rho_{\rm i}
 ({\bf v}_{\rm c} - {\bf v}_{\rm i})^2
 \sim C_{\rm D} d^2 \rho_{\rm i} v_{\rm c}^2 \;,
 \end{equation}
where $C_{\rm D}$ ($\sim$1) is the ballistic coefficient. A
clump of mean density $\rho_{\rm c}$ has momentum $d^3
\rho_{\rm c} v_{\rm c}$ and hence  persistence of geodesic
motion for $Q$ orbital periods would require a density ratio
 \begin{equation}
 \label{rho.cont}
 \frac{\rho_{\rm c}}{\rho_{\rm i}} \gg
 \frac{rQ C_{\rm D}}{d} \gg 10^5\, C_{\rm D}
\left(\frac{Q}{100}\right)^2 \;,
 \end{equation}
where in writing the last inequality on the right we have
used condition~(\ref{tidal}).

The requirement that the clumps not collide too frequently
leads to an even more stringent lower bound on the density
of the gas in the clumps relative to the mean density near
the star. The mean clump-clump collision rate near the star
is $R = n_c \sigma v_{\rm rel} \approx n_c 4\pi d^2 v_{\rm
rel}$, where $n_c$ is the density of clumps near the star
and $v_{\rm rel}$ is their typical relative velocity. Even
if the clumps all move on geodesics with the same values of
$\rp$ and $\ra$, their periastron phases will be different
(see Fig.~\ref{fig.geodesics}) and hence they will approach
each other with relative velocities that are comparable to
$v_{\rm K}$. Now $n_c = 3{\bar\rho}_c/(4\pi\rho_c d^3)$,
where ${\bar\rho}_c$ is the mean mass density in the clumped
region. Hence $R \approx 3 (\bar\rho_c/\rho_c)(v_{\rm
K}/d)$. The observed coherence of the kilohertz QPOs
requires $2\pi QR\rp/v_{\rm K} \alt 1$, or 
 \begin{equation}
 \frac{\rho_c}{\bar\rho_c} \agt \frac{6\pi\rp Q}{d} \gg 
 2\ee6 \left(\frac{Q}{100}\right)^2,
 \end{equation}
where in the last inequality we have again used
condition~(\ref{tidal}).

These dynamical constraints show that the density contrast
in the clumps required by the geodesic precession hypothesis
is enormously greater than the density contrasts and
magnetic field fluctuations considered in the magnetospheric
(Alpar \& Shaham 1985; Lamb et al.\ 1985) and sonic-point
(Miller et al.\ 1998) beat-frequency models of QPOs; in the
latter models, the fluctuations have relative amplitudes
$\alt1$ and move with the mean flow. How clumps with such
extremely high densities would form or persist in a gas
dynamical flow and how they would be replenished are
unknown. A large number of clumps would have to be produced
to carry a substantial fraction of the accretion flow and
produce significant X-ray modulation. Clumps would have to
be formed from the gas in the region where the orbits are
nearly circular and then ``shot'' quickly into the required
highly eccentric orbits in an inner, highly rarefied region
near the star, in order to avoid interaction with the gas
moving on nearly circular orbits. However, any forces
capable of producing the required sudden deviation from
circular motion would also be strong enough to disrupt the
clumps and cause them to deviate strongly from purely
geodesic motion. As noted earlier, what picks out the
particular geodesics required by the geodesic precession
hypothesis is also unknown.

\subsection{Implications for QPO amplitudes}

The small size necessary for clumps to persist as long as
required by the observed coherence of the kilohertz QPOs
means that the projected area of a clump is only a very
small fraction of the projected area of the stellar surface.
The fractional area of a clump is
 \begin{equation}
 f = \frac{\pi d^2}{\pi R^2} < \frac{d^2}{\rp^2}
 \ll 7\ee{-7} \;.
 \label{SolidAngle}
 \end{equation}
This is also an upper bound on the fractional modulation of
the X-ray flux that can be produced by clumps, whether the
modulation is caused by occultation of the surface of the
star, reflection of radiation from the star, or emission
from a clump at about the same temperature as the stellar
surface.

\section{Discussion and conclusions}
\label{discussion}

If it could be established that the frequencies of the most
prominent humps, bumps, and QPOs observed in power spectra
of the X-ray emission from neutron star and black hole LMXBs
are simply the characteristic frequencies of gas clumps
orbiting these objects and that hydrodynamic, radiation, and
magnetic forces have no significant effects, so that their
motion is purely geodesic motion, this would be an
extraordinarily important result. In particular, it would
mean that measurements of these frequencies would provide a
very simple way to explore the effects of strong-field
gravity and the properties of compact objects. This is why
we have carefully investigated this suggestion. Here we
discuss our principal results and summarize our conclusions.

\subsection{Geodesics and frequency relations}
\label{correlationDiscussion}

We first explored whether sequences of geodesics could be
found with azimuthal and apsidal precession frequencies
$\nuap$ and $\nuk$ that agree, respectively, with the
observed frequencies $\nu_1$ and $\nu_2$ of the lower and
upper kilohertz QPOs. We also investigated whether geodesic
sequences could be found for which the first overtone
$2\nunp$ of the nodal precession frequency and $\nuk$ and
$\nuap$ agree, simultaneously, with the frequencies of the
low-frequency and kilohertz QPOs. Here we briefly summarise
the results of these investigations.

\subsubsection{Infinitesimally and moderately eccentric
geodesics}

In Section~\ref{CompareMEGs} we explored whether it is
possible to construct sequences of infinitesimally or
moderately eccentric geodesics that give $\nuap$-$\nuk$
relations consistent with the observed correlations between
$\nu_1$ and $\nu_2$. 

We first investigated infinitesimally eccentric geodesics.
The geodesic precession hypothesis makes no predictions
about the apastron and periastron radii $\ra$ and $\rp$ of
the geodesics or how they vary, except that they must
(obviously) be larger than the equatorial radius of the
star. In exploring infinitesimally eccentric geodesics, we
set $\rp=\ra$ and treated $\ra$ as a free parameter, subject
only to the requirement that it exceed the radius of the
innermost stable circular orbit $\risco$. The parameters in
this model are therefore the EOS, mass, and spin frequency
of the star, and an orbital radius for each pair of
kilohertz QPO frequencies that was fit.

We found that the $\nuap$-$\nuk$ relations given by the
best-fitting sequences of infinitesimally eccentric
geodesics are much steeper than the observed $\Dnu \equiv
\nu_2 - \nu_1$ vs.\ $\nu_2$ correlation. Our $\nuap$-$\nuk$
relations are consistent with the $\nuap$-$\nuk$ relations
reported by SV, which are also much steeper than the
frequency data presented in their paper. The qualitative
disagreement between the best-fitting frequency relations
for infintesimally eccentric geodesics and the kilohertz QPO
frequency data is indicated by the very large values of
$\chisqdof$ found ($\gta35$ for the Z source \sco1 and
$\gta10$ for the atoll source \fu{1728$-$34}).

We note that the qualitative disagreement between the
frequency relations for infinitesimally eccentric geodesics
and the frequency data is a serious difficulty for the
resonant ring model recently proposed by Psaltis \& Norman
(2000) to explain the QPOs in both neutron star and black
hole LMXBs. Not only does this disc model predict a dense
spectrum of power spectral peaks, in contrast to the small
number of peaks observed, the frequencies are necessarily
the frequencies of nearly circular geodesics. In order for
``hydrodynamic corrections'' (i.e., gas pressure forces) to
be large enough to shift the frequencies by $\sim$15\% to bring them into agreement with the data, the disc
height would have to be $\sim0.2\,r$, invalidating the
narrow ring approximation on which the model is based. The
behaviour of the disc frequencies would then differ
qualitatively from the behaviour predicted by the narrow
ring model.

Next, we investigated geodesics with finite but moderate
eccentricities. Once geodesics with finite eccentricities
are considered, the geodesic precession hypothesis does not
by itself predict a relation between $\nuap$ and $\nuk$ (see
the discussion in Section~\ref{intro}). In order to create a
relation between $\nuap$ and $\nuk$, SV proposed that some
physical mechanism keeps $\rp$ constant in a given source
while $\ra$ varies. They suggested that $\rp$ may remain
constant because it is equal to the radius $\risco$ of the
innermost stable circular orbit, the radius $r_m$ at which
the magnetic field of the star first couples to the
accreting gas, or the equatorial radius $\re$ of the neutron
star. However, there is no physical reason why the
periastron radius of an eccentric geodesic cannot be
significantly smaller than $\risco$ (see Markovi\'c 2000).
The magnetic coupling radius decreases as the mass accretion
rate increases (Lamb 1989; Ghosh \& Lamb 1992; Miller et
al.\ 1998; Psaltis et al.\ 1999a). Moreover, if the stellar
magnetic field is strong enough to truncate orbits at $r_m$,
the motion of the accreting gas will not be purely geodesic
motion for a substantial distance outside $r_m$. The
equatorial radius $\re$ provides a lower bound on $\rp$, but
otherwise provides no constraint.

In exploring sequences of geodesics with finite
eccentricities, we simply followed SV, who assumed that
$\rp$ is constant in a given source but that it can be
chosen freely for each source to give the best possible fit
of the $\nuap$-$\nuk$ and $2\nunp$-$\nuk$ relations to the
observed $\nu_1$-$\nu_2$ and $\nul$-$\nuk$ correlations. We
also assumed that for each source $\ra$ can be varied freely
to give the best possible agreement between the allowed
frequency relation and the observed frequency correlation,
requiring only that it exceed $\rp$. Therefore the
parameters in this model are the EOS, mass, and spin
frequency of the neutron star, the periastron radius $\rp$,
and an apastron radius for each pair of kilohertz QPO
frequencies that was fit.

We found that the $\nuap$-$\nuk$ relations given by the
best-fitting sequences of moderately eccentric geodesics,
defined as geodesics with $\ra/\rp \lta 1.5$, also disagree
qualitatively with the kilohertz QPO frequency data. As we
explained in detail in Section~\ref{CompareMEGs} and the
Appendix, SV came to a different conclusion in their paper
proposing the geodesic precession model because they
computed the azimuthal frequency $\nuk$ incorrectly. When
computed correctly, the azimuthal frequencies of the
geodesics cited by SV as giving agreement with the \sco1
frequency data miss most of the data points by
$\sim$100\,Hz. {\em Our investigation shows that the
best-fitting sequences of moderately eccentric geodesics
disagree qualitatively not only with the kilohertz QPO
frequency data on \sco1 but also with the data on other
sources, regardless of the neutron star model and spin rate
assumed}. The reason sequences of such geodesics are unable
to fit the data is that although the frequency difference
$\Delta\nu \equiv \nu_2 - \nu_1$ decreases with increasing
$\nu_2$, the radial epicyclic frequency $\nu_r = \nuk-\nuap$
given by such geodesics decreases much more rapidly with
increasing $\nuk$ for any realistic neutron star and any
allowed spin rate.

\subsubsection{Highly eccentric geodesics}

In Section~\ref{CompareHEGs} we explored whether dropping
the requirement that geodesics be only moderately eccentric
would allow acceptable fits to the kilohertz QPO frequency
data. We were able to find sequences of highly eccentric
geodesics that give $\nuap$-$\nuk$ relations qualitatively
similar to the frequency data on some sources, but only by
using geodesics with very high eccentricities ($\ra/\rp \gta
3$). We were able to find geodesic sequences that agree
qualitatively with the frequencies observed in \sco1 and
\fu{1728$-$34}, provided that the EOS of neutron star matter
is similar to the UU or A18$+$UIX$+\delta v_b$ equations of
state, although the fits are not formally acceptable. In
general, frequency relations similar to the observed
frequency correlations are possible only if values of $\rp$
smaller than $\risco$ are allowed. We were able to find
sequences of highly eccentric geodesics that give formally
acceptable fits to the \gx{17$+$2} and \gx{340$+$0}
kilohertz QPO frequencies; however, only a few frequency
measurements have been reported for these sources and the
uncertainties are large. The best-fitting geodesic sequence
appears to be inconsistent with the \gx{5$-$1} frequency
data. In Section~\ref{otherFreqs} we showed that geodesic
frequencies other than $\nuap$ and $\nuk$ are inconsistent
with the frequencies of the kilohertz QPOs.

In order to test the suggestion of SV and SVM that the
low-frequency QPOs in the atoll sources and the HBOs in the
Z sources are the first overtones of nodal precession
frequencies, in Section~\ref{CompareNP} we investigated
whether the nodal precession frequencies of the sequences of
best-fitting highly eccentric geodesics found in
Section~\ref{CompareHEGs} are consistent with the frequency
behaviour of the low-frequency QPOs. We found that the
predicted $\nunp$-$\nuk$ and $2\nunp$-$\nuk$ relations are
qualitatively different from the $\nul$-$\nuk$ and
$\nuhbo$-$\nuk$ correlations observed in \fu{1728$-$34},
\gx{5$-$1}, and \sco1. The predicted $2\nunp$-$\nuk$
relation is somewhat similar to the frequency correlation
observed in \gx{17$+$2}, if the spin rate of this neutron
star is 660~Hz. However, the value of $\chisqdof$ is large
(4.3) and the required stellar mass is equal to the maximum
stable mass for this spin rate.

Finally, in Section~\ref{CompareCir1} we considered the
unusual source \cir1, which is thought to be a neutron star
(see van der Klis 2000). The frequencies $\nul$ and $\nuh$
of its low-frequency QPO and high frequency `hump' both vary
by more than an order of magnitude (their reported ranges
are 1--12~Hz and 20--200~Hz, respectively; see Tennant,
Fabian \& Shafer 1986; Tennant 1987; Shirey 1998; Shirey et
al.\ 1996, 1998, 2000). This makes Cir~X-1 a crucial testing
ground for any model that seeks to explain the neutron star
and black hole QPOs within a single framework. Indeed,
without the Cir~X-1 data, the evidence for a connection
between the black hole and neutron star QPOs would be much
weaker (see Fig.~2 of Psaltis et al.\ 1999b).

We first investigated whether a sequence of moderately
eccentric geodesics can be constructed that gives a
$2\nunp$-$\nuap$ relation similar to the $\nul$-$\nuh$
correlation observed in \cir1 ($\nunp$ is much too low to be
consistent with $\nul$). We find that sequences of
moderately eccentric geodesics around rapidly spinning
($\nus \gta 650\,$Hz) neutron stars can give
$2\nunp$-$\nuap$ relations that are roughly consistent with
the observed $\nul$-$\nuh$ correlation for $\nuh<100\,$Hz.
However, only moderately eccentric geodesics around very low
mass ($M \alt 1.3\,\msun$ for EOS UU) neutron stars spinning
near breakup give relations roughly consistent with both the
lower-frequency data and the $\nul \approx 12\,$Hz, $\nuh
\approx 200\,$Hz measurement by Tennant (1987). No
quantitative comparison is possible, because no errors have
been reported for most of the frequency measurements. We
then explored whether using sequences of highly eccentric
geodesics would make it easier to construct a frequency
relation similar to that observed in \cir1. We found that
the situation for highly eccentric geodesics is similar to
that for moderately eccentric geodesics. The lower-frequency
data favour moderately eccentric geodesics that are
extremely eccentric ($\ra/\rp \agt 6$) and far from the
neutron star ($\ra/\re \agt 10$). Inclusion of the
higher-frequency measurement of Tenant (1987) again makes it
much more difficult to find a sequence of geodesics that
gives a frequency relation similar to the observed
correlation. Only highly eccentric geodesics around stars
with very low masses and very high spin rates give frequency
relations that pass near all the measurements.

In the majority of cases, the values of $\rp$ required to
produce $\nuap$-$\nuk$ relations similar to the observed
kilohertz QPO frequency corelations are substantially larger
than the equatorial radius $\re$ of the corresponding
neutron star model (for example, in \gx{340$+$0} the
required value of $\rp$ is more than $2\re$). Where a
marginally bound geodesic exists, $\rp$ is also
significantly larger than $\rmb$ for the best-fit HEG
sequences. Hence the geodesic precession hypothesis does not
provide any obvious explanation of why the periastron radius
should have the required value or of why it should remain
fixed in a given source. As discussed in
Section~\ref{frequencyRelations}, it seems impossible to
overcome this difficulty for any realistic EOS.

The situation for \cir1 is even more serious, because
geodesics with apastron radii $\ra \sim 5$--$50\,\re$ are
required to fit the frequency data. Clumps so far from the
star are unlikely to produce significant X-ray modulation.

\subsection{Expected frequencies and power spectra}
\label{spectraDiscussion}

Up to this point in our analysis we simply assumed that the
dominant modulation frequencies produced by clumps moving on
geodesics around a neutron star would be $\nuk$, $\nuap$,
and $2\nunp$, as proposed by SV, SVM, and Karas (1999). In
Section~\ref{spectra} we carried out extensive numerical
simulations to determine the waveforms and power spectra
that would actually be produced by geodesic motion of
clumps. We considered six effects: occultation of the whole
stellar surface, occultation of a bright equatorial band,
reflection of radiation by clumps, two models of emission of
radiation by clumps, and interaction of clumps with a
rarified gaseous disc in the rotation equator of the star.
Any orbiting clump model will generally produce a power
spectrum that is a superposition of the power spectra
produced by these effects.

We find that, contrary to what was simply assumed
previously, {\em $\nuk$, $\nuap$, and $2\nunp$ are not the
most prominent frequencies generated by orbiting clumps}.
Instead, most of the power generated by such clumps is
typically in peaks of roughly equal power at the radial
epicyclic frequency $\nur$, at $\nuk+\nur$, and at
$\nuk+2\nur$ (see Table~\ref{table.frequencies}). As
explained in Section~\ref{spectra}, the reason for this is
that for the general relativistic geodesics that are
relevant, the only strictly periodic motion is the radial
epicyclic motion. In particular, the azimuthal motion is not
periodic. There is usually power at $\nuk$, but it is always
$\alt25$\% of the power in each of the dominant peaks and
there are often many other peaks as strong or stronger. For
relevant geodesics, the power at $\nuap$ and at $\nunp$
produced by most mechanisms is $\alt10$\% of the power in
the dominant peaks. We were unable to find any effect or
geometry that produces a significant peak at $2\nunp$. Thus,
the frequencies of the significant peaks in the power
spectra that would be generated by orbiting clumps are
qualitatively different from the frequencies of the peaks
observed in power spectra of the kilohertz QPO sources.

Periodic modulation of the X-ray emission by orbiting clumps
is possible only if the inclination $i$ of the system, which
is expected to be the same as the inclination of the star's
spin axis and the normal to the accretion disc, is
sufficiently large. More generally, the frequency structure
and the relative strengths of the oscillations at different
frequencies produced by this mechanism depend sensitively on
the inclination of the system. However, there is no evidence
for any inclination-dependence of the frequency pattern in
the low- or high-frequency QPOs, or even in their relative
amplitudes. This is difficult to understand in orbiting
clump models.

\subsection{Dynamical considerations}
\label{constraintsDiscussion}

In Section~\ref{dynamics} we investigated the dynamical
constraints on orbiting clump models and the X-ray
modulation generated by such models. We find that in order
for clumps to persist in geodesic motion for $\sim$100
orbits, as required by the coherence of the kilohertz QPOs,
their radii must be $\ll 10^{-3}$ of the periastron radius.
In order to avoid rapid inspiral caused by drag forces, the
density of the gas in the clumps must be $\gg 10^5$ times
the density of the interclump medium. In order for clumps to
avoid colliding with each other too frequently, the density
of the gas in the clumps must be $\gg 10^6$ times the mean
density near the star. These densities are extreme.

As discussed in Section~\ref{dynamics}, the small size
required for clumps to persist as long as required by the
observed coherence of the kilohertz QPOs means that the
fractional modulation of the X-ray flux that can be produced
by clumps is $\ll10^{-6}$.  This upper bound is many orders
of magnitude smaller than the observed fractional
amplitudes, which are $\sim$10$^{-2}$--10$^{-1}$ (van der
Klis 2000).

The very low interclump gas density and the small covering
factor by clumps that is required by the geodesic precession
hypothesis means that radiation can escape from the clumps
and from the stellar surface without suffering significant
scattering or absorption. Hence scattering or absorption can
have little effect on the frequency content of the waveforms
produced by the clumps. Therefore all the peaks at all the
frequencies we have discussed should be observable. However,
most have not been detected (see, e.g., Jonker et al. 1998).

A related point is that even weakly nonuniform emission from
the stellar surface should produce readily observable
periodic X-ray oscillations at the stellar spin rate. Again,
such oscillations have not been detected (see van der Klis
2000).

A major deficiency of the geodesic precession model is that
no attempt has been made to explain what causes the gas in
the accretion disc, which is moving along nearly circular
streamlines, to break up into clumps moving on highly
eccentric geodesics. This would require the sudden removal
of a substantial fraction of the angular momentum of the gas
in the clumps. Nor has there been any attempt to explain why
and how clumps are formed only on geodesics with nearly the
same values of $\ra$ and $\rp$, why these special geodesics
have similar frequencies and obey similar frequency
relations in sources with very different properties (e.g.,
accretion rates that differ by factors $\sim$30), or why the
properties of the special geodesics change with the
accretion rate in just the way needed to explain the
observed changes in the frequencies of the QPOs. Finally, no
explanation has been given of how the clumps dissipate and
how the accreting gas reaches the stellar surface (injection
of clumps onto geodesics that do not intersect the stellar
surface does not by itself produce any accretion onto the
star).

\subsection{Conclusions} \label{conclusions}

We conclude that there are significant difficulties in
interpreting the kilohertz and low-frequency QPOs in the
atoll and Z sources as consequences of gas clumps moving along geodesics.

The best-fitting geodesic sequences that can be constructed
using moderately eccentric geodesics give frequency
relations that differ qualitatively from the kilohertz QPO
frequency data. Frequency relations qualitatively similar to
the kilohertz QPO data on some sources can be constructed,
but only if highly eccentric geodesics are used. Such
geodesics are very difficult to accommodate in any
physically consistent gas dynamical picture of the accretion
flow near the star. The best-fitting $2\nunp$-$\nuk$ and
$\nuap$-$\nuk$ relations disagree qualitatively with the
low-frequency and kilohertz QPO frequency data, for any
realistic neutron star models and allowed spin frequencies.
Other geodesic frequencies are qualitatively inconsistent
with the observed frequencies.

Modulation by orbiting clumps would not produce strong peaks
in the power spectra at the frequencies $\nuap$, $\nuk$, and
$2\nunp$ as required by the geodesic precession
hypothesis. Instead, the strongest peaks would be at the
radial epicyclic frequency $\nur$ and at $\nuk+\nur$ and
$\nuk+2\nur$, frequencies that are inconsistent with the
frequencies observed.

Finally, the dynamical constraints on orbiting clump models
make it very difficult to see how clumps of the required
density could be formed on highly eccentric geodesics, or
how X-ray modulation with the amplitudes observed could be
generated by orbiting clumps.

\subsection*{Acknowledgements} The authors thank Mariano
M\'endez for providing observational data and for many
stimulating discussions. The authors are grateful to
Dimitrios Psaltis and Luca Zampieri for extensive, detailed
discussions of the precession hypothesis and fits to the
data, and to Mariano M\'endez and Cole Miller for comments
on the paper. The authors also thank Scott Hughes, Michiel
van der Klis, Tom Loredo, Cole Miller, Sharon Morsink, Luigi
Stella, Jean Swank, and William Zhang for helpful
conversations. The authors gratefully acknowledge the
hospitality of the Aspen Institute for Physics, where many
of these discussions took place. This work was supported in
part by NSF grant AST~96-18524 and by NASA grants NAG~5-2925
and NAG~5-8424.

\appendix
\section{Keplerian (Azimuthal) frequencies of eccentric
geodesics}

In order to clarify the difference between the definition of
the Keplerian (azimuthal) frequency of eccentric geodesics
and the quantity used by Stella \& Vietri (2000; hereafter
SV), we first discuss the definition of the Keplerian
frequency, then derive expressions for the radial epicyclic
and Keplerian frequencies of eccentric geodesics in the
Schwarzschild spacetime, and finally explain how we were
able to reproduce the curve shown by SV in their Fig.~2.
The expressions needed are given by Markovic (2000).
However, in order to facilitate comparison with the
expressions given by SV, in this appendix we use the
notation of SV, whose derivation follows closely the
discussion in Chandrasekhar (1983, pp. 100--105).

\subsection{Definition of the Keplerian frequency}

For a test particle moving on an eccentric geodesic, the
rate of azimuthal phase advance varies with its radius, and
hence only the {\em average\/} rate of azimuthal phase
advance is meaningful. This is the quantity conventionally
denoted $2\pi\nuphi$ or $ 2\pi\nuk$. In the Schwarzschild
spacetime, eccentric geodesics are not closed (because of
apsidal precession), the azimuthal motion of a particle
moving on such a goedesic is not periodic, and the average
rate of azimuthal phase advance is not proportional to the
inverse of the time required for the particle to traverse a
$2\pi$ interval in azimuthal phase (see, e.g., Section~14 of
Landau \& Lifshitz 1976). However, the radial motion is
periodic, with period $1/\nu_r$, where $\nu_r$ is the radial
epicyclic frequency. The rate at which a particle's
azimuthal phase advances depends only on its radius and
hence is also periodic with period $1/\nu_r$. The average
rate at which a particle's azimuthal phase advances can
therefore be computed by calculating the phase advance
during any time interval that is a multiple of half the
radial epicyclic period and then dividing by the time
interval. For example, if $\Delta\phi$ is the azimuthal
phase advance during each radial epicyclic period, then
$\nuphi = (\Delta\phi/2\pi)\, \nu_r$.

Instead of using the usual definition of $\nuphi$, SV appear
to have assumed that $\nuphi$ is the inverse of the time
required for the azimuthal phase to advance by $2\pi$.
However, as just explained, the rate of azimuthal phase
advance of a particle moving on an eccentric geodesic in the
Schwarzschild spacetime depends on the instantaneous radius
of the particle and the azimuthal motion is not periodic.
Hence the time required for the particle's azimuthal phase
to increase by $2\pi$ depends on the assumed initial
position of the particle and is not even well-defined; it is
not the inverse of the Keplerian frequency. As Fig.~A1
shows, the time required for the azimuthal phase to advance
by $2\pi$ depends sensitively on the starting position used
in computing it, even when the eccentricity is relatively
small. The non-uniqueness of this time is a consequence of
the nonperiodic nature of the azimuthal motion. (The
nonperiodic nature of the azimuthal motion is also the
reason why the peak at $\nuphi$ in power spectra of the
X-ray flux modulation produced by eccentric geodesic motion
is not very prominent.)

\begin{figure*}
\vspace{-0.7 truecm}
\centerline{\epsfxsize=14 truecm
              \hspace{+4.0 truecm}
 \epsfbox{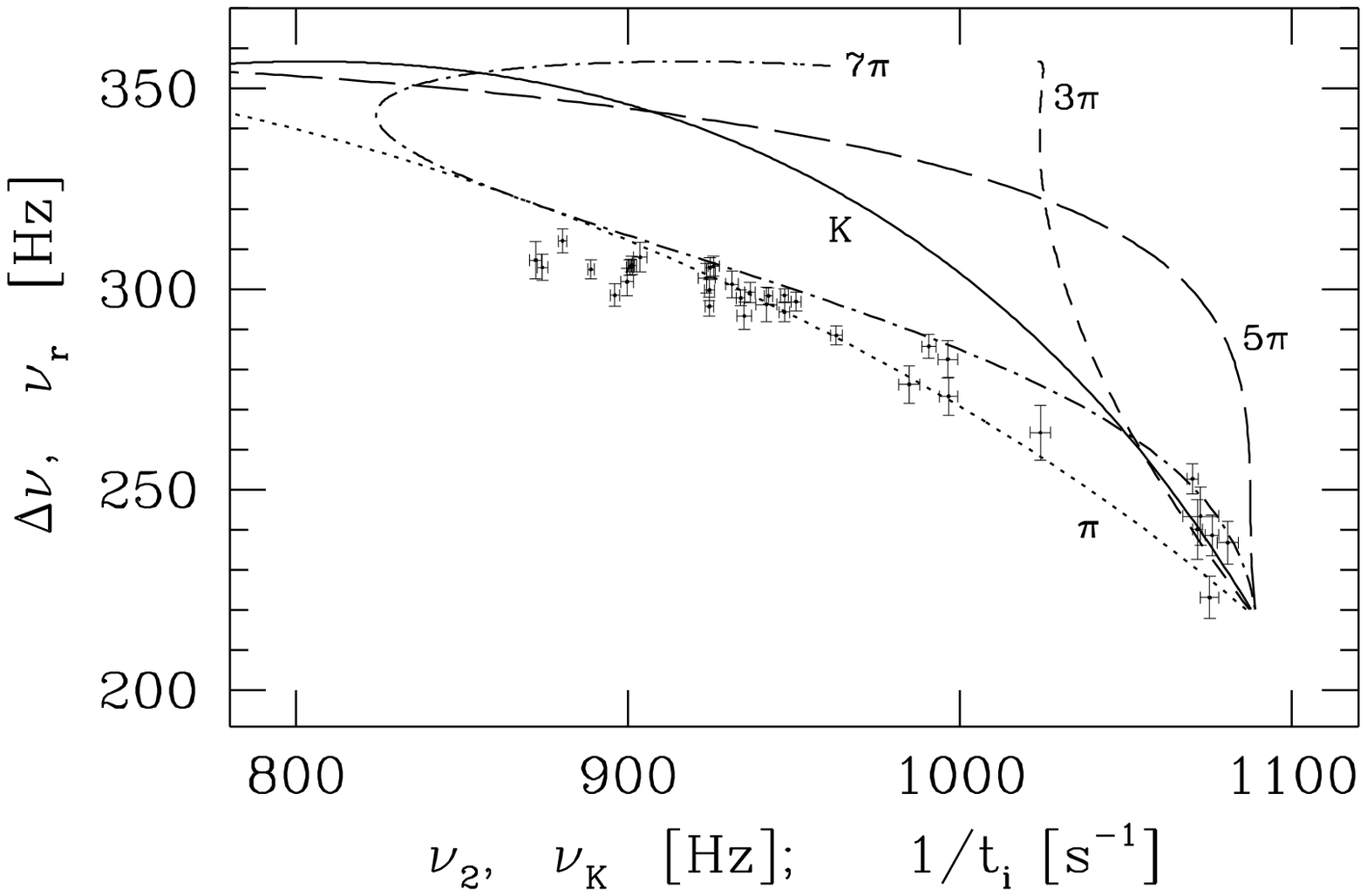}\hspace{-4.6 truecm}
 \epsfxsize=14 truecm
 \epsfbox{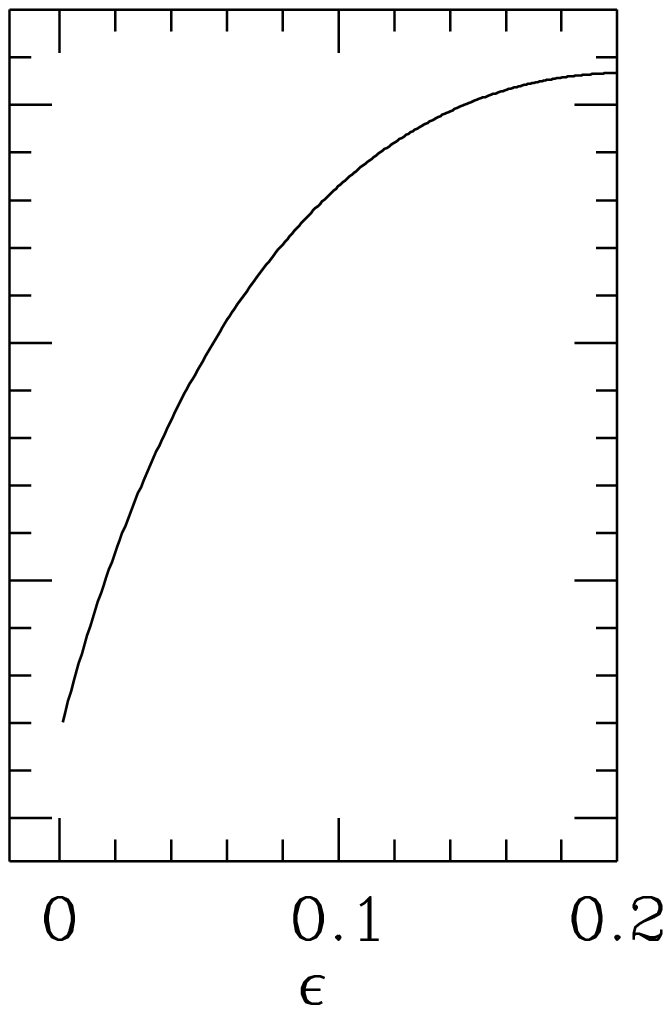}}
\vspace{-5.5 truecm}
\caption[]{The data points in the left panel show the
$\Delta\nu$-$\nu_2$ correlation observed in \hbox{Sco\,X-1}.
The solid curve labeled K in the left panel shows how the
radial epicyclic frequency $\nu_r$ varies with the Keplerian
(azimuthal) frequency $\nu_{\rm K}$ for the sequence of
geodesics specified in the caption of Fig.~2 of Stella \&
Vietri (1999). For these geodesics, the eccentricity
$\epsilon$ increases with increasing $\nur$, as shown in the
right panel. The dotted curve labeled $\pi$ in the left
panel shows $\nu_r$ as a function of $1/t_\pi$, the inverse
of the time required for a particle to advance in azimuthal
phase from $\pi$ radians before apastron to $\pi$ radians
after apastron. The short dashed curve shows $\nu_r$ as a
function of $1/t_{3\pi}$, the inverse of the time required
to advance from $\pi$ radians after apastron to $3\pi$
radians after apastron. The long dashed curv shows $\nu_r$
as a function of $1/t_{5\pi}$, the inverse of the time
required to advance from $3\pi$ radians after apastron to
$5\pi$ radians after apastron. The dash-dotted curve shows
$\nu_r$ as a function of $1/t_{7\pi}$, the inverse of the
time required to advance from $5\pi$ radians after apastron
to $7\pi$ radians after apastron.}
\end{figure*}

\subsection{Calculation of the Keplerian and radial
epicyclic frequencies $\nuk$ and $\nur$}

For a non-rotating ($j=0$) spherical star,
equations~(20)--(22) of Markovic (2000), which describe the
geodesic motion of a particle of unit proper mass, simplify
considerably. Following Chandrasekhar (1983), we use the
radial variable $u\equiv 1/r$, where $r$ is the radius in
Schwarzschild coordinates. Then (see Chandrasekhar 1983,
Section~19; Shapiro \& Teukolsky 1983, Ch.~12)
 \begin{eqnarray}
 \label{motion}
 \dot{u}^2 &=& \left(2 L^2 u^3 - L^2 u^2 + 2u +E^2 -1\right) u^4
                \nonumber \\
 \dot{t} &=& \frac{E}{1-2u}
                \nonumber \\
 \dot{\phi} &=& L u^2,
 \end{eqnarray}
where $E$ and $L$ are, respectively, the conserved energy
and angular momentum of the particle. Here all distances are
in units of $GM/c^2$ and all times, in units of $GM/c^3$.

From equations~(\ref{motion}) it follows that
 \begin{equation}
 \label{dudphi}
 \left(\frac{du}{d\phi}\right)^2 =
                    2 (u-u_1) (u-u_2) (u-u_3),
 \end{equation}
where $u_1 = 1/\ra$, $u_2 = 1/\rp$, and  $u_3 > u_2 > u_1$.
Introducing $\mu \equiv 1/l$ and evaluating the coefficients
of the polynomials in $u$ in equations~(\ref{motion}) and
(\ref{dudphi}), one finds
 \begin{equation}
 \label{us}
 u_1 = \frac{1-\epsilon}{l}\;, \hspace{0.6cm}
 u_2 = \frac{1+\epsilon}{l}\;, \hspace{0.6cm}
 u_3 = \frac{1}{2} - \frac{2}{l}\;,
 \end{equation}
and
 \begin{eqnarray}
 \label{coeffs}
 \frac{1}{L^2} &=& \frac{1}{l} \left[ 1 -\mu(3 +\epsilon^2)\right]
                        \nonumber \\
 \frac{E^2}{L^2} &=& \frac{1}{l}\left[ (1-2\mu)^2 - 4\epsilon^2 \mu^2\right],
 \end{eqnarray}
where $l \equiv 2/(u_1 + u_2)$ and $\epsilon \equiv (\ra -
\rp)/(\ra + \rp) = (u_2 - u_1)/(u_2 + u_1)$ is the
eccentricity.

From equations~(\ref{motion})--(\ref{coeffs}), we obtain
\begin{eqnarray}
 \label{dchidphi}
 \left(\frac{d\chi}{d\phi}\right)^2
  &=& 1- 2\mu (3 + \epsilon\cos\chi) \nonumber \\
  &=& (1 -6\mu + 2\epsilon\mu) \left[ 1 - \kappa^2  \cos^2 (\chi/2)\right]\;,
 \end{eqnarray}
where $\kappa^2 \equiv4\epsilon\mu/(1 - 6\mu +2\epsilon\mu)$
and $\chi$ is defined implicitly by
 \begin{equation}
 \label{chi.def}
 u \equiv \frac{1 + \epsilon\cos\chi}{l} \;.
 \end{equation}
Equations~(\ref{motion}) and~(\ref{dchidphi}) yield
 \begin{eqnarray}
 \label{dtdchi}
 \frac{dt}{d\chi}
  &=& \frac{dt}{d\phi}\frac{d\phi}{d\chi}
      \nonumber \\
  &=& \left(\frac{E}{1-2u}\right)\left(\frac{1}{Lu^2}\right)
      \frac{1}{\left[1- 2\mu (3 +\epsilon \cos\chi)\right]^{1/2}}
      \nonumber \\
  &=& \frac{l^{3/2}\left[(1-2\mu)^2 - 4\epsilon^2 \mu^2\right]^{1/2}
      (1 + \epsilon\cos\chi)^{-2}}
      { [1- 2\mu (1 +\epsilon \cos\chi)]
       [1- 2\mu (3 +\epsilon \cos\chi)]^{1/2}} \;.
      \nonumber \\
\end{eqnarray}

As $\chi$ increases from $0$ to $2\pi$, the particle
executes one full period of its {\em radial\/} motion,
moving from its periastron radius $r_{\rm p} = 1/u_2$ to the
apastron radius $r_{\rm a} = 1/u_1$ and back to $r_{\rm p}$
again. Hence the radial epicyclic frequency $\nur$ is given
by
 \begin{equation}
 \label{nur.def}
 \frac{1}{\nu_r}
  = \int_0^{2\pi} \left(\frac{dt}{d\chi}\right)d\chi \;.
 \end{equation}
One can relate $\nu_r$ to the Newtonian frequency $\nu_{\rm
N} = a^{-3/2}/2\pi$ using the expression $a = (1/u_1 +
1/u_2)/2 = l/(1-\epsilon^2)$ for the semi-major axis $a$ of
the geodesic. The result is
 \begin{equation}
 \label{nur}
 \nu_r = 2\pi\nu_{\rm N} \frac{1}{(1 - \epsilon^2)^{3/2}}
        \frac{1}{\left[ (1-2\mu)^2 - 4\epsilon^2 \mu^2\right]^{1/2}}
            \frac{1}{f(\epsilon,\mu)} \;,
 \end{equation}
where
 \begin{eqnarray}
 \label{f.def}
 f(\epsilon,\mu) \hspace{-0.2cm} &\equiv& \hspace{-0.2cm}
 \int^{2\pi}_0
 \frac{(1 + \epsilon\cos\chi)^{-2} d\chi}
        { [1- 2\mu (1 +\epsilon \cos\chi)]
           [1- 2\mu (3 +\epsilon \cos\chi)]^{1/2}}\;.
  \nonumber \\ & &    \hspace{-2cm}
 \end{eqnarray}
As the particle executes one period of its radial epicyclic
motion, its azimuth $\phi$ increases by (see
eq.~\ref{dchidphi})
 \begin{equation}
 \label{delta.phi}
 \Delta\phi =
 \frac{1}{\sqrt{1 -6\mu + 2\epsilon\mu}}
   \int_0^{2\pi} \frac{d\chi}{
     \left[ 1 - \kappa^2 \cos^2 (\chi/2)\right]^{1/2}} \;.
 \end{equation}
The mean rate of advance of the particle's azimuth $\phi$ is
$2\pi\nu_{\phi} = \nu_{r} \Delta\phi$. Hence
 \begin{equation}
 \label{nuphi}
 \nu_{\phi} = \nu_r \frac{\Delta\phi}{2\pi}\;.
 \end{equation}
Quite generally, $\Delta\phi > 2\pi$, so the geodesics are
not closed.

\subsection{Calculation of SV}

Equations~(\ref{nur}) and ~(\ref{f.def}) for the radial
epicyclic frequency $\nur$ are the same as equations~(2)
and~(3) given by SV. However, instead of computing the
azimuthal frequency $\nu_\phi$ given by
equations~(\ref{delta.phi}) and~(\ref{nuphi}), SV appear to
have computed the time required for a particle to advance
through an azimuthal interval of $2\pi$ centered about the
azimuthal phase of its apastron and then to have identified
the inverse of this time with $\nuphi$, which it is not.

If a particle has $\chi=\chi_0$ initially, its azimuthal
phase will have increased by $2\pi$ at $\chi=\chi_1$, where
$\chi_1$ is given implicitly by
 \begin{equation}
 \label{chic.def}
 2\pi = \frac{1}{\sqrt{1 - 6\mu + 2\epsilon\mu}}
        \int_{\chi_{_0}}^{\chi_{_1}} \frac{d\chi}{
        \left[ 1 - \kappa^2 \cos^2 (\chi/2)\right]^{1/2}}\;.
 \end{equation}
In the Schwarzschild spacetime, the radial motion of a
particle is always slower than its azimuthal motion and
hence $\chi_1 - \chi_0 < 2\pi$. As noted above, the rate at
which the particle's azimuthal phase advances depends on its
radius and its motion in the azimuthal direction is
nonperiodic. Hence, the time required for the azimuthal
phase of a particle to advance by $2\pi$ depends on its
initial position, i.e., on $\chi_0$.

According to SV, the azimuthal frequency is given by the
inverse of their equation~(2) (our eq.~[\ref{nur}]), but
with $f(\epsilon,\mu)$ given by their equation~(3) (our
eq.~[\ref{f.def}]), the integral in their equation~(3) taken
over 0 to $\chi_c$ rather than 0 to $2\pi$, and $\chi_c$
given implicitly by their eq.~(4). This `azimuthal
frequency' is the inverse of the time required for the
particle to traverse the azimuthal phase interval of $2\pi$
centered on the particle's apastron, if the integration in
SV's eq.~(3) is meant to be over $\pi - 2\chi_c$ to $\pi +
2\chi_c$, rather than 0 to $\chi_c$, and their expression
for $k$ is instead an expression for $k^2$.\footnote{SV
follow Chandrasekhar's (1983) notation, but use $x$ instead
of $\chi$. Hence they use the variable $\gamma \equiv \pi/2
- \chi/2$ rather than $\chi$ in their eq.~(4), which
corresponds to our eq.~(\ref{chic.def}). Also, they write
their eq.~(4) in terms of the integral over only the first
half of the $2\pi$ phase interval centered at the particle's
apastron.} Indeed, we were able to reproduce the $\nur$ vs.\
`azimuthal frequency' curve shown in Fig.~2 of SV (see
below) by using as the `azimuthal frequency' the inverse of
the time required for $\chi$ to increase from $\chi_0^*$ to
$\chi_1^*$, where $\chi_1^* - \pi = \pi - \chi_0^*$.
($\chi_0^*$ is related to $\chi_c$ by $\chi_0^* = \pi -
2\chi_c$.) This is the inverse of the time required for the
particle to traverse the $2\pi$ phase interval centered on
its apastron phase (see eq.~[\ref{chi.def}]). Fig.~A1 shows
that the times required for the azimuthal phase of the
particle to advance by successive additional intervals of
$2\pi$ are all different from one another.

The rate of azimuthal phase advance is smallest at apastron,
and hence the azimuthal phase interval used by SV gives the
smallest possible inverse time and thus the flattest
possible $\nu_r$-inverse time curve. This flattest possible
curve is shown as the dotted curve in Fig.~A1, for the
parameters Stella \& Vietri list in the caption of their
Fig.~2. Fig.~A1 shows that any other choice of inverse time
would have given a $\nu_r$--inverse time curve much further
from the \sco1 data. These inverse times are not relevant in
any case, because none of them are equal to the azimuthal
frequency of the particle. As shown in
Section~\ref{CompareMEGs}, the best-fitting $\nu_r$-$\nuphi$
relation is qualitatively different from the correlation
between the kilohertz QPO frequencies observed in \sco1.


\begin{thebibliography}{99}

\bibitem[\protect\citename{Akmal, Pandharipande \& Ravenhall
}1998]{Akmal98} Akmal, A., Pandharipande, V. R. \&
Ravenhall, D. G. 1998, Phys. Rev. C, 58, 1804

\bibitem[\protect\citename{Alpar \& Shaham }1985]{Alpar85}
Alpar, A., \& Shaham, J. 1985, Nature, 316, 239

\bibitem[\protect\citename{Armitage \& Natarajan
}1999]{Armitage98} Armitage, P.J., Natarajan, P. 1999, ApJ,
525, 909

\bibitem[\protect\citename{Arnett \& Bowers }1977]{Arnett}
Arnett, W.D., Bowers, R.L. 1977, ApJ Suppl, 33, 415

\bibitem[\protect\citename{Bethe \& Johnson }1974]{C} Bethe
H.A., Johnson, M.B. 1974, Nucl. Phys. A, 230, 1

\bibitem[\protect\citename{Bildsten }1998]{Bildsten98}
Bildsten, L. 1998, in The Many Faces of Neutron Stars, NATO
ASI, Lipari (Dordrecht: Kluwer), C515, 419

\bibitem[\protect\citename{Chandrasekhar }1983]{Chan83}
Chandrasekhar, S. 1983 The Mathematical theory of Black
Holes.  Oxford University Press, Oxford

\bibitem[\protect\citename{Cook, Shapiro \& Teukolsky
}1992]{Cook92} Cook, G.B., Shapiro, S.L., Teukolsky, S.A.
1992, ApJ, 398, 203

\bibitem[\protect\citename{Cook, Shapiro \& Teukolsky
}1994a]{Cook94a} Cook, G.B., Shapiro, S.L., Teukolsky, S.A.
1994a, ApJ, 422, 227

\bibitem[\protect\citename{Cook, Shapiro \& Teukolsky
}1994b]{Cook94b} Cook, G.B., Shapiro, S.L., Teukolsky, S.A.
1994a, ApJ, 424, 823

\bibitem[\protect\citename{Di Salvo et al.\
}2000]{DiSalvo00} Di Salvo, T., M\'endez, M., van der Klis,
M., Ford, E., Robba, N.R. 2000, preprint

\bibitem[\protect\citename{Jonker et al.\ }1998]{GX340}
Jonker, P.J., et al. 1998, ApJ, 499, L191

\bibitem[\protect\citename{Kaaret et al.\ }1999]{Kaaret99}
Kaaret., P., et al. 1999, ApJ, 523, 197

\bibitem[\protect\citename{Kalogera \& Psaltis} 1999]{kp99}
Kalogera, V., Psaltis, D. 1999, Phys. Rev. D, 61, 024009

\bibitem[\protect\citename{Karas }1999]{Karas99a} Karas, V.
1999, ApJ, 526, 953 (K)

\bibitem[\protect\citename{Komatsu, Eriguchi \& Hachisu
}1989a]{Kom89a} Komatsu, H., Eriguchi, Y., Hachisu, I 1989,
MNRAS, 237, 355

\bibitem[\protect\citename{Komatsu, Eriguchi \& Hachisu
}1989b]{Kom89b} Komatsu, H., Eriguchi, Y., Hachisu, I. 1989,
MNRAS 239, 153

\bibitem[\protect\citename{Lamb \& Miller }2000]{LM2000}
Lamb, F.K, Miller, M.C. 2000, in preparation

\bibitem[\protect\citename{Lamb et al.\ }1985]{Lamb85} Lamb,
F.K., Shibazaki, N., Alpar, A., \& Shaham, J. 1985, Nature,
317, 681

\bibitem[\protect\citename{Landau \& Lifshitz }1983]{LL76}
Landau, L.D., Lifshitz, E.M.  1976 Mechanics.
Butterworth-Heinemann

\bibitem[\protect\citename{Markwardt, Strohmayer \& Swank
}1998]{1702} Markwardt, C.B., Strohmayer, T.E., Swank, J.H.
1998, ApJ, 512, L125

\bibitem[\protect\citename{Markovi\'c }1999]{M1999}
Markovi\'c, D. Companion paper.

\bibitem[\protect\citename{Markovi\'c \& Lamb }1998]{ML1998}
Markovi\'c, D., Lamb, F.K. 1998, ApJ, 507, 316

\bibitem[\protect\citename{M\'endez \& van der Klis
}1998]{4U1608} M\'endez, M., et al. 1998, ApJ, 505, L23

\bibitem[\protect\citename{M\'endez \& van der Klis
}1999]{4U1728} M\'endez, M., van der Klis, M. 1999, ApJ,
517, L51

\bibitem[\protect\citename{Miller }1999a]{Miller99a} Miller,
M.C. 1999a, ApJ, 515, L77

\bibitem[\protect\citename{Miller }1999b]{Miller99b} Miller,
M.C. 1999b, ApJ, 520, 256

\bibitem[\protect\citename{Miller \& Lamb }1993]{ML93}
Miller, M.C., Lamb, F.K. 1993, ApJ, 413, L43.

\bibitem[\protect\citename{Miller \& Lamb }1996]{ML96}
Miller, M.C., Lamb, F.K. 1996, ApJ, 470, 1033.

\bibitem[\protect\citename{Miller, Lamb \& Psaltis
}1998]{MLP98} Miller, M.C., Lamb, F.K., Psaltis, D. 1998,
ApJ, 508, 791

\bibitem[\protect\citename{Morsink \& Stella
}1999]{Sharon99} Morsink, S., Stella, L. 1999, ApJ, 513, 827

\bibitem[\protect\citename{Nozawa et al. }1998]{Nozawa98}
Nozawa, T., Stergioulas, N., Gourgoulhon, E., Eriguchi, Y.
1989, A\&A, 132, 431

\bibitem[\protect\citename{Pandharipande, Akmal \& Ravenhall
}1998]{PAR.98} Pandharipande, V. R., Akmal, A., Ravenhall,
D. G. 1998, in Nuclear Astrophysics, Proc. International
Workshop XXVI on Gross Properties of Nuclei and Nuclear
Excitations, ed. M. Buballa, N. No\={o}renberg J. Wambach,
\& A. Wirzba (Darmstadt: GSI), 11

\bibitem[\protect\citename{Pandharipande \& Smith }1975]{L}
Pandharipande, V.R., Smith, R.A. 1975, Phys. Lett., 59B, 15

\bibitem[\protect\citename{Psaltis et al. }1999]{PBvdK99}
Psaltis, D., Belloni, T., van der Klis, M. 1999b, ApJ, 520,
L262; astro-ph/9902130

\bibitem[\protect\citename{Psaltis, et al. }1998]{Psaltis98}
Psaltis, D., et al. 1998, ApJ, 501, L95

\bibitem[\protect\citename{Psaltis \& Norman }2000]{PN00}
Psaltis, D., Norman, C. 2000, {\tt astro-ph/0001391}

\bibitem[\protect\citename{Psaltis, et al. }1999]{Psaltis99}
Psaltis, D., et al. 1999a, ApJ, 520, 763

\bibitem[\protect\citename{Schaab \& Weigel
}1999]{Schaab.99} Schaab, C, Weigel, M.K. 1999, MNRAS 308,
718

\bibitem[\protect\citename{Shirey }1998]{Shirey.98} Shirey,
R.E. 1998, Ph.D. Thesis, MIT

\bibitem[\protect\citename{Shirey et al.
}1996]{Shirey.etal.96} Shirey, R.E., Bradt, H.V., Levine,
A.M., Morgan, E.H. 1996, ApJ, 469, L21

\bibitem[\protect\citename{Shirey et al.
}1998]{Shirey.etal.98} Shirey, R.E., Bradt, H.V., Levine,
A.M., Morgan, E.H. 1998, ApJ, 506, 374

\bibitem[\protect\citename{Stella \& Vietri }1998]{SV.98}
Stella, L., Vietri, M. 1998, ApJ, 492, L59

\bibitem[\protect\citename{Stella \& Vietri }1999]{SV.99}
Stella, L., Vietri, M. 1999, Phys. Rev. Lett., 82, 17 (SV)

\bibitem[\protect\citename{Stella, Vietri \& Morsink
}1999]{SVM.99} Stella, L., Vietri, M., Morsink, S. 1999,
ApJ, 524, L63 (SVM)

\bibitem[\protect\citename{Stergioulas \& Friedman
}1995]{Sterg.95} Stergioulas, N., Friedman, J.L. 1995, ApJ,
444, 306

\bibitem[\protect\citename{Strohmayer }1999]{Stroh99}
Strohmayer, T.E. 1999, ApJ, 523, L51

\bibitem[\protect\citename{Strohmayer et al. }1996]{Stroh96}
Strohmayer, T.E., Zhang, W., Swank, J.H., Smale, A.,
Titarchuk, L., Day, C. 1996, ApJ, 469, L9

\bibitem[\protect\citename{Strohmayer et al. }1998]{Stroh98}
Strohmayer, T.E., Zhang, W., Swank, J.H., Lapidus, I. 1998,
ApJ, 503, L147

\bibitem[\protect\citename{Tennant }1987]{Tennant.87}
Tennant, A.F. 1987, MNRAS, 226, 971

\bibitem[\protect\citename{Tennant, Fabian \& Shafer
}1986]{Tennant.86} Tennant, A.F., Fabian, A.C. Shafer, R.A.
1986, MNRAS, 219, 871

\bibitem[\protect\citename{van der Klis }1995]{vdK.95} van
der Klis, M. 1995, in X-Ray Binaries, ed. W.H.G. Lewin, J.
van Paradijs \& E.P.J. van den Heuvel (Cambridge Univ.
Press), 252

\bibitem[\protect\citename{van der Klis }1998]{vdK.98} van
der Klis, M. 1998 in the Proceedings of the Third William
Fairbank Meeting, Rome June 29--July 4 1998, {\tt
astro-ph/9812395}

\bibitem[\protect\citename{van der Klis }2000]{vdK.2000} van
der Klis, M. 2000, Ann. Rev. Astr. \& Astrophys., {\tt
astro-ph/0001167}

\bibitem[\protect\citename{van der Klis et al. }1997]{ScoX1}
van der Klis, M., Wijnands, R.A.D., Horne, K., Chen W. 1997,
ApJ, 481, L97

\bibitem[\protect\citename{Wijnands \& van der Klis
}1997]{1731} Wijnands, R., van der Klis 1997, ApJ, 482, L65

\bibitem[\protect\citename{Wijnands et al. }1997a]{1636b}
Wijnands, R. et al. 1997a, ApJ, 479, L141

\bibitem[\protect\citename{Wijnands et al. }1997b]{GX17}
Wijnands, R. et al. 1997b, ApJ, 490, L157

\bibitem[\protect\citename{Wijnands et al. }1998]{GX5}
Wijnands, R. et al. 1998, ApJ, 504, L35

\bibitem[\protect\citename{Wiringa, Fiks \& Fabrocini
}1988]{UU} Wiringa, R.B., Fiks, V., Fabrocini, A. 1988,
Phys. Rev. C, 38, 1010

\bibitem[\protect\citename{Zhang et al. }1996]{1636a} Zhang,
W. et al. 1996, ApJ, 469, L17

\bibitem[\protect\citename{Zhang et al. }1998]{1820a} Zhang,
W., Strohmayer, T.E., Swank, J.H. , ApJ, 500, L167

\end{thebibliography}
\end{document}